\documentclass[12pt]{article}
\usepackage{graphics,cite,amssymb,epsfig,float,psfrag}
\usepackage[usenames,dvips]{color}
\usepackage{rotating}

\newenvironment{appendletterA}
 {
  \setcounter{section}{0}
  \setcounter{equation}{0}
  
 }{
 }

\newcommand{\lsim}{\raisebox{-0.13cm}{~\shortstack{$<$ \\[-0.07cm] $\sim$}}~} 
\newcommand{\gsim}{\raisebox{-0.13cm}{~\shortstack{$>$ \\[-0.07cm] $\sim$}}~} 
\newcommand{\ra}{\rightarrow} 
 
\newcommand{\ee}{e^+e^-} 
\newcommand{\tb}{\tan \beta} 
 
\newcommand{\non}{\nonumber} 
\newcommand{\beq}{\begin{eqnarray}} 
\newcommand{\eeq}{\end{eqnarray}} 
\newcommand{\s}{\smallskip} 

\def\msbar{\overline{\rm MS}}
\def\be{\begin{equation}}
\def\ee{\end{equation}}

\def\gtilu{{\tilde g}_u}
\def\gtild{{\tilde g}_d}
\def\gtilup{{\tilde g}_u^{\,\prime}}
\def\gtildp{{\tilde g}_d^{\,\prime}}
\def\mgut{M_{\rm \scriptscriptstyle GUT}}
\def\msusy{M_S}
\def\mewsb{Q_{\rm \scriptscriptstyle W}}
\def\spsd{Split-SUSY}
\def\sps{Split SUSY}

\begin{document}
\baselineskip=15.5pt

\thispagestyle{empty}

\begin{flushright}
LPT--07-25\\
CERN--PH--TH/2007-074\\
\end{flushright}

\vspace{.5cm}

\begin{center}

{\Large\sc{\bf The MSSM with heavy scalars}}

\vspace*{9mm}
\setcounter{footnote}{0}
\setcounter{page}{0}
\renewcommand{\thefootnote}{\arabic{footnote}}

\mbox{ {\sc Nicol\'as BERNAL}$^1$, {\sc Abdelhak DJOUADI}$^{\,1}$ and 
{\sc Pietro SLAVICH}$^{\,2,3}$}

\vspace*{0.9cm}

{\it $^1$ Laboratoire de Physique Th\'eorique d'Orsay, UMR8627--CNRS,\\
Universit\'e Paris--Sud, B\^at. 210, F--91405 Orsay Cedex, France.}\s

{\it $^2$ LAPTH, 9 Chemin de Bellevue, F--74941 Annecy-le-Vieux, France.}\s

{\it $^3$ CERN, Theory Division,  CH--1211 Geneva 23, 
Switzerland.} 

\end{center}

\vspace{1cm}

\begin{abstract}

We perform a comprehensive analysis of the Minimal Supersymmetric
Standard Model (MSSM) in the scenario where the scalar partners of the
fermions and the Higgs particles (except for the Standard-Model-like
one) are assumed to be very heavy and are removed from the low-energy
spectrum.  We first summarize our determination of the mass spectrum,
in which we include the one-loop radiative corrections and resum to
all orders the leading logarithms of the large scalar masses, and
describe the implementation of these features in the {\sc Fortran}
code {\tt SuSpect} which calculates the masses and couplings of the
MSSM particles.  We then study in detail the phenomenology of the
model in scenarios where the gaugino mass parameters are non-universal
at the GUT scale, which leads to very interesting features that are
not present in the widely studied case of universal gaugino mass
parameters.  We discuss the constraints from collider searches and
high-precision measurements, the cosmological constraints on the relic
abundance of the neutralino candidate for the Dark Matter in the
Universe -- where new and interesting channels for neutralino
annihilation appear -- and the gluino lifetime. We then analyze, in
the case of non-universal gaugino masses, the decays of the Higgs
boson (in particular decays into and contributions of SUSY particles),
of charginos and neutralinos (in particular decays into Higgs bosons
and photons) and of gluinos, and highlight the differences from the
case of universal gaugino masses.

\end{abstract}

\newpage

\section{Introduction}

The main reason for introducing low energy Supersymmetry (SUSY)
\cite{SUSY} in particle physics was its ability to solve the
naturalness and hierarchy problems \cite{I-Hierarchy-SUSY}. Indeed,
SUSY prevents the Higgs boson mass from acquiring very large radiative
corrections: the quadratically divergent loop contributions of the
Standard Model (SM) particles to the squared Higgs masses are exactly
canceled by the corresponding loop contributions of their
supersymmetric partners.  This cancellation stabilizes the huge
hierarchy between the Grand Unification (GUT) and the electroweak
symmetry breaking (EWSB) scales and no extreme fine tuning of
parameters relevant to the mechanism of EWSB is required for this
purpose. However, SUSY is not an exact symmetry and there is still a
residual contribution to the Higgs masses that is proportional to the
mass differences between the SM particles and the new SUSY
particles. Therefore, for the fine-tuning problem not to be
reintroduced in the theory, the mass $M_S$ of the new particles should
be at most of the order of the TeV.  The requirement of no fine tuning
is the main reason for expecting SUSY particles to be accessible at
the next generation of high-energy colliders, in particular at the
Large Hadron Collider (LHC) and the International Linear Collider
(ILC). \s

Nevertheless, there is no compelling criterion to define the maximal
acceptable amount of fine tuning \cite{fine-tune} and the choice of
the upper bound on the SUSY scale $\msusy$ is somewhat
subjective. Thus, it might well be that $\msusy$ is significantly
larger than one TeV, in which case the SUSY particles would be too
heavy and not observable at the LHC. However, there are two more
important motivations for SUSY that also call for some light SUSY
particles: the consistent unification of the three gauge coupling
constants at the GUT scale \cite{Gauge-Unif} and the solution to the
Dark Matter (DM) problem \cite{DM-review}.  Indeed, the SM slope of
the evolution of the three gauge couplings has to be modified early
enough by some SUSY particle contributions to achieve unification, and
the DM problem calls for the existence of a new stable, neutral and
weakly interacting particle that is not too heavy in order to have the
required cosmological relic density. However, it has been pointed out
\cite{AHD,GR,wells} that, for SUSY to provide solutions to the
unification and DM problems, only gauginos and higgsinos, the
spin--$\frac12$ superpartners of the gauge and Higgs bosons, need to
be relatively light, with masses of the order of the EWSB scale. The
scalar partners of the SM fermions sit in complete irreducible
representations of SU(5), therefore they could be very heavy without
spoiling gauge coupling unification. For $\msusy \gg 1$ TeV the model
would be extremely fine-tuned, though, and one would have to give up
SUSY as the solution to the hierarchy problem. \s

An interesting feature of such a scenario, commonly known as Split
Supersymmetry \cite{AHD,GR}, is that it is much more predictive than
the general Minimal Supersymmetric Standard Model (MSSM) \cite{MSSM}.
Indeed, it is well known that, in the most general case, the MSSM has
a very large number of free parameters, ${\cal O}(100)$, most of them
related to the sfermion sector. Even if one constrains the model to
have a viable phenomenology, as in the so-called phenomenological MSSM
\cite{pMSSM} where minimal flavor mixing and CP violation as well as
universality of the first- and second-generation sfermion masses are
assumed, there are still more than 20 free parameters left. On the
other hand, if the masses of all the scalars (except for one SM-like
Higgs doublet) are pushed to large values only a handful of parameters
are needed to describe the low-energy SUSY theory. As a by-product,
such an assumption cures many problems of the general MSSM (such as
the excess of flavor and CP violation, fast proton decay with
dimension-5 operators, etc.) while still allowing for gauge coupling
unification and a good DM candidate, the lightest of the
neutralinos.\s

Indeed, in the MSSM with heavy scalars, besides a common value
$\msusy$ of the soft SUSY-breaking sfermion mass parameters, the basic
inputs are essentially the three gaugino masses $M_1,M_2,M_3$, which
can be unified to a common value at the high-energy GUT scale, the
Higgs--higgsino mass parameter $\mu$ which is no longer fixed by the
requirement of proper EWSB as in the general MSSM, and the parameter
$\tan\beta$. The trilinear Higgs--sfermion couplings are forced to be
small by the same symmetry that protects the higgsino and gaugino
masses, and they play a very minor role. One can then derive in an
exhaustive manner the relationship between this small number of inputs
and the physical parameters, i.e.~the (super)particle masses and
couplings.  However, if the scalars are very heavy, they will lead to
significant quantum corrections in the Higgs and gaugino--higgsino
sectors, enhanced by potentially large logarithms of the ratio between
the EWSB scale and the scalar mass scale, $\log( M_{\rm
EWSB}/\msusy)$. In order to have reliable predictions, one has to
properly decouple the heavy states from the low-energy theory and
resum the large logarithmic corrections by means of Renormalization
Group Equations (RGEs).\s

In fact, from a more practical point of view, in most (if not all) of
the numerical RGE codes that calculate the Higgs and SUSY particle
spectrum of the MSSM \cite{suspect,RGEcodes}, one cannot assume too
large masses for the sfermions, $\msusy \gsim$\,a few TeV, as the
programs become unreliable. Indeed, one has to perform major
modifications to the general algorithms of such programs, which
involve three main steps: the RG evolution, the consistent
implementation of EWSB and the calculation of the mass spectra
including the radiative corrections. In particular, one needs to
properly decouple the heavy scalars and include the intermediate scale
$\msusy$ in the RG running of parameters back and forth between the
low-energy scales $M_Z$ and $M_{\rm EWSB}$ and the high-energy scale
$\mgut$; ignore the implementation of (radiative) EWSB as a large fine
tuning is tolerated; resum the large logarithms in the radiative
corrections to the calculated physical masses and couplings as
mentioned above.\s

The purpose of this paper is to analyze, in the most comprehensive
way, the SUSY and Higgs particle spectrum in the MSSM with heavy
scalars, with $\msusy$ ranging from a few TeV to $\sim 10^{12}$
GeV. We start by summarizing our determination of the particle
spectrum of the low-energy theory in which the heavy scalars have been
integrated out. We determine the masses and couplings of the Higgs
boson and of the higgsinos and gauginos at the weak scale by solving
the one-loop RGEs of \sps\ \cite{GR,ADGW}, thereby resumming to all
orders the leading corrections that involve logarithms of the large
scalar masses. We also include one-loop radiative corrections to the
Higgs boson, chargino and neutralino masses, by adapting to the \spsd\
case the MSSM formulae presented in ref.~\cite{PBMZ}; we investigate
the variation of the masses with the renormalization scale, which can
be considered as a rough estimate of the higher-order corrections. We
also describe the implementation of these features in {\tt SuSpect}
\cite{suspect}, one of the public computer codes that calculate the
MSSM mass spectrum.\s

We then study the regions in the parameter space that are compatible
with the presently available collider data as well as with the
constraints on the relic density of DM and on the gluino lifetime. In
the former case we reemphasize that a new possibility \cite{DM-split},
which is not present in the usual MSSM, has to be considered: the
efficient annihilation of the DM lightest neutralinos through the
exchange of the Higgs boson which decays into a real and a virtual $W$
boson, the latter subsequently decaying into two massless
fermions. Finally, we have adapted to the \spsd\ scenario the programs
{\tt HDECAY} \cite{HDECAY} and {\tt SDECAY} \cite{SDECAY} for the
calculation of the decay widths and branching ratios of the MSSM Higgs
bosons and of the SUSY particles, respectively. We briefly discuss
some features in the decays of the Higgs boson and the gauginos,
focusing on interesting channels such as Higgs decays into invisible
neutralinos and into two photons, chargino/neutralino decays into
lighter ones and a Higgs boson and a photon, and gluino decays into a
gluon and a neutralino. \s

Since the number of input parameters in the low-energy theory is
rather limited, one can relax the usual assumption \cite{mSUGRA} of
unified gaugino masses at the GUT scale, $M_1=M_2=M_3\equiv m_{1/2}$,
and still have a rather predictive model. Therefore, for the sake of
generality, we will also consider specific patterns of non-universal
gaugino masses \cite{non-universal} and, for illustration, we will
discuss two different models: one in which SUSY-breaking occurs via an
F--term that is not a SU(5) singlet \cite{non-singlet} and another
based on an orbifold string model \cite{OII}. In these models, the
gaugino mass parameters at the electroweak scale can be widely
different from the pattern of the universal scenario. In particular,
the neutralino that is the lightest SUSY particle (LSP) can be, most
of the time, either wino-like (as in anomaly mediated SUSY-breaking
models for instance) or higgsino-like (which also occurs in the
universal scenario for small values of the parameter $\mu$), leading
to a near degeneracy of the LSP neutralino mass with the lightest
chargino mass. In addition, in some of these scenarios, the LSP can
also be close in mass with the gluino and/or a bino with a very small
mass.  This leads to several interesting phenomenological features
that we will discuss in some detail, such as rapid annihilation of
neutralino DM through the exchange of a $Z$ boson or co-annihilation
of the gluinos which can be the next-to-lightest SUSY particle
(NLSP).\s

The rest of the paper is organized as follows. In the next section we
summarize the model with heavy scalars and discuss the implementation
of the RGEs, the radiative corrections to the Higgs mass as well as to
the gluino, chargino and neutralino masses, and the boundary
conditions on the soft SUSY-breaking gaugino masses. In section 3 we
summarize the various constraints on the model parameter space from
collider data and cosmology. In section 4 we present some results on
the Higgs boson decays involving SUSY particles and the decays of the
chargino and neutralino states as well as the gluinos. Conclusions are
given in section 5. Finally, a set of useful formulae is collected in
the appendix.


\section{The low-energy effective theory}

In this section we summarize our determination of the mass spectrum of
the effective theory valid below the scale $\msusy$ at which all the
heavy scalars of the MSSM, i.e. the sfermions and one Higgs doublet,
are removed from the spectrum. We also discuss the different boundary
conditions on the soft SUSY-breaking gaugino mass parameters that will
be considered in the phenomenological analysis of sections 3 and
4. Finally, a brief summary is given of how the model is implemented
in the {\sc Fortran} code {\tt SuSpect}.

\subsection{Determination of the mass spectrum}

If the common mass of the scalars is assumed to be very large, $\msusy
\gg 1$ TeV, the low-energy theory contains, besides the SM particles,
the higgsinos ${\tilde H}_{u},{\tilde H}_{d}$, the gluino $\tilde g$,
the wino $\tilde W$ and the bino $\tilde B$. Omitting the
gauge-invariant kinetic terms, as well as non-renormalizable operators
suppressed by powers of the heavy scale $\msusy$, the Lagrangian of
the effective theory reads \cite{GR}
\beq
{\cal L}&\supset&m^2 H^\dagger H-\frac{\lambda}{2}\left( H^\dagger H\right)^2
-\biggr[ h^u_{ij} {\bar q}_j u_i\epsilon H^* 
+h^d_{ij} {\bar q}_j d_iH
+h^e_{ij} {\bar \ell}_j e_iH  \nonumber \\
&&\nonumber\\
&&+\frac12 {M_3}\, {\tilde g}^A {\tilde g}^A
+\frac12 {M_2}\, {\tilde W}^a {\tilde W}^a
+\frac12 {M_1}\, {\tilde B} {\tilde B}
+\mu\, {\tilde H}_u^T\epsilon {\tilde H}_d \nonumber \\
&&\nonumber\\
\label{lagrangian}
&&\left. +\frac{H^\dagger}{\sqrt{2}}\left( \gtilu \sigma^a {\tilde W}^a
+\gtilup {\tilde B} \right) {\tilde H}_u
+\frac{H^T\epsilon}{\sqrt{2}}\left(
-\gtild \sigma^a {\tilde W}^a
+\gtildp {\tilde B} \right) {\tilde H}_d +{\rm h.c.}\right]\,,
\eeq
where $\sigma^a$ are the Pauli matrices, $\epsilon =i\,\sigma^2$ and 
$i,j$ are generation indices. The SM-like Higgs doublet $H$ is a linear 
combination of the two MSSM Higgs doublets $H_u$ and $H_d$, fine-tuned 
to have a small mass term $m^2$:
\be
H = -\cos\beta\,\epsilon\,H_d^* + \sin\beta \,H_u~.
\ee

At the high scale $\msusy$ the boundary conditions on the quartic
Higgs coupling and on the Higgs--higgsino--gaugino couplings of the
effective theory are determined by Supersymmetry:
\begin{eqnarray}
\label{boundlam}
\lambda(\msusy) &=& \frac{1}{4}\left[ g^2(\msusy )+g^{\prime 2}(\msusy )
\right] \,\cos^22\beta + \Delta_{\rm th} \lambda~, \\
\gtilu (\msusy )&=& g (\msusy )\sin\beta~, \hspace*{1cm}
\gtild (\msusy )= g (\msusy )\cos\beta~, \\
\label{boundgt}
\gtilup (\msusy ) &=& g^{\,\prime} (\msusy ) \sin\beta~, \hspace*{1cm}
\gtildp (\msusy )= g^{\,\prime} (\msusy )\cos\beta~.
\end{eqnarray}
where $g$ and $g'$ are the SU(2) and U(1) gauge couplings. Note that
$\tan\beta$ is not a parameter of the low-energy effective theory, and
it enters only the boundary conditions on the couplings. In fact,
contrary to what happens in the usual MSSM, $\tan\beta$ is not
interpreted here as the ratio of two Higgs vacuum expectation values,
but rather as the fine-tuned angle that rotates the two Higgs doublets
into one heavy and one light, SM-like doublet. \s

In the boundary condition to the quartic Higgs coupling, we include
also a one-loop threshold correction of ${\cal O}(h_t^4)$, where $h_t=
m_t/v$ is the top quark Yukawa coupling,
\be
\label{threshold}
\Delta_{\rm th} \lambda =
\frac{3\,h_t^4}{8\,\pi^2}\,\left[
\left(1 -\frac{g^2+g^{\prime 2}}{8\,h_t^2}\right)\,\frac{X_t^2}{M_S^2}
- \frac{X_t^4}{12\,M_S^4}\right]~.
\ee
$X_t = A_t - \mu/\tan\beta$ is the left--right mixing of the stop
squarks, with $A_t$ being the trilinear Higgs--stop coupling; here,
$M_S$ is interpreted as the average of the two stop masses. Note,
however, that in \spsd\ models the trilinear coupling $A_t$ cannot be
too large, since it is protected by the same symmetry that keeps the
gaugino and the higgsino mass parameters small. The threshold
correction in eq.~(\ref{threshold}) can thus be relevant only for
relatively small values of $\msusy$. Beyond tree level the boundary
conditions in eqs.~(\ref{boundlam})--(\ref{boundgt}) are valid only in
the $\overline{\rm DR}$ renormalization scheme. The one-loop
electroweak corrections to eqs.~(\ref{boundlam})--(\ref{boundgt}) that
account for the shift to the $\overline{\rm MS}$ scheme are given in
the appendix.\s

The gauge and third-family Yukawa couplings as well as the vacuum
expectation value (vev) of the SM Higgs field (normalized as $v
\approx 174$ GeV) are extracted from the following set of physical
inputs \cite{PDG,topmass}: the strong gauge coupling $\alpha_s(M_Z) =
0.1187$; the electromagnetic coupling $\alpha(M_Z) = 1/127.918$; the
$Z$--boson mass $M_Z = 91.1876$ GeV; the Fermi constant $G_F = 1.16637
\times 10^{-5}$ GeV$^{-2}$; the physical top and tau masses $M_t =
170.9 \pm 1.8$ GeV and $M_\tau = 1.777$ GeV and the running bottom
mass $m_b(m_b) = 4.25$ GeV. We use one-loop formulae from
ref.~\cite{PBMZ} to convert all the physical inputs into running
parameters evaluated in the $\overline{\rm MS}$ scheme at a reference
scale that we choose as equal to $M_Z$. To this purpose we need to
adapt to the \spsd\ scenario the formulae of ref.~\cite{PBMZ}, which
were originally derived for the MSSM in the $\overline{\rm DR}$
scheme: we remove the contributions of the heavy scalars and insert
appropriate $\overline{\rm DR}$--$\overline{\rm MS}$ conversion
factors. A summary of the relevant formulae is given in the appendix.
\s

The gaugino masses are given as input at the GUT scale, defined as the
scale where the two gauge couplings $g_1 \equiv
\sqrt{5/3}\,g^{\,\prime}$ and $g_2 \equiv g$ unify, and evolved down
to the scale $\msusy$ by means of the one-loop RGEs of the MSSM. In
addition to the minimal case where the three gaugino masses unify at
the GUT scale, i.e.~$M_i(\mgut) = m_{1/2}$, our analysis will consider
scenarios of SUSY breaking in which the boundary conditions at the GUT
scale are different. Finally, the $\mu$ parameter has to be provided
as an independent input, contrary to the constrained MSSM case in
which it can be extracted from the EWSB conditions, and we take it as
a running parameter evaluated at the scale $M_Z$.\s

The parameters of the Lagrangian in eq.~(\ref{lagrangian}) are then
evolved to a common renormalization scale $\mewsb$ of the order of
the weak scale, by means of the one-loop RGEs of the \spsd\ model,
which we take from ref.~\cite{GR}. Since some of the boundary
conditions on the parameters are given at the SUSY scale $\msusy$ and the
others are given at the weak scale $M_Z$ an iterative procedure is
necessary. The resulting couplings evaluated at the weak scale
$\mewsb$ account for the all-order resummation of the leading
logarithmic corrections involving powers of $\log(\msusy/\mewsb)$.\s

Once the iteration for the determination of the Lagrangian parameters
converges, the physical masses of the Higgs boson, the charginos and
the neutralinos are computed at the scale $\mewsb$ including one-loop
radiative corrections; the gluino mass is computed separately at the
scale $M_3$. The relation between the physical Higgs boson mass $M_H$
and the quartic coupling $\lambda$ computed at the generic scale $Q$
reads
\be
\label{hmass}
M_H = \sqrt{\frac{\lambda(Q)}{\sqrt{2}\,G_F}}\,
\left[1+\delta^{\rm SM}(Q) + \delta^{\chi}(Q)\right]~.
\ee
The SM contribution $\delta^{\rm SM}$ can be found in ref.~\cite{SZ},
and contains terms enhanced by $M_t^4$ coming from top-quark
loops. The remaining term $\delta^{\chi}$ is the contribution of the
diagrams involving charginos and neutralinos, and reads
\be
\label{deltachi}
\delta^\chi = \frac{1}{2}\,\left[
\frac{T_H^{\,\chi}}{\sqrt{2}\,m_H^2 \,v}- \frac{\Pi^\chi_{HH}(m_H^2)}{m_H^2}
+\frac{\Pi^\chi_{WW}(0)}{m_W^2}\right]~,
\ee
where $m_H^2 = 2\,\lambda\,v^2$ and $m_W^2 = \frac12 g^2 v^2$ are tree-level
Higgs and $W$ masses, while $T_H^{\,\chi}$, $\Pi^\chi_{HH}$ and
$\Pi^\chi_{WW}$ denote the chargino and neutralino contributions to
the Higgs boson tadpole, the Higgs boson self-energy and the $W$ boson
self-energy at zero external momentum, respectively. The explicit
dependence of tadpole and self-energies on the renormalization scale
compensates, up to higher-order (i.e.~two-loop) effects, the implicit
scale dependence of the Higgs quartic coupling in eq.~(\ref{hmass}).
We extracted the formulae for the Higgs tadpole and the self-energies
from the MSSM results of ref.~\cite{PBMZ}, by appropriately rotating
the Higgs fields, dropping the contributions of heavy scalars and
expressing the couplings of charginos and neutralinos in terms of the
effective Higgs--higgsino--gaugino couplings $\tilde g_{u,d}$ and
$\tilde g^{\,\prime}_{u,d}$. We have also checked that our results for
$\delta^\chi$ agree with the recent computation of
ref.~\cite{binger}. The relevant formulae can be found in the
appendix.\s

\begin{figure}[tp]
\vspace*{-3mm}
\begin{center}
\mbox{~\epsfig{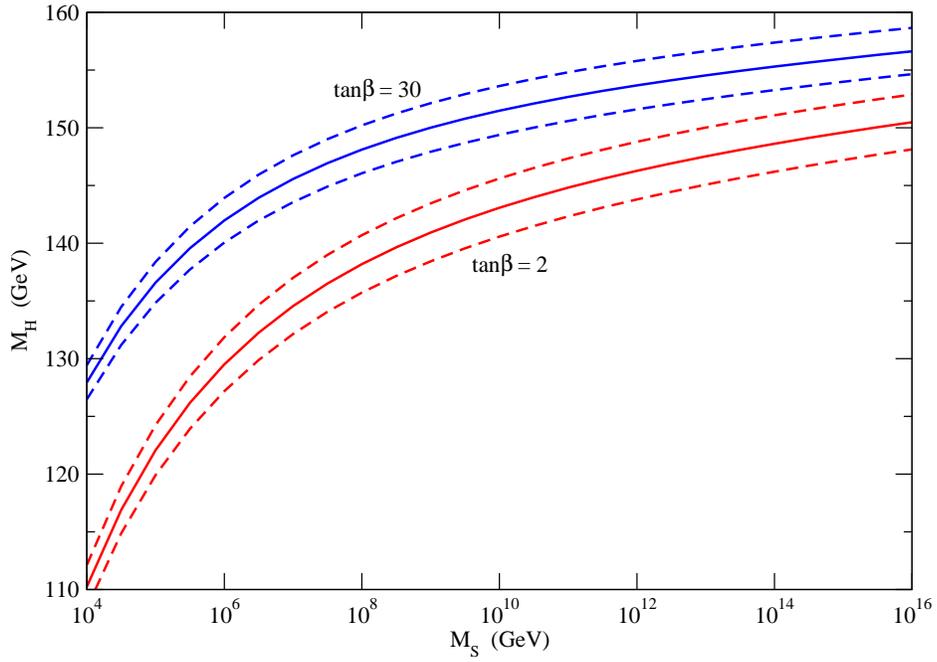}}
\end{center}
\vspace*{-3mm}
\caption{The Higgs boson mass as a function of the scalar mass
$\msusy$ for $A_t = 0,~\mu=m_{1/2}=500$ GeV and two values of
$\tan\beta$. The dashed lines correspond to a $\pm 1\sigma$ variation
of the top quark mass.}
\label{fig-higgs1}
\end{figure}

\begin{figure}[tp]
\begin{center}
{\epsfig{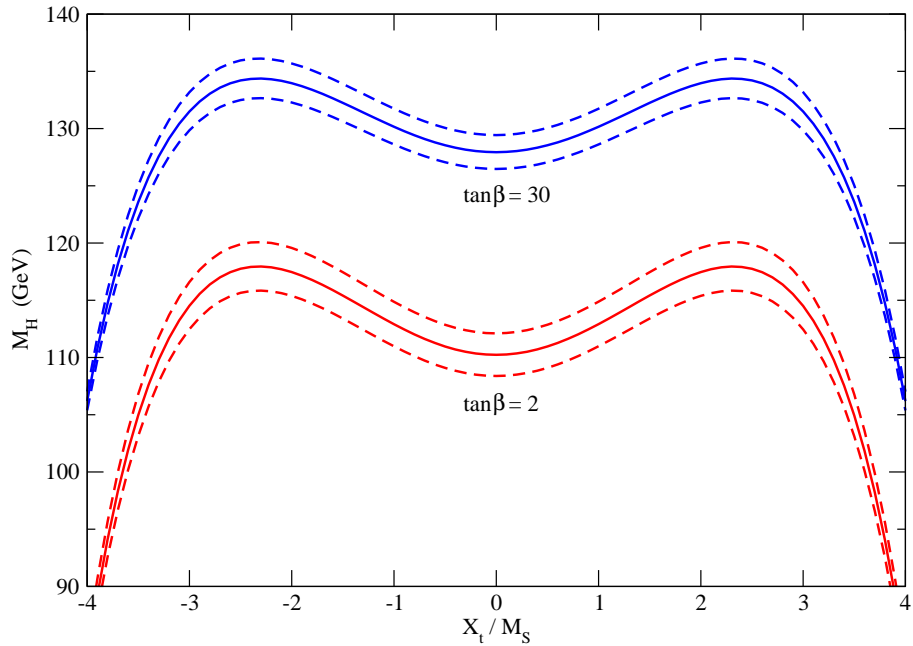}}
\end{center}
\vspace*{-3mm}
\caption{The Higgs boson mass as a function of the ratio $X_t/\msusy$
for $\msusy = 10$ TeV, $\mu=m_{1/2}=500$ GeV and two values of
$\tan\beta$. The dashed lines correspond to a $\pm 1\sigma$ variation
of the top quark mass.}
\label{fig-higgs2}
\vspace*{-3mm}
\end{figure}

\begin{figure}[t]
\vspace*{-3mm}
\begin{center}
{\epsfig{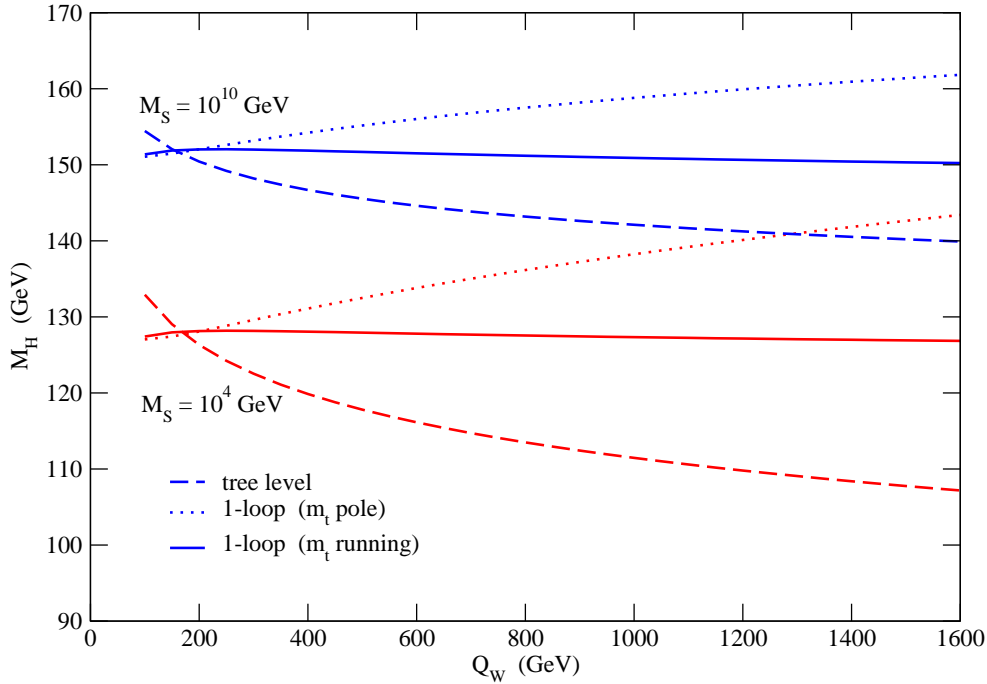}}
\end{center}
\vspace*{-5mm}
\caption{The Higgs boson mass as a function of the scale $\mewsb$ 
in various approximations for $\msusy=10^4$ and $10^{10}$
GeV. For the meaning of the different curves see the text.}
\label{fig-higgs3}
\vspace*{-3mm}
\end{figure}

The logarithmic dependence of $M_H$ on the sfermion mass scale
$\msusy$ has been discussed in earlier papers
\cite{AHD,ADGW,GR,binger}, but we show it for completeness in
fig.~\ref{fig-higgs1}, choosing the SUSY input parameters as
$A_t=0,~\mu(M_Z) = m_{1/2}= 500$ GeV and $\tan\beta = 2$ or $30$. The
top Yukawa coupling entering the one-loop corrections to the Higgs
mass, eqs.~(\ref{threshold}) and (\ref{hmass}), is expressed in terms
of the running top mass, which is in turn extracted from the physical
top mass via suitable threshold corrections. For each choice of
$\tan\beta$ the solid lines in fig.~\ref{fig-higgs1} are obtained with
the central value of the physical top mass, $M_t=170.9$ GeV, and the
dashed lines correspond to a $\pm 1 \sigma$ variation in the top
mass. It can be seen that, for relatively low values of $\msusy$,
going from small to large $\tan\beta$ can change the Higgs boson mass
by nearly 20 GeV. This is the most important way in which the
parameter $\tan\beta$ affects the low-energy phenomenology of the
model, and it is due to the $\cos^2 2\beta$ term in the tree-level
part of the boundary condition on the Higgs quartic coupling
$\lambda$, eq.~(\ref{boundlam}). When $\msusy$ gets larger, however,
the low-energy value of $\lambda$ is dominated by the radiative
correction induced by the RG evolution, and the effect of the
variation in $\tan\beta$ becomes less important.\s

Fig.~\ref{fig-higgs2} shows the dependence of $M_H$ on the stop mixing
term $X_t$, for $\msusy = 10$ TeV and the other SUSY parameters chosen
as in fig.~\ref{fig-higgs1}. It can be seen that, for not too large
values of $X_t/\msusy$, the dependence of $M_H$ on $X_t$ is milder
than in the usual MSSM plots with stop masses of the order of 1 TeV
(see, e.g., fig.~1.4 of the second paper in ref.~\cite{anatomy}). This
is due to the fact that the threshold correction to the Higgs quartic
coupling in eq.~(\ref{threshold}) is computed in terms of the running
top Yukawa coupling $h_t(\msusy)$, and the RG evolution up to the
large scale $\msusy$ has the effect of suppressing $h_t$.\s

It is also useful to investigate how the one-loop result for $M_H$
depends on the renormalization scale $\mewsb$ at which we stop the RG
evolution of the Lagrangian parameters and compute the physical mass
of the Higgs boson.  In an ideal, all-loop calculation the result for
the physical mass would not depend at all on the choice of such
scale. However, since we are truncating our calculation at the
one-loop order there will be a residual scale dependence, that we can
take as a lower bound on (but not necessarily as a full estimate of)
the uncertainty associated with higher-order corrections. In
fig.~\ref{fig-higgs3} we show the dependence of $M_H$ on the scale
$\mewsb$ for the two choices of the heavy scalar mass $\msusy = 10^4$
GeV and $\msusy = 10^{10}$ GeV (the other relevant parameters are
chosen as $\tan\beta = 30,~A_t=0,~\mu(M_Z) = m_{1/2} = 500$ GeV). In
each set of curves the dashed line represents the tree-level result,
as in eq.~(\ref{hmass}) with $\delta^{\rm SM}$ and $\delta^\chi$ set
to zero; the dotted line represents the one-loop result in which
$\delta^{\rm SM}$ is expressed in terms of the physical top mass
$M_t$; the solid line represents the one-loop result in which
$\delta^{\rm SM}$ is expressed in terms of the running top mass
$m_t(\mewsb)$ (as is done in figs.~\ref{fig-higgs1} and
\ref{fig-higgs2}). It can be seen that the tree-level result shows a
marked dependence on $\mewsb$, due to the scale dependence of
$\lambda$ in eq.~(\ref{hmass}). A non-negligible dependence is also
present in the one-loop result based on the physical $M_t$. On the
other hand, the residual scale dependence is very small when the
running top mass is used in the one-loop correction. This is
reminiscent of the situation in the MSSM, where the use of the
physical top mass in the one-loop corrections leads to an excessively
high estimate of the light Higgs boson mass, compensated for by large
negative two-loop corrections, whereas the two-loop corrections are
much smaller when the running top mass is used in the one-loop
part. We also note that the three determinations of $M_H$ in
fig.~\ref{fig-higgs3} are in good agreement with each other for the
particular choice $\mewsb \approx M_t$.\s

\begin{figure}[p]
\begin{center}
{\epsfig{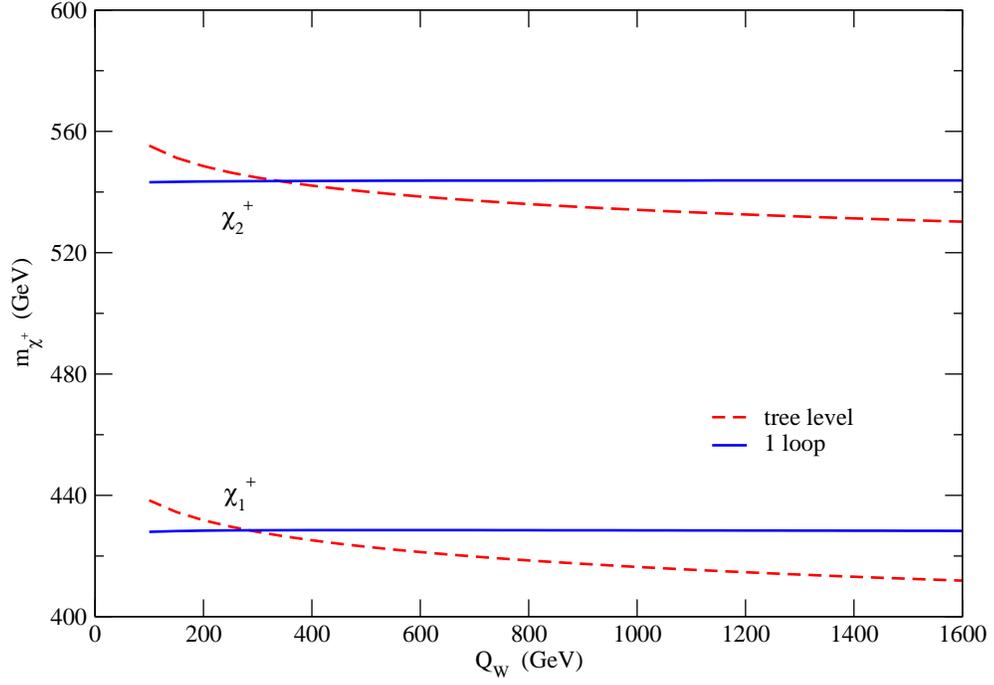}}
\end{center}
\vspace*{-3mm}
\caption{The chargino masses as a function of the scale $\mewsb$ for
$\msusy=10^4$ GeV, $\tan\beta = 30,~\mu = m_{1/2} = 500$ GeV. The
dashed lines are tree-level results and the solid lines are one-loop
results. }
\label{fig-char}
\end{figure}

\begin{figure}[p]
\begin{center}
{\epsfig{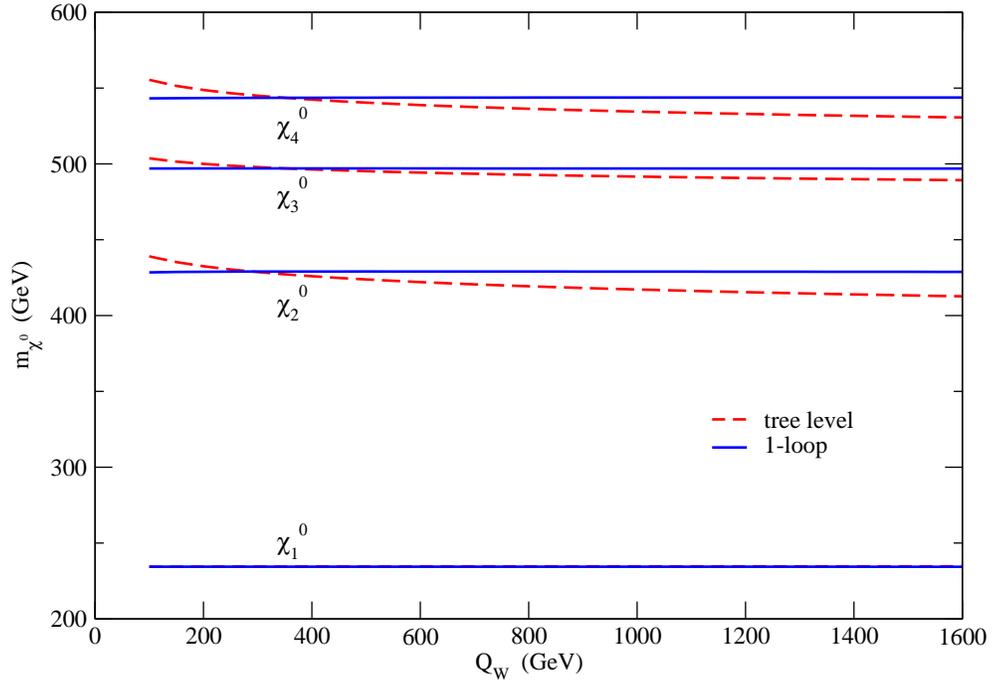}}
\end{center}
\vspace*{-3mm}
\caption{Same as fig.~\ref{fig-char} for the four neutralino masses.}
\label{fig-neut}
\end{figure}

The tree-level chargino and neutralino mass matrices read
\be
\label{neutchar}
{\cal M}_C = \left( \begin{array}{cc} M_2 & {{\tilde g}_u\,v}
\\ {{\tilde g}_d\,v}& \mu \end{array} \right)\,,~~~~~~~~~
{\cal M}_N = \left( \begin{array}{cccc}
M_1 & 0 & -\frac{{\tilde g}'_d\,v}{\sqrt 2} 
& \frac{{\tilde g}'_u\,v}{\sqrt 2} \\
0   & M_2 & \frac{{\tilde g}_d\,v}{\sqrt 2} 
& -\frac{{\tilde g}_u\,v}{\sqrt 2} \\
-\frac{{\tilde g}'_d\,v}{\sqrt 2} 
&\frac{{\tilde g}_d\,v}{\sqrt 2} & 0 & -\mu \\
\frac{{\tilde g}'_u\,v}{\sqrt 2} & 
-\frac{{\tilde g}_u\,v}{\sqrt 2} & -\mu & 0
\end{array} \right)~.
\ee
The values of the tree-level chargino ($\chi^+_{1,2}$) and neutralino
($\chi^0_{1,2,3,4}$) masses depend on the renormalization scale at
which the various parameters in eq.~(\ref{neutchar}) are computed. To
reduce this scale dependence we include the one-loop corrections to
the chargino and neutralino masses, once again adapting the MSSM
formulae of ref.~\cite{PBMZ} to the \spsd\ case (see the appendix for
details). In figs.~\ref{fig-char} and \ref{fig-neut} we show the
masses of charginos and neutralinos, respectively, as function of the
scale $\mewsb$, for $\msusy = 10^4$ GeV, $\tan\beta = 30$ and
$\mu(M_Z) = m_{1/2} = 500$ GeV. In each plot the dashed lines
represent the tree-level result and the solid lines represent the
one-loop result. It can be seen that the inclusion of the radiative
correction improves the scale dependence of the masses of charginos
and neutralinos. For the choice of parameters considered in
figs.~\ref{fig-char} and \ref{fig-neut} the only exception is the mass
of the lightest neutralino, which is mostly bino: the tree-level value
of $m_{\chi_1^0}$ has already very little dependence on the scale, and
the one-loop corrections are negligible.\s

Finally, the physical gluino mass $M_{\tilde g}$ is related to the
${\rm \overline{MS}}$ parameter $M_3$ by
\be
M_{\tilde g} = M_{3}(Q)\,\left[1+\frac{\alpha_s}{4\,\pi}\,\left(
12 + 9\,\log\frac{Q^2}{M_3^2}\right)\,\right]~.
\ee
 
In the analysis of the constraints on the \spsd\ parameter space that
we present in sections \ref{section3} and \ref{section4} we will be
mostly interested in phenomena that involve charginos and
neutralinos. For this reason we find it convenient to choose by
default a value of $\mewsb$ that is representative of the masses in
the chargino/neutralino sector, i.e. $\mewsb = \sqrt{\mu\,M_2}$. While
a priori this might not be the best choice of renormalization scale
for phenomena involving the Higgs boson, fig.~\ref{fig-higgs3} shows
that the Higgs boson mass is not significantly affected by the choice
of $\mewsb$ when we compute the one-loop corrections in terms of the
running top mass. On the other hand, in many models of SUSY breaking
the gluino can be quite heavier than the other gauginos, thus we will
choose $\mewsb = M_3$ when discussing the gluino mass and decays.

\subsection{Patterns of soft SUSY-breaking gaugino masses} 

\begin{figure}[t]
\vspace*{-1mm}
\begin{center}
\rotatebox{-90}{\epsfig{figure=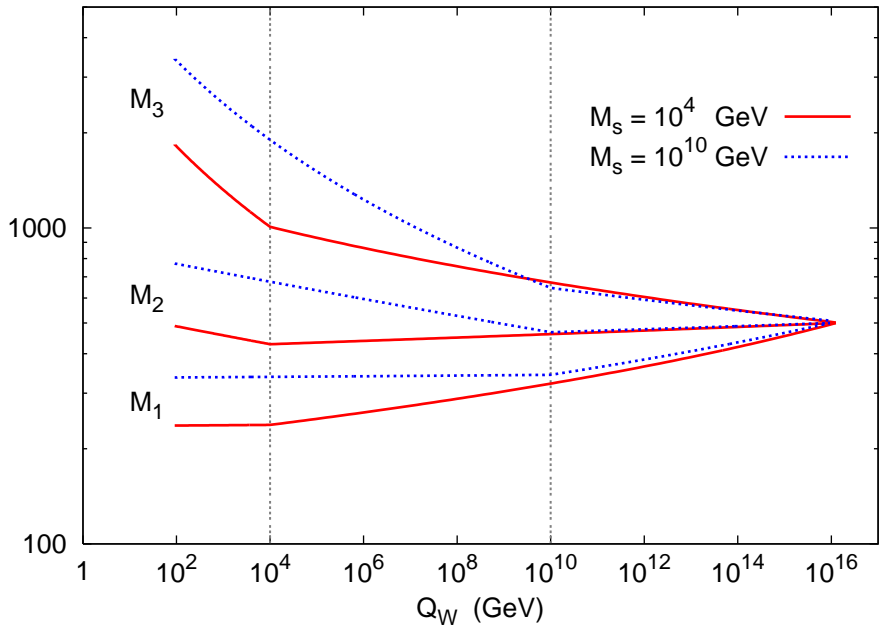,width=9cm}}
\end{center}
\vspace*{-3mm}
\caption{The evolution of the gaugino mass parameters from the GUT
scale to the weak scale in the universal scenario for $\msusy=10^4$
and $10^{10}$ GeV.}
\label{fig-Mevol}
\end{figure}

The soft SUSY-breaking gaugino mass parameters $M_{1,2,3}$ entering
the chargino, neutralino and gluino masses are determined via one loop
RGEs once their values at the GUT scale are fixed. If one assumes
universality of these parameters, $M_1=M_2=M_3\equiv m_{1/2}$ at
$\mgut$, as is done e.g.~in the minimal Supergravity (mSUGRA) model
\cite{mSUGRA}, the ratios between the weak scale values are simply
related to the squares of the gauge coupling constants $\alpha_i\equiv
g_i^2/(4\pi)$ by $M_1 \!:\!M_2\!:\!M_3
=\alpha_1\!:\!\alpha_2\!:\!\alpha_s$.  These values also depend on the
intermediate scale $M_S$ below which the contributions of the
sfermions and of the heavy Higgs doublet are decoupled from the
RGEs. The evolution of $M_{1,2,3}$ is displayed in
fig.~\ref{fig-Mevol}, where a common GUT value $m_{1/2}= 500$ GeV has
been assumed and two values for the scalar masses, $M_S=10^4$ GeV and
$M_S=10^{10}$ GeV, have been used. At the scale $M_Z$ one obtains the
ratios $M_1\!:\!M_2\!:\!M_3 = 1.0\!:\!2.0\!:\!7.8$ for $M_S=10^4$ GeV,
while one has $M_1\!:\!  M_2\!:\! M_3 = 1.0\!:\!2.3\!:\!10.2$ for
$M_S=10^{10}$ GeV. If the scale $M_S$ had been set to 1 TeV, as in the
usual MSSM with scalar masses of the same order as the gaugino masses,
one would have obtained $M_1\!:\! M_2\!:\! M_3 \simeq
1\!:\!2\!:\!7$.\s

Since for heavy scalars the number of basic input parameters of the
model is rather small, one can relax the assumption of a universal
gaugino mass at the GUT scale and still have a rather predictive model
which, in many instances, could lead to a rather different
phenomenology. In this paper, rather than performing a general scan
with 4 or 5 input parameters (i.e.~$M_1,M_2,M_3, \mu$ and possibly
$M_S$) we will consider two specific SUSY models in which the boundary
conditions for the gaugino masses at the GUT scale are different from
those of the universal scenario.  This will simplify our numerical
analysis and, at the same time, allow us to address the new
interesting phenomenology induced by the non-universality of the
gaugino masses. \s

The first scenario that we consider is a gravity-mediated
SUSY-breaking scenario in which, to lowest order, the gaugino masses
arise from a dimension--5 operator
\be
{\cal L} \propto \frac{\langle F_\Phi
\rangle _{ab} }{M_{\rm Planck}} \cdot \lambda^a \lambda^b~,
\ee
where $\lambda^{a,b}$ are the gaugino fields and $F_\Phi$ the
auxiliary component of a left-handed chiral superfield $\Phi$ which
couples to the SUSY field strength. In the usual mSUGRA model with
SU(5) grand unification the SUSY-breaking field $F_\Phi$ is a singlet
under the unifying gauge group, leading to universal gaugino
masses. However, the chiral superfield $\Phi$ can sit in any
representation of the symmetric product of the adjoint group
\cite{non-singlet}. In the case of SU(5) symmetry, $F_\Phi$ could
belong to an irreducible representation which results from the
symmetric product of two adjoints
\beq  
(\mathbf{24 \otimes 24})_{\rm symmetric}  = \mathbf{1 \oplus 24 \oplus 75
\oplus 200} 
\eeq
Once the neutral component of $F_\Phi$ has acquired a vacuum
expectation value, $\langle F_\Phi \rangle _{ab} = V_{a} \delta_{ab}$,
the vevs $V_a$ determine the relative magnitude of the soft
SUSY-breaking gaugino mass parameters $M_a$ at the GUT scale
\cite{non-universal}. These are shown in the left-hand side of table
\ref{tab-gauginos} and, as can be seen, only in the singlet case {\bf
1} are these parameters universal.\s

Another set of scenarios that we will consider are four-dimensional
string models in which the source of SUSY breaking resides
predominantly in the moduli sector. In particular, in the orbifold
{\bf OII} scenario in which all chiral fields have modular weight
equal to unity, the boundary conditions for the gaugino mass
parameters at the GUT scale are \cite{OII}
\beq
M_a ~\approx~  \pm\,10^{-3} \,( b_a + \delta_{GS} ) \,m_{3/2} \ , \ \ a=1,2,3
\eeq
where $m_{3/2}$ is the gravitino mass, $b_a$ the coefficient of the
MSSM beta function for the gauge coupling constant $g_a$, and
$\delta_{GS}$ the Green-Schwartz mixing parameter which is a negative
integer number in this case. In these scenarios the scalars, with
masses $m_0^2 \approx 10^{-3}~(-\delta_{GS})\,m_{3/2}^2 $, are expected
to be much heavier than the gauginos and the pattern of gaugino
masses, compared to the universal case, is driven by the parameter
$\delta_{GS}$. For the choice $\delta_{GS}=-4$ one obtains at $M_{\rm
GUT}$ the mass pattern $M_1\!: \!M_2\!:\! M_3 \sim \frac{53} {5}\!:\!
5\!: \! 1$; see table \ref{tab-gauginos}.\s

Other mass patterns can be obtained by choosing different values of
the integer parameter $\delta_{GS}$, and some of them are in fact
similar to those of the non-singlet representation models shown in
table 1. For instance, for $\delta_{GS}=-2$, one obtains the ratios
$M_1\!: \!M_2\!:\! M_3 \sim \frac{43}{5} \!:\!  3\!: \! -1$ which are
close to those of the scenario {\bf 200}. In the orbifold {\bf
OI}--type scenario \cite{OII}, which differs from the previous one in
the fact that the modular weights are not all equal to unity, one
could obtain different gaugino mass ratios. However, in many cases,
the pattern is similar to that of the scenarios discussed above 
and, for instance, ratios that are close to those of the scenario {\bf
75} can be obtained for $\delta_{GS}=-5$.\s

The relations between the gaugino masses $M_{1,2,3}$ at the scale
$\mgut$ and at the weak scale $M_Z$ are summarized in table
\ref{tab-gauginos} for the different scenarios. We have used the
one-loop RGEs for the couplings and assumed a common scalar mass
$\msusy$ of either $10^4$ GeV or $10^{10}$ GeV.  As can be seen, the
pattern of the gaugino mass parameters at $M_Z$ (and hence the gluino,
neutralino and chargino masses) can be quite different from the
universal case (scenario {\bf 1}) in which for $\msusy = 10^4$ GeV one
has approximately $M_1\!:\!M_2\!:\!M_3= \alpha_1\!:\!\alpha_2\!:
\!\alpha_s \simeq 1\!:\!2\!:\!8$.\s

In particular, in the scenario {\bf 200} where $M_2 <M_1$, the LSP is
wino-like for large values of the parameter $\mu$, implying that
$\chi_1^0$ and $\chi_1^\pm$ are nearly degenerate in mass. Again for
large $\mu$ values, the neutralinos $\chi_1^0$ and $\chi_2^0$ and the
charginos $\chi_1^\pm$ masses are very close to each other in the
scenario {\bf 75} since $|M_1| \sim |M_2|$, while in the scenario {\bf
24} the mass splitting between the LSP and the states $\chi_2^0$ and
$\chi_1^\pm$ can be very large since $M_2 \sim 6 M_1$. Finally, in the
{\bf OII} model one has $M_3 <M_1, M_2$ and the gluino tends to be the
LSP unless $\mu$ is very small, in which case $\chi_1^0, \chi_2^0$ and
$\chi_1^+$ are higgsino-like and almost degenerate in mass. Note that,
in general, the weak-scale ratios among the three gaugino masses show
a dependence on the scale $\msusy$ at which the sfermions are
integrated out.\s

\begin{table}[t]
\vspace*{2mm}
\begin{center}
\renewcommand{\arraystretch}{1.5}
\begin{tabular}{|c||c|c|c|} \hline
$\ \ \ ~ \ \ \ $ & $\ \ Q=\mgut \ \ $ & 
$\ \ Q=M_Z~~ [\msusy = 10^4\,{\rm GeV}] \ \ $ 
& $\ \ Q=M_Z~~ [\msusy = 10^{10}\, {\rm GeV}] \ \ $ \\ \hline
{\bf 1} & $1\,:\,1\,:\,1$ & $1.0\,:\,2.0\,:\,7.8$ 
& $1.0\,:\,2.3\,:\,10.2$ \\ \hline
{\bf 24} & $1\,:\,3\,:\,-2$ & $1.0\,:\,6.3\,:\,-15.2$ & 
$1.0\,:\,6.9\,:\,-19.7$ \\ \hline
{\bf 75} & $5\,:\,-3\,:\,-1$ & $1.0\,:\,-1.2\,:\,-1.5$ & 
$1.0\,:\,-1.4\,:\,-2.0$ \\ \hline
{\bf 200} & $10\,:\,2\,:\,1$ & $2.4\,:\,1.0\,:\,1.9$ & 
$2.2\,:\,1.0\,:\,2.2$ \\ \hline \hline
{\bf OII} & $53/5\,:\,5\,:\,1$ & $1.4\,:\,1.3\,:\,1.0$ 
& $1.0\,:\,1.1\,:\,1.0$
\\ \hline
\end{tabular}
\end{center}
\vspace*{0mm}
\caption{The ratios of gaugino mass parameters, $M_1\!:\!M_2\!:\!M_3$,
at the renormalization scales $\mgut$, $M_Z$ (with $\msusy= 10^4$ GeV)
and again $M_Z$ (with $\msusy=10^{10}$ GeV), for the different
patterns of soft SUSY breaking.}
\label{tab-gauginos}
\end{table}

There are many other SUSY models that lead to non-universal gaugino
masses, and a review has recently been given in
ref.~\cite{Nilles}. However, in most cases one is very close in
practice to the patterns that have been introduced above, and the
phenomenology of the gaugino sector is quite similar. For instance,
the weak-scale pattern of gaugino masses that emerges from anomaly
mediated SUSY-breaking (AMSB) models \cite{AMSB}, $M_1\!:\! M_2 \sim
3\!:\!  1$, is similar to that of the model {\bf 200}. In mirage
gaugino mediation \cite{Mirage}, where SUSY breaking is realized in
higher-dimensional brane models (a scheme that is realized naturally
in the so-called KKLT-type moduli stabilization models \cite{KKLT}),
one has \cite{Nilles} $M_1\!: \!M_2\!\sim\! 1\!:\! 1.3$ which, again,
is similar to the pattern of the scenario {\bf 75}. Thus, we believe
that the scenarios discussed above, with the patterns of gaugino
masses of table 1, are representative of a wide spectrum of
non-universal models.

\subsection{Implementation of the MSSM with heavy scalars in {\tt SuSpect}}

We have implemented this MSSM scenario with heavy scalars into the RGE
code {\tt SuSpect} \cite{suspect}. This model can be chosen by
selecting at the very beginning of the input file {\tt suspect2.in} or
alternatively {\tt suspect2\_lha.in}, the option {\tt SHeavy} by
putting {\tt ichoice(1)~=~3}.  The only two sets of basic input
parameters needed to be set are:\s

\non 
-- The SM basic input parameters, i.e. the electromagnetic, strong and 
weak couplings, the $Z$ boson mass  and the third-family fermion masses:
$$\alpha (M_Z), \alpha_s (M_Z), G_F, M_Z, M_t, m_b (m_b), M_\tau$$ 
-- The additional input parameters specific for this model: 
$$\mu (M_Z),\, M_1 (M_{\rm GUT}),\, M_2 (M_{\rm GUT}),\, M_3 (M_{\rm 
GUT}),\, M_S,\, \tan\beta (M_S),\, A_t (M_S).$$
All the other parameters of these files are irrelevant.\s

The routine {\tt SHeavy.f} performs the RG evolution for the gauge
couplings, the third generation fermion Yukawa couplings, the
gaugino-higgsino-Higgs boson couplings ($g_{u,d}, \,{\tilde
g}_{u,d}$), the gaugino mass parameters, the $\mu$ parameter, and the
quartic Higgs coupling $\lambda$.  This routine also contains all the
relevant one-loop radiative corrections.\s

The output file contains the physical chargino and neutralino masses
and the elements of the mixing matrices $U,\,V$ and $N$ (see the
appendix for details), as well as the physical masses of the gluino
and of the lightest Higgs boson. They are computed at a scale $Q_W$
set by default to $Q_W =\sqrt {\mu\,M_2}$, except for the gluino mass
which is computed at the scale $M_3$.  The masses of all the sfermions
and of the heavy Higgs bosons are considered to be degenerate and set
to $M_S$, while the various mixing angles (in the
third-generation-sfermion and Higgs sectors) are set to zero.\s

Note that, for the phenomenological analyses that will be presented in
the next two sections, we also needed to adapt to the \spsd\ scenario
the two programs {\tt HDECAY} \cite{HDECAY} and {\tt SDECAY}
\cite{SDECAY}, which compute the decay widths and branching ratios of
the MSSM Higgs bosons and of the SUSY particles, respectively.  These
programs use the output given by {\tt SuSpect} for the soft
SUSY-breaking parameters, the mixing matrix elements and the sparticle
and Higgs masses, but they calculate internally the various
couplings. In particular, the Higgs couplings to neutralinos and
charginos (which are different from the usual MSSM case) are
hard-coded and need to be adapted.

\section{Collider and Dark Matter constraints}
\label{section3}

In this section we analyze the constraints on the MSSM with heavy
scalars, first from collider searches and high-precision data
\cite{PDG} and then from cosmological data, in particular the relic
density measurement of DM by the WMAP satellite \cite{WMAP} and the
gluino lifetime \cite{Toharia,ADGPW,GGS}. A special attention will be
given to the non-universal gaugino mass scenarios discussed in the
previous section, where several new features compared to the universal
case appear. The DM and some collider constraints for non-universal
gaugino masses have been discussed in ref.~\cite{DM-non-uni} in the
usual MSSM with light scalars, while some implications of DM in the
Split-SUSY scenario have been studied in
Refs.~\cite{DM-split,DM-split2}.

\subsection{Constraints from collider data}

\begin{figure}[t]
\begin{center} 
\epsfig{figure=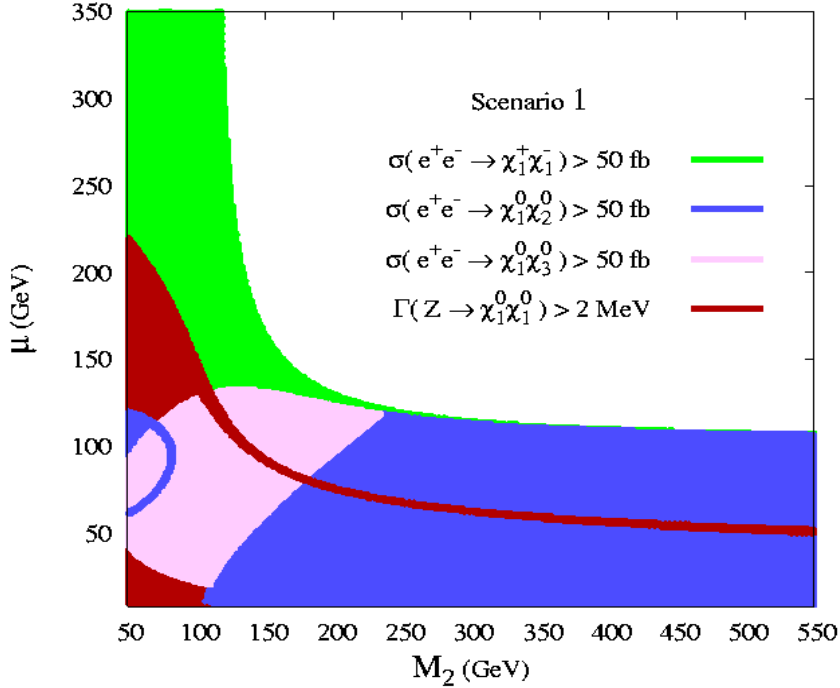,width=15cm,height=10cm} 
\vspace*{-.5cm}
\caption{Constraints on the $[M_2, \mu]$ parameter space for the
scenario {\bf 1} with universal gaugino masses at the GUT scale and
with $\msusy=10^4$ GeV, $A_t=0$ and $\tan\beta = 30$; the different
colors correspond to the regions excluded by the LEP bounds on the
partial decay width $\Gamma( Z \to \chi_1^0 \chi_1^0)$ and on the
production cross sections $e^+ e^- \to \chi_1^\pm
\chi_1^\mp,~ \chi_1^0 \chi_2^0,~ \chi_1^0 \chi_3^0$.}
\label{fig:LEP2cons1}
\end{center} 
\vspace*{-.7cm}
\end{figure}

In the scenario in which a universal gaugino mass $m_{1/2}$ is assumed
at the GUT scale, leading to the approximate relation
$M_1\!:\!M_2\!:\!M_3 \sim 1\!:\!2\!:\!8$ at the weak scale, the
strongest experimental bound is due to the negative search of
charginos at LEP2 up to energies of $\sqrt{s} \simeq 208$ GeV. From
pair production of the lightest chargino, $e^+ e^- \to \chi_1^\pm
\chi_1^\mp$, one obtains the mass bound \cite{PDG}
\beq
m_{\tilde\chi_1^\pm} \gsim 103~{\rm GeV}~.
\label{eq:charginop}
\eeq
The chargino mass bound in eq.~(\ref{eq:charginop}) is valid only if
the mass splitting between the lightest chargino and neutralino is
large enough, $\Delta M\equiv m_{\chi_1^\pm}- m_{\chi_1^0} \gsim$ a
few GeV.  For small $\Delta M$ values, as is the case when the LSP
neutralino is higgsino- or wino-like, $m_{\chi_1^\pm} \simeq
m_{\chi_1^0} \simeq |\mu|$ or $M_2$, the bound becomes weaker and, for
very heavy scalar fermions, one has $m_{\tilde \chi_1^\pm} \gsim
92$~GeV \cite{PDG}. Using the weak-scale gaugino mass relations above,
the bound on the lightest chargino mass translates into a lower bound
on the LSP mass, $m_{\tilde\chi_1^0} \gsim 50$ GeV if the LSP
neutralino is bino-like, in which case one has $m_{\chi_1^0} \simeq
M_1 \simeq \frac12 M_2 \simeq \frac12 m_{\chi_1^ \pm}$; $\chi_1^0$ is
thus too heavy to be kinematically accessible in invisible $Z$ boson
decays at LEP1, $Z \to \chi_1^0 \chi_1^0$. In the higgsino- and
wino-like regions for the LSP neutralino the mass bound is higher,
$m_{\tilde\chi_1^0} \gsim 92$~GeV, as $\chi_1^0$ is almost degenerate
with the chargino ${\chi_1^\pm}$ as discussed above.  Furthermore, the
constraint on $M_2$ from the bound on $m_{\chi_1^\pm}$ translates into
a constraint on the gluino mass, $m_{\tilde{g}} \sim M_3 \gsim 350$
GeV, which is higher than the direct bound from Tevatron searches when
scalar quarks are very heavy \cite{PDG}
\beq
m_{\tilde{g}} \gsim 195~{\rm GeV}  ~.
\label{eq:gluinop}
\eeq 
The constraints on the $[M_2, \mu]$ parameter space from LEP2 and
Tevatron negative searches of SUSY particles are summarized in
fig.~\ref{fig:LEP2cons1}. Here and in the following $M_2$ and $\mu$
have to be interpreted as $\msbar$ parameters evaluated at the
renormalization scale $Q=\mewsb$. Besides the process $Z \to \chi_1^0
\chi_1^0$, which contributes to the invisible $Z$ boson decay that is
tightly constrained by LEP1 data, $\Gamma_Z^{\rm inv} \lsim 2$ MeV
\cite{PDG}, we have imposed that the cross sections for the three
processes $e^+e^- \to \chi_1^\pm \chi_1^\mp,~ e^+e^- \to \chi_1^0
\chi_2^0$ and $e^+e^- \to \chi_1^0 \chi_3^0$ are smaller than 50 fb
which, given the collected luminosity of ${\cal L} \sim 100$ pb$^{-1}$
at the c.m.~energy $\sqrt s\sim 208$ GeV, corresponds to less than 5
events.  As can be seen, the strongest constraint is by far due to
eq.~(\ref{eq:charginop}), i.e.~the green (light grey) area. The
process $e^+ e^- \to \chi_1^0 \chi_2^0$ where the LSP and the
next-to-lightest neutralino are pair produced, although more favored by
phase space, does not add much information as the cross section is
generally much smaller. In the small blue (dark grey) oval region on
the left of the plot $\chi_2^0$ is mostly wino, while in the region
with large $M_2$ and small $\mu$ on the right of the plot $\chi_2^0$
is mostly higgsino.\s

\begin{figure}[p]
\begin{center} 
\mbox{\hspace*{-1.3cm} \epsfig{figure=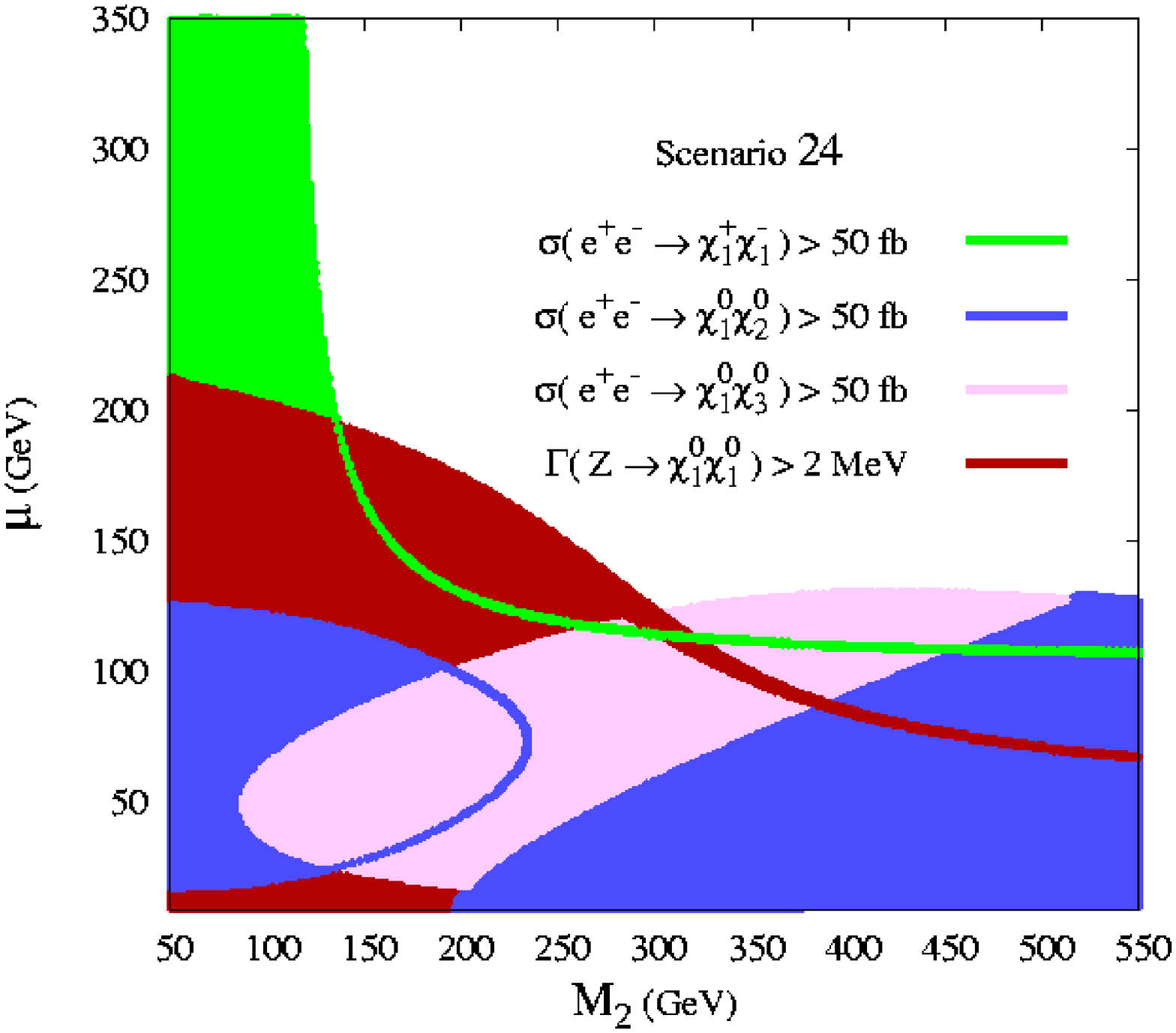,width=10.5cm,height=9.5cm}
      \hspace*{-2.cm} \epsfig{figure=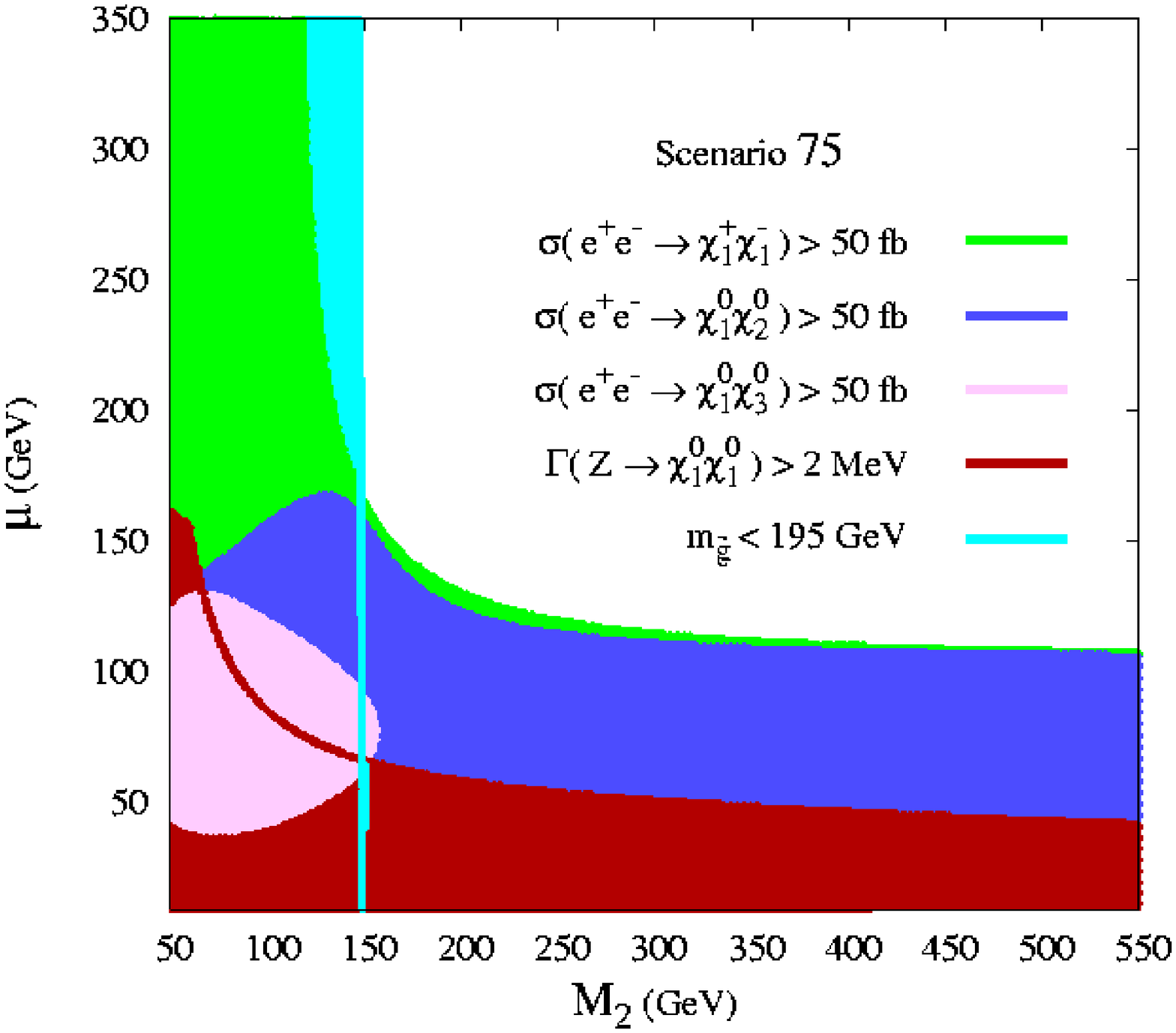,width=10.5cm,height=9.5cm}} 
\mbox{\hspace*{-1.3cm} \epsfig{figure=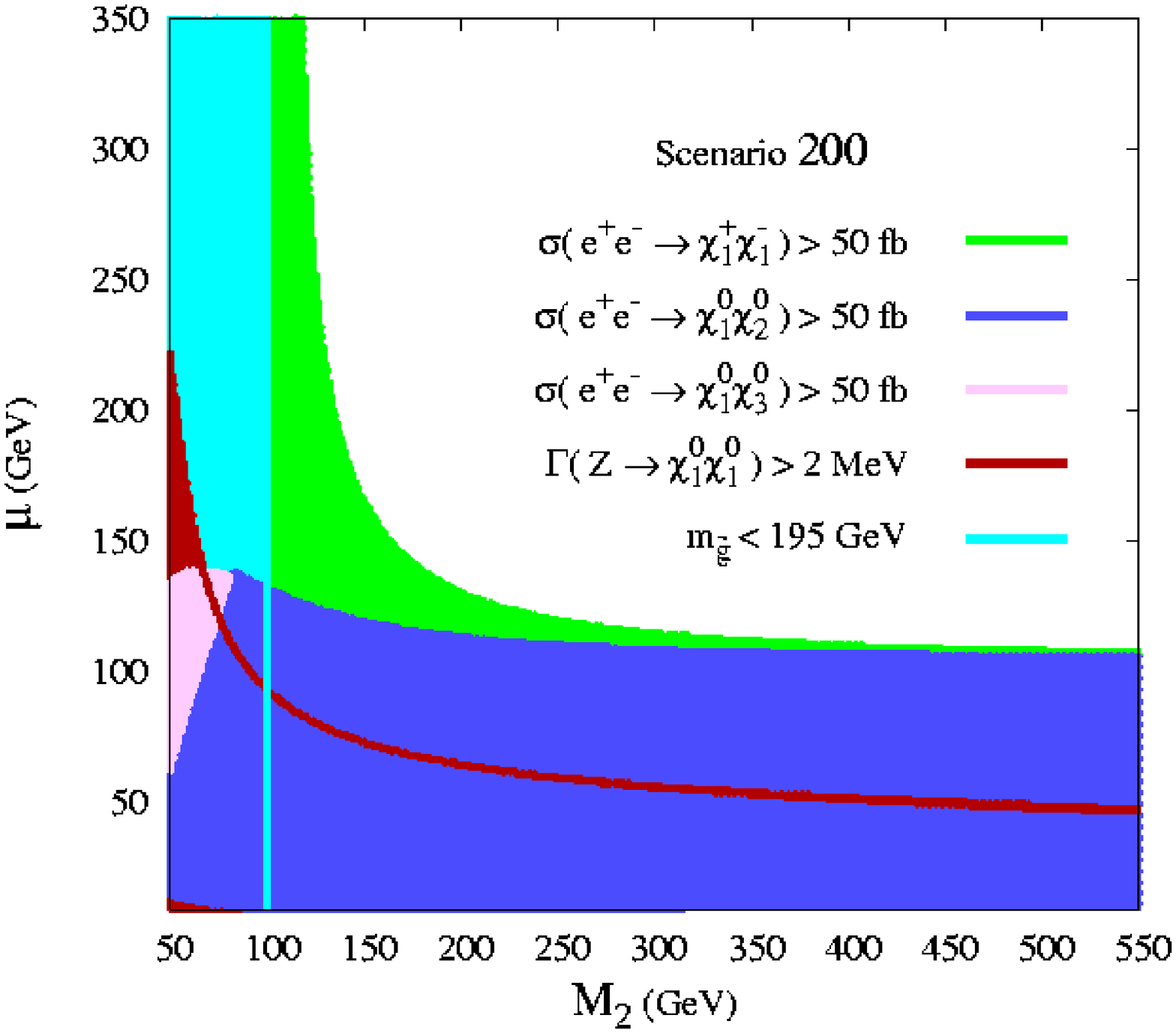,width=10.5cm,height=9.5cm}
      \hspace*{-2cm} \epsfig{figure=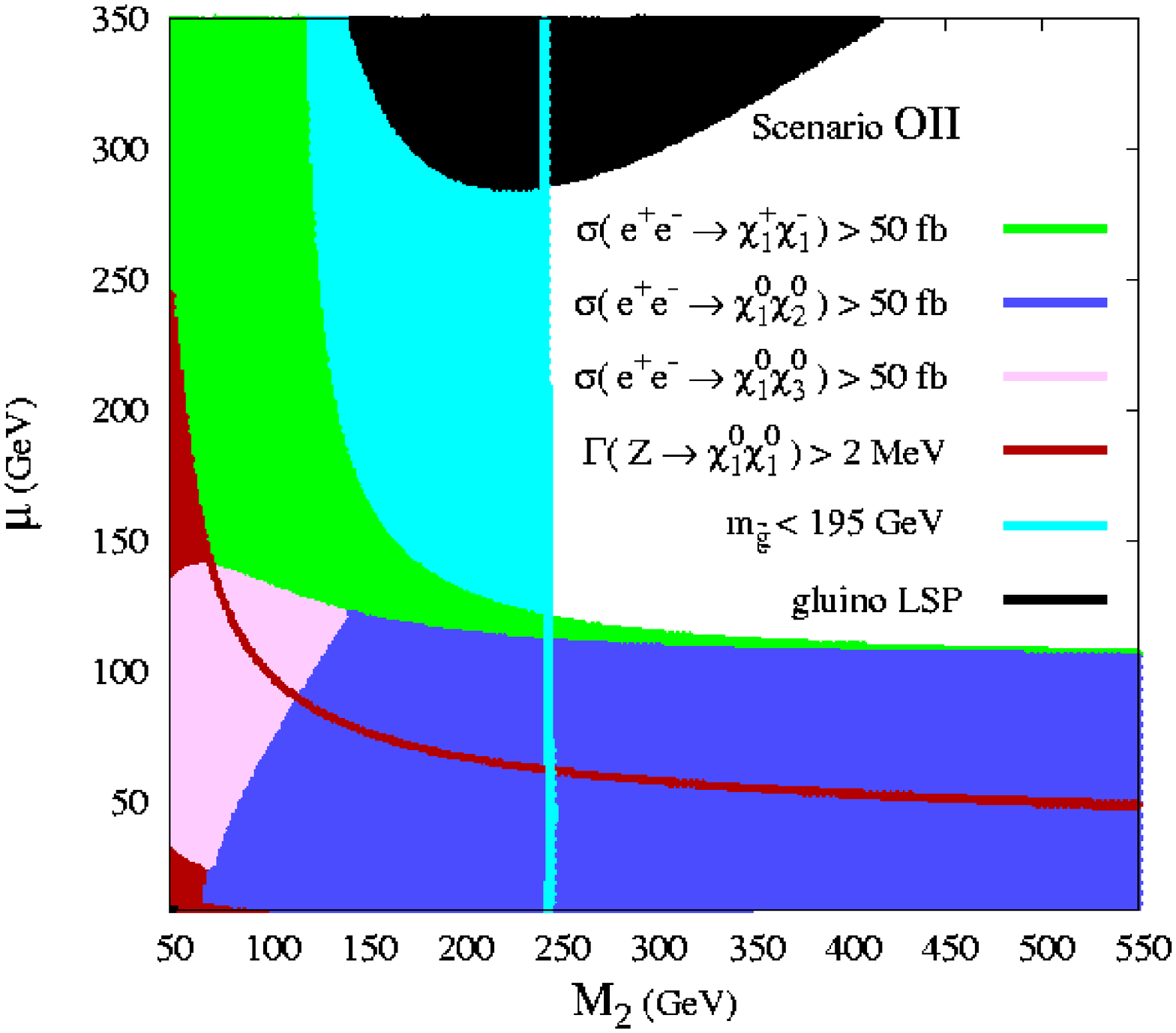,width=10.5cm,height=9.5cm}}
\caption{The same as in fig.~\ref{fig:LEP2cons1} but for the
non-universal scenarios. Where relevant, the effect of the gluino mass
bound from Tevatron searches is also shown.}
\label{fig:LEP2cons2}
\end{center} 
\end{figure}

All these bounds can be adapted to scenarios in which the boundary
conditions for the gaugino masses at the high scale are different.
However, in some cases, their impact can be widely different compared
to the universal scenario, as shown in fig.~\ref{fig:LEP2cons2} where
the constrained $[M_2,\mu]$ parameter space with $\msusy=10^4$ GeV is
displayed for the four scenarios {\bf 24, 75, 200} and {\bf OII}. We
have assumed $\mu>0$ but a similar pattern is obtained for $\mu
<0$. \s

In the non-universal scenario {\bf 24} one has $M_2\!:\! M_1 \sim 6\!:
\!1$ for the wino and bino masses at the weak scale and, as in the
scenario {\bf 1}, the chargino mass bound in eq.~(\ref{eq:charginop})
leads to the strongest constraint and rules out the entire $\mu, M_2
\gsim 100$ GeV range. For large values of the wino mass, $M_2 \gsim
300$ GeV, and small values of $\mu$, the phase space for the process
$e^+e^- \to \chi_1^0 \chi_3^0$ is still open and a small additional
region is ruled out. The process where the second neutralino is
produced in association with the LSP, $e^+e^- \to \chi_1^0 \chi_2^0$ ,
also plays a role at even larger values of $M_2$ and rules out another
portion of the parameter space. If the lighter chargino and
neutralinos are gaugino-like, $M_2, M_1 \ll |\mu|$, the LSP neutralino
is bino-like and the chargino mass bound in eq.~(\ref{eq:charginop})
translates into the relatively weak bound $m_{\chi_1^0} \gsim 17$ GeV.
This opens the possibility that the decay mode $Z \to \chi_1^0
\chi_1^0$ contributes to the invisible decay of the $Z$ boson. For
moderate values of $\mu$, for which the LSP has a higgsino component
and hence sizable couplings to the $Z$ boson, the constraint
$\Gamma_{\rm inv} \lsim 2$ MeV rules out a region in the $[M_2, \mu]$
plane that is not excluded by any other process.\s

In the scenario {\bf 75} the masses of the lightest chargino and of
the two lightest neutralinos are very close to each other, since
$M_1:|M_2| \sim 1:1.2$. Therefore, chargino pair production, which has
the largest cross section among the three processes with a similar
phase space $e^+e^- \to \chi_1^\pm \chi_1^\mp,~ \chi_1^0 \chi_2^0$ and
$\chi_2^0 \chi_2^0$, leads to the strongest constraint on the $[M_2,
\mu]$ parameter space.  However, since the weak-scale value of $M_3$
is also close to the values of $M_1$ and $M_2$, the constraint
$m_{\tilde g} >195$ GeV from negative searches of gluinos at the
Tevatron plays an important role, in contrast to the scenarios {\bf 1}
and {\bf 24} in which $M_3 \gsim 3$--$4 \,M_2$ and the bound of
eq.~(\ref{eq:gluinop}) is superseded by that of
eq.~(\ref{eq:charginop}). Since $M_3 \sim 1.3 \,M_2$, the entire area
$M_2 \gsim 150$ GeV is ruled out by eq.~(\ref{eq:gluinop})
independently of the value of $\mu$. \s

A similar pattern occurs in the scenario {\bf 200} where the mass
parameters $M_1,M_2$ and $M_3$ are also close to each other. However,
since here the ratio $M_3/M_2$ is larger than in the scenario {\bf
75}, the chargino mass bound is stronger than the bound from gluino
searches. Note also that, in this scenario, the LSP is wino-like for
large values of $\mu$ and the chargino mass bound translates to
$m_{\chi_1^0} \gsim 100$ GeV for the LSP and the two particles are
almost degenerate in mass (this also holds in the higgsino-like region
for the LSP).\s

Finally, the string-inspired scenario {\bf OII} is interesting in many
respects.  As in the universal scenario {\bf 1} chargino pair
production is the most constraining of all LEP production processes
and rules out the same area of the $[M_2,\mu]$ parameter
space. However, since here the smallest of the gaugino masses is the
gluino mass $M_3$, the Tevatron bound in eq.~(\ref{eq:gluinop}) rules
out a significant portion of the parameter space, namely $M_2 \lsim
230$ GeV. Furthermore, the gluino tends to be the LSP for large values
of $\mu$ when the lightest neutralino is gaugino-like. A strongly
interacting stable particle is disfavored by cosmological data,
therefore this area of the parameter space has to be excluded.\s

Another important collider constraint comes from the negative search
of Higgs bosons at LEP2 where a lower bound, $M_H \gsim 114$ GeV, has
been set on the mass of a SM-like Higgs boson \cite{PDG}.  As
previously discussed, the Higgs boson in the MSSM with heavy scalars
is SM-like but its mass is generally larger than 130 GeV (for $M_S
\gsim 10^4$ GeV and $\tan\beta$ large enough), therefore this
constraint does not apply in our case. Note, however, that there is a
chance that such a Higgs particle could be observed at the Tevatron,
either in the processes $qq \to HW \to b\bar b \ell \nu$ for $M_H \sim
130$ GeV or in the process $gg \to H \to WW^{(*)}$ for $M_H \sim 160$
GeV, if a large integrated luminosity is collected; see for instance
ref.~\cite{anatomy} for details.\s

Finally, we summarize the constraints from high-precision data and
rare decays. Because the sfermions are very heavy, the SUSY-particle
contributions to the anomalous magnetic moment $(g-2)_\mu$ of the muon
(which occur essentially through smuon--neutralino and
sneutrino--chargino loops, with possible very small contributions from
Higgs bosons) and to the rare decay of the $b$--quark into a strange
quark and a photon BR$(b \to s\gamma)$ (which at leading order occur
via loops involving the charged Higgs bosons and top quarks as well as
charginos and top squarks) are extremely small, and one is left only
with the SM contributions.  The effect of SUSY particles on the
high-precision electroweak observables measured at LEP, SLC, Tevatron
and elsewhere is also very tiny in the case of heavy scalars. Indeed,
the dominant contributions to these observables, in particular the $W$
boson mass and the effective weak mixing angle $\sin^2\theta_W$, enter
via a deviation from unity of the $\rho$ parameter (which measures the
relative strength of the neutral to charged current processes at zero
momentum transfer).  The sfermions and the non-SM Higgs bosons are
heavy enough that their contribution to this quantity is negligible.
The chargino and neutralino contributions are small because the only
terms in the mass matrices that could break the custodial SU(2)
symmetry are proportional to $M_W$. This has been verified explicitly
in the case of universal gaugino masses in ref.~\cite{Martin}, where
it has been shown that only when charginos and neutralinos have masses
very close to the experimental lower bounds can they affect, and only
slightly, the electroweak observables.

\subsection{The Dark Matter constraint}
\label{sec:DM}

As deduced from the WMAP satellite measurement of the temperature
anisotropies in the Cosmic Microwave Background, cold Dark Matter
makes up approximately 25\% of the energy of the
Universe~\cite{WMAP}. The DM cosmological density is precisely
measured to be
\beq \Omega_{\rm DM}\, h^2 ~=~ 0.111^{+0.006}_{-0.008}~,
\label{eq:omegah2}
\eeq
which leads to $0.088 \leq \Omega_{\rm DM}\, h^2 \leq 0.128$ at the
3$\sigma$ level. The accuracy is expected to be improved to the
percent level by future measurements at Planck.\s

As is well known, the LSP neutralino is an ideal candidate for the
weakly interacting massive particle that is expected to form this
cold DM \cite{DM-review} and in some areas of the SUSY parameter space
the $\chi_1^0$ cosmological relic density, which is inversely
proportional to the neutralino annihilation cross section $\sigma_{\rm
ann} \equiv \, \sigma(\chi_1^0 \chi_1^0 \rightarrow {\rm SM\,
particles})$, falls in the range required by WMAP. In the MSSM with
heavy scalars, there are essentially only three regions (see later for
a fourth possibility) in which this constraint is satisfied:

\begin{itemize}

\item[--] The ``mixed region" in which the LSP is a higgsino--gaugino
mixture, $M_1 \sim |\mu|$, which enhances (but not too much) its
annihilation cross sections into final states containing gauge and/or
Higgs bosons and top quarks, $\chi_1^0 \chi_1^0 \to W^+ W^-, ZZ, HZ,
HH$ and $t\bar t$.

\item[--] The ``pure higgsino" and ``pure wino" regions, in which the
LSP is almost (but not exactly) degenerate in mass with the lightest
chargino and the next-to-lightest neutralino, leading to and enhanced
destruction of sparticles since the $\chi_1^+, \chi_2^0$
co-annihilation cross sections are much larger than that of the LSP;
this solution generally requires LSP masses beyond 1 TeV.

\item[--] The ``$H$-pole'' region in which the LSP is rather light,
$m_{\chi_1^0} \sim \frac12 M_H$, and the $s$-channel $H$ exchange is
nearly resonant allowing the neutralinos to annihilate efficiently.

\end{itemize}

The latter scenario has been discussed in the usual MSSM \cite{hpole}
in which the Higgs boson, which has a mass below $\sim 130$ GeV,
decays mostly into $b\bar b$ pairs. However, if the common scalar mass
$\msusy$ is very large, as is the case in the present discussion, the
Higgs mass will be larger reaching, for instance, the value $M_H \sim
150$ GeV for $\msusy \sim 10^{10}$ GeV. In this case the dominant
decay mode of the Higgs boson is a pair of $W$ bosons, one being on
mass-shell while the other is virtual and decays into two massless
fermions, $H \to WW^* \to W f\bar f$.\s

The partial decay width for this decay is given by \cite{HWWdecay}
\beq 
\Gamma (H \ra WW^*) &=& \frac{3 G_\mu^2 M_W^4}{16 \pi^3} M_H \left[ 
\frac{3(1-8x+20x^2)}{(4x-1)^{1/2}} \arccos \left( \frac{3x-1} {2x^{3/2}} 
\right) \right. \non \\
&-& \left. \frac{1-x}{2x} (2-13x+47x^2) - \frac{3}{2}(1-6x+4x^2) \log x 
\right]~,
\eeq
with $x=M_W^2/M_H^2$.  Using the two-body-like formula given above and
the approximation of ignoring the kinematical effects induced by the
three-body final-state phase space, we have implemented this
additional channel in a DM {\sc Fortran} code based on
ref.~\cite{DN-routine} which is linked to the program {\tt
SuSpect}. Using this routine we have scanned the $[M_2, \mu]$
parameter space and determined the regions in which the WMAP
constraint is fulfilled\footnote{For some selected points of the
parameter space, we have verified that the results that we obtain for
$\Omega h^2$ using the routine of ref.~\cite{DN-routine} are in a
relatively good agreement with those obtained with the program {\tt
micrOMEGAs} \cite{micromegas}.}. This is performed not only in the
scenario with a universal gaugino mass $m_{1/2}$ at the GUT scale but
also in scenarios with different boundary conditions.\s

\begin{figure}[t]
\begin{center} 
\epsfig{figure=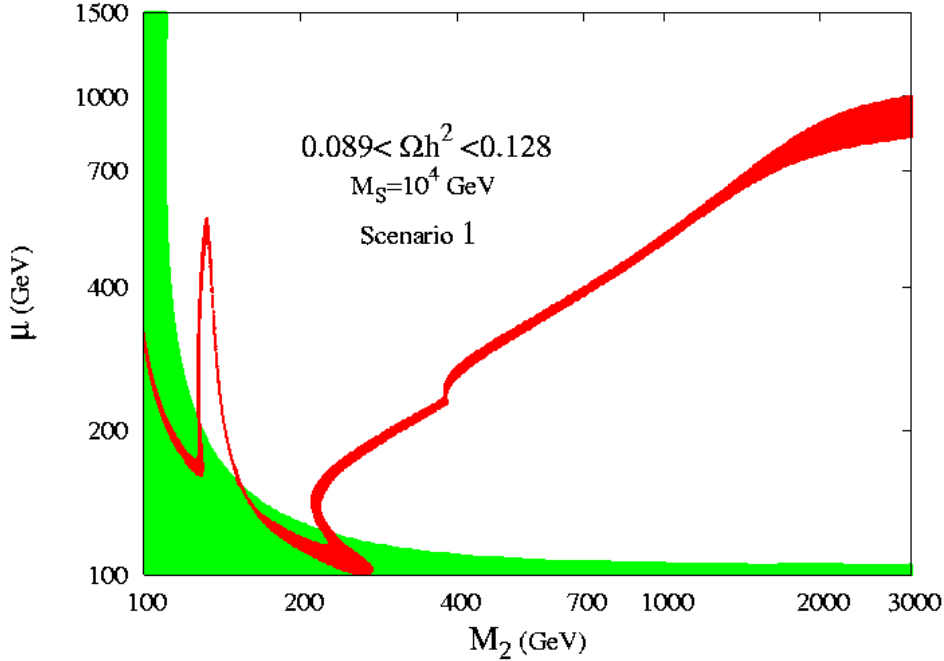,width=15cm}
\vspace*{-.3cm}
\caption{The regions of the $[M_2,\mu]$ parameter space in which the
WMAP constraint is fulfilled (red area) for a common scalar mass value
$\msusy=10^4$ GeV, $\tb=30$ and the assumption of a universal gaugino mass at
the GUT scale. The green area on the left and the bottom is the one
excluded by direct searches of SUSY particles.}
\label{fig:DM1}
\end{center} 
\vspace*{-.3cm}
\end{figure}

In fig.~\ref{fig:DM1} we display the area in the $[M_2,\mu]$ parameter
space in which the WMAP constraint is fulfilled; a common scalar mass
$\msusy=10^4$ GeV is chosen and a universal gaugino mass at $M_{\rm
GUT}$ is assumed; here and in the subsequent discussions we will fix
the value of $\tb$ to $\tb=30$.  The green (light grey) area in the
left and bottom parts of the figure denotes the region excluded by the
collider data discussed in the previous subsection.  The peak for
small $M_2$ values, $M_2 \sim 2M_1 \sim M_H$, is due to the
$s$-channel exchange of the Higgs boson, $\chi_1^0 \chi_1^0 \to
H$. For the mass value obtained here, $M_H \sim 130$ GeV, the Higgs
boson mainly decays into $b\bar b$ final states while the $H \to WW^*$
channel, which has also been included, has a smaller branching ratio
and does not play a leading role.  Between the two bands of the peak
one is too close to the Higgs mass pole, and the LSP annihilation is
too efficient leading to a too small $\Omega h^2$. The peak reaches up
to $\mu \sim 600$ GeV, a value beyond which the LSP is almost
bino-like and its coupling to the Higgs boson is too small (the Higgs
prefers to couple to a higgsino--gaugino mixture) to generate a
sizable annihilation cross section.\s

For larger $\mu$ and $M_2$ values there is an almost straight band in
which $\mu \sim M_1$ and the LSP is a bino--higgsino mixture with
sizable couplings to $W,Z$ and Higgs bosons, allowing for reasonably
large rates for neutralino annihilation into $\chi_1^0 \chi_1^0 \to
W^+W^-, ZZ, HZ$ and $HH$ final states. For instance, for $M_2\sim 300$
GeV and $\mu=200$ GeV, the annihilation cross section is mostly due to
the $WW$ and $HH$ final states ($\sim 40\%$ for both channels) and, to
a lesser extent, the $ZZ$ and $ZH$ final states ($\sim 10\%$
each). For slightly larger $\mu$ and $M_2$ values there is a jump due
to the opening of the $\chi_1^0 \chi_1^0 \to t \bar t$ channel, which
then dominates the annihilation cross section. Above the band and
below the band, the LSP couplings to the various final states are
either too strong or too weak to generate the relevant relic
density. For $\mu$ values close to 1 TeV and even larger values of
$M_2$ there is a wider area in which the WMAP constraint is also
fulfilled. In this region the LSP is almost a pure higgsino and a
correct $\Omega h^2$ can also be obtained thanks to the
co-annihilation of the LSP with the $\chi_1^\pm$ and $\chi_2^0$
states.  For lower $\mu$ values and $M_2$ still very large the LSP
co-annihilation with $\chi_1^\pm$ and $\chi_2^0$ is too strong and
leads to a too small $\Omega h^2$. \s

\begin{figure}[t]
\begin{center} 
\mbox{\hspace*{-1.1cm}
\epsfig{figure=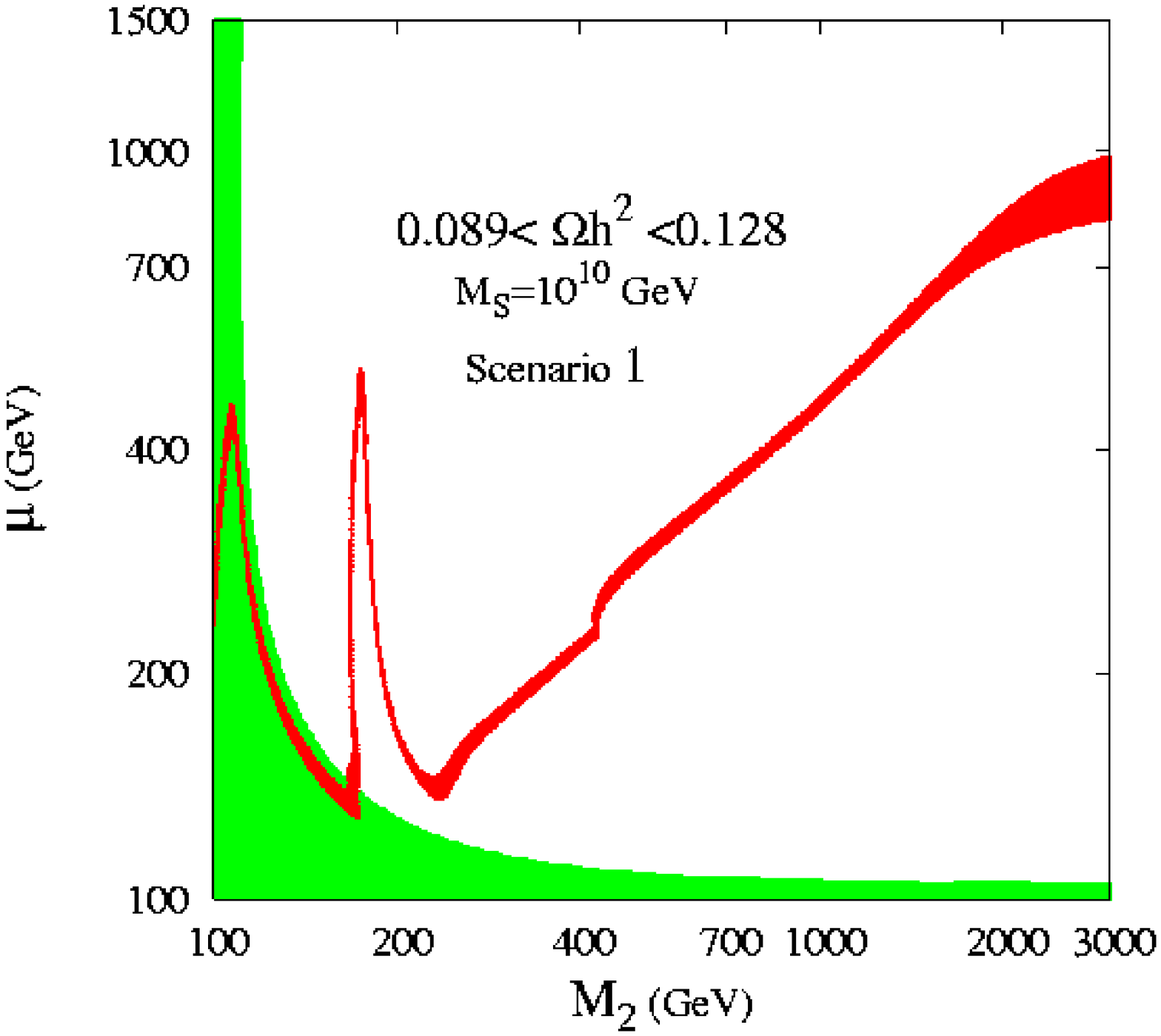,width=12cm,height=9.5cm} \hspace*{-3.5cm}
\epsfig{figure=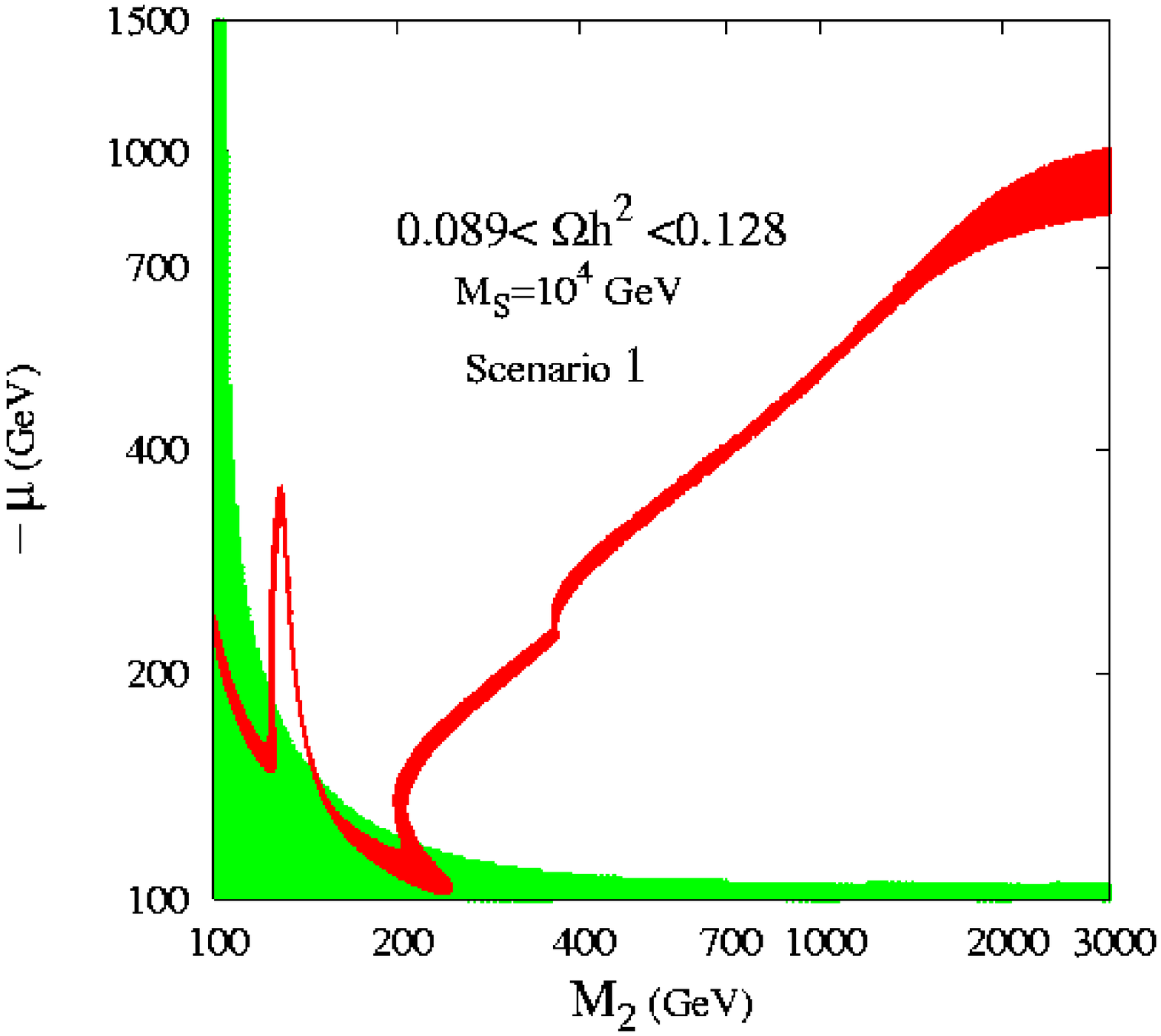,width=12cm,height=9.5cm} }
\vspace*{-1.cm}
\caption{The same as in fig.~\ref{fig:DM1} but with $\msusy=10^{10}$
GeV (left) or $\mu<0$ (right)}
\label{fig:DM2}
\end{center} 
\vspace*{-4mm}
\end{figure}

Figure \ref{fig:DM2} is similar to the previous one, with the
difference that $\msusy=10^{10}$ GeV (left pane) or the sign of the
parameter $\mu$ is reversed (right pane). The figure in the left-hand
side shows similar features as fig.~\ref{fig:DM1}, except that the
peak due to the $s$-channel Higgs boson exchange is shifted to a
slightly higher $M_2$ value, $M_2 \sim 2 M_1 \sim M_H \sim 150$
GeV. In this case the new annihilation channel $\chi_1^0 \chi_1^0 \to
H \to WW^* \to Wf \bar f$ discussed before gives a significant
contribution to the total cross-section.  For negative $\mu$ values
the $[M_2,\mu]$ area of the parameter space that leads to the desired
$\Omega h^2$ is also similar to fig.~\ref{fig:DM1}. The only
difference is that the Higgs peak reaches up to $|\mu| \sim 400$ GeV
only, as for negative $\mu$ values the LSP becomes bino-like more
quickly than in the positive case and its couplings to the Higgs boson
are thus smaller.\s

\begin{figure}[!t]
\vspace*{-5mm}
\begin{center}
\mbox{\hspace*{-.9cm}
\hspace*{-.4cm}\epsfig{figure=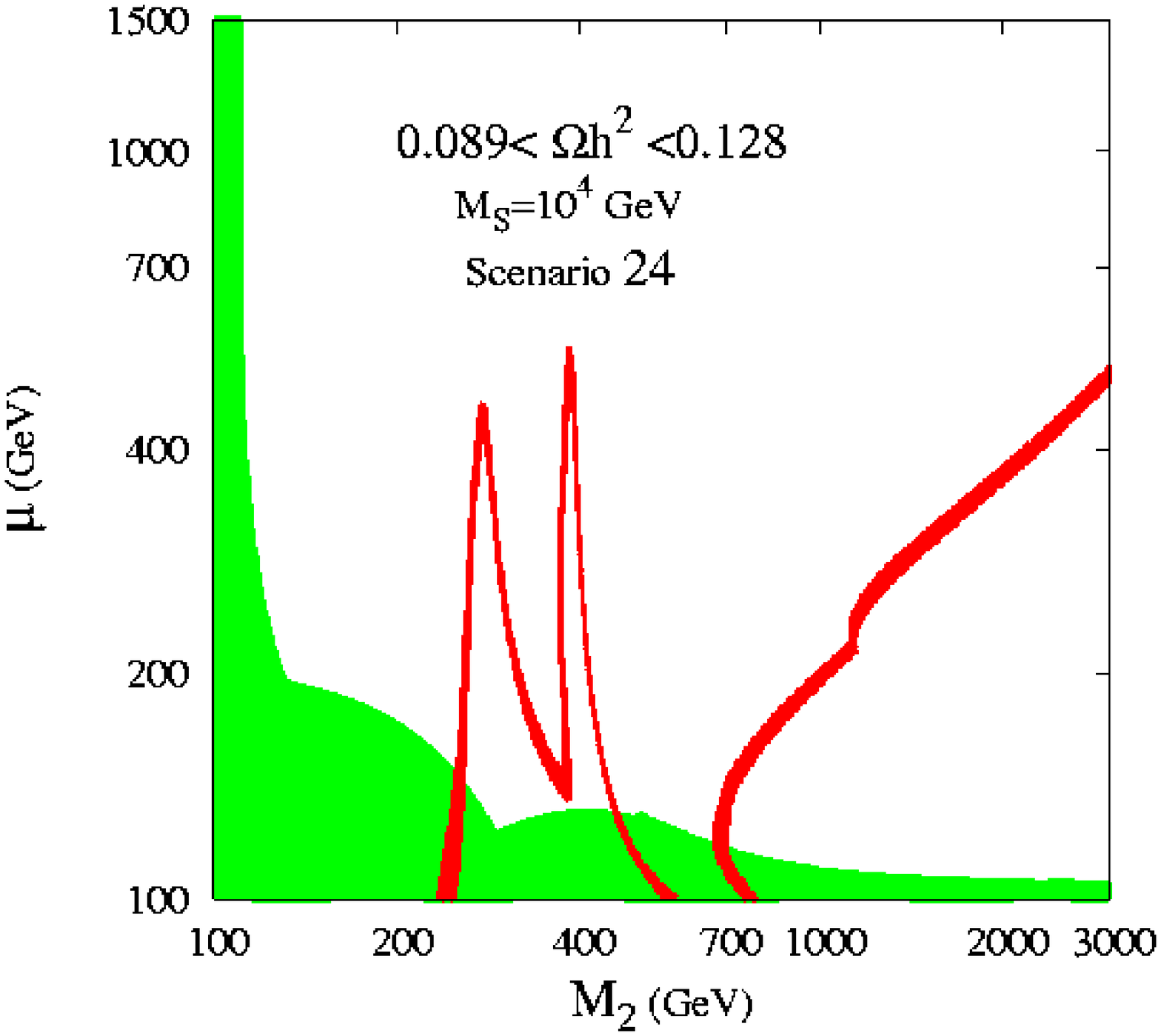,width=12cm,height=9.5cm} \hspace*{-3.5cm}
\epsfig{figure=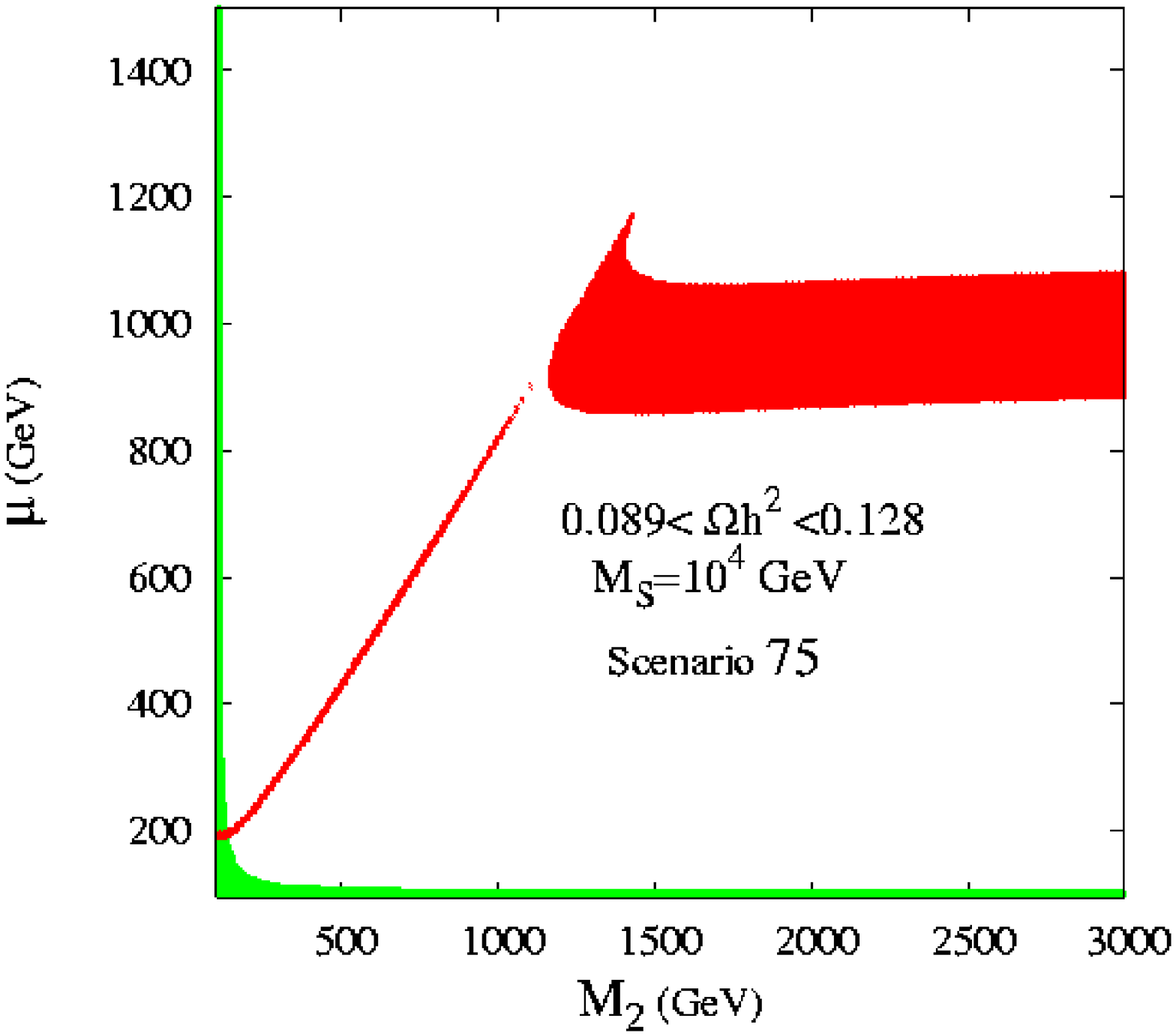,width=12cm,height=9.5cm} }
\vspace*{-1.5cm}
\mbox{\hspace*{-.9cm}
\hspace*{-.4cm}\epsfig{figure=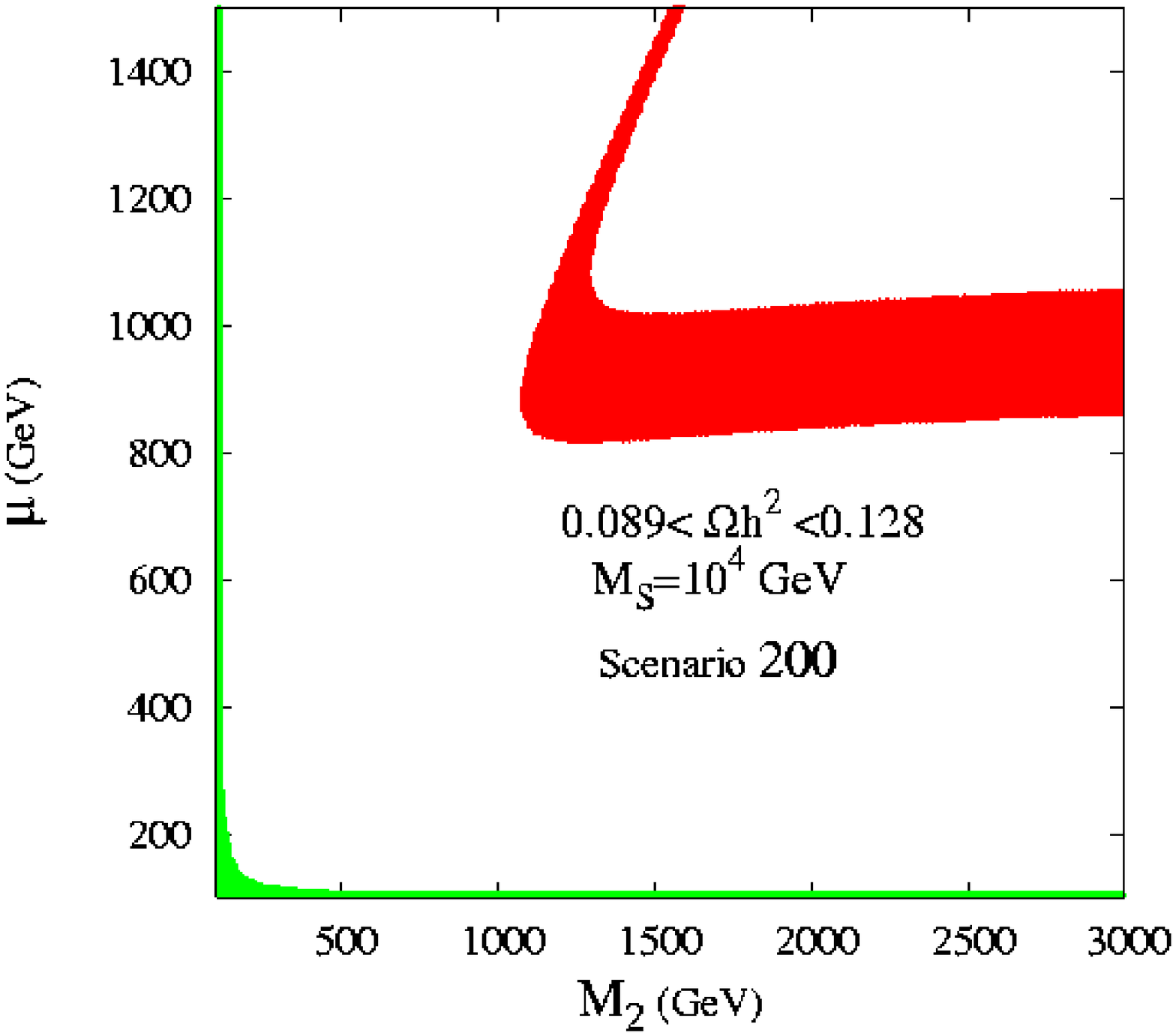,width=12cm,height=9.5cm} \hspace*{-3.5cm}
\epsfig{figure=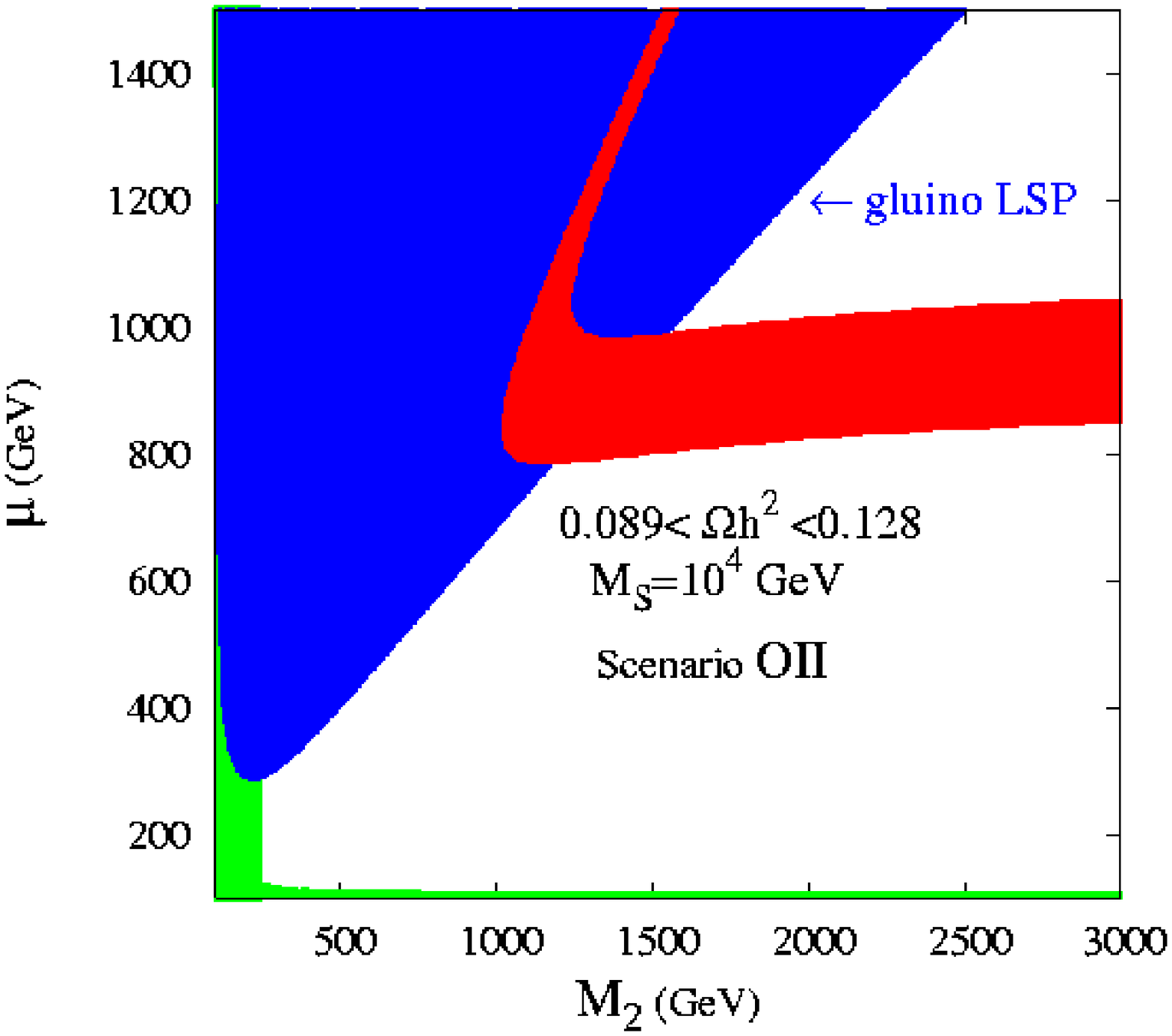,width=12cm,height=9.5cm} }
\end{center}
\vspace*{.5cm}
\caption{The regions of the $[M_2, \mu]$ parameter space in which the
WMAP constraint is fulfilled for $\msusy=10^4$ GeV in the various models
with non-universal GUT-scale boundary conditions for the gaugino mass
parameters.}
\label{fig:DM3}
\vspace*{-3mm}
\end{figure}

Figure \ref{fig:DM3} shows the $[M_2,\mu]$ area that is compatible
with WMAP results for $\msusy=10^4$ GeV in the case where the boundary
conditions for the gaugino masses at the GUT scale are not
universal. In the scenario {\bf 24} the same trend as for the
universal case occurs but with two major differences. First, the Higgs
peak is now shifted to $M_2 \sim 400$ GeV, a mere consequence of the
fact that the ratio of the weak-scale wino-to-bino masses is much
larger in this model, $M_2\!:\! M_1 \sim 6$, than in the universal
case, $M_2\!:\!  M_1 \sim 2$; for the same reason, the band at large
$M_2$ values is shifted downward compared to fig.~\ref{fig:DM1}. The
other major difference with the universal scenario is that, despite
the constraint on the invisible $Z$ decay width $\Gamma (Z \to
\chi_1^0 \chi_1^0) \lsim 2$ MeV and the one from chargino and
neutralino production at LEP2, the possibility that the LSP mass is
close to $\frac12 M_Z$ is not excluded. In this case, the LSP
annihilation channel $\chi_1^0 \chi_1^0 \to Z \to f\bar f$ can become
resonant, thus generating the required cosmological relic
density. This leads to a peak similar to the one due to Higgs boson
exchange but at smaller $M_2$ values, $M_2 \simeq 6 \,M_1 \simeq
3\,M_Z \sim 300$ GeV.\s

In the scenario {\bf 75} one has $M_1\!:\!M_2\!:\!M_3 \sim
1\!:\!-1.2\!:\!-1.5$ at the weak scale, so that the LSP is in general
close in mass to the lightest chargino and the next-to-lightest
neutralino, and co-annihilation of these states plays a very important
role. Indeed, in the thin straight line below $\mu, M_2 \lsim 1$ TeV
the LSP is dominantly bino-like (e.g. $\sim 85\%$ for $\mu=M_2 \sim
600$ GeV) with a small higgsino component, and the WMAP $\Omega h^2$
range is obtained with the efficient co-annihilation of $\chi_2^0
\chi_1^\pm, \chi_1^\mp \chi_1^\pm$ ($\sim 50\%$) and $\chi_1^0
\chi_1^\pm, \chi_1^0 \chi_2^0$ ($\sim 25\%)$ with the remaining part
due to $\chi_1^0 \chi_1^0$ annihilation.  In the large band with
$\mu=900$--1100 GeV for $M_2 \gsim 1.2$ TeV the LSP is very close to
be a pure higgsino state (e.g. $\sim 99\%$ for $M_2\sim 2\mu \sim 2$
TeV), but the co-annihilation cross sections are still reasonably
small, the mass difference between the LSP and the other
chargino/neutralino states being large enough.  All other areas,
including the Higgs peak and the mixed gaugino--higgsino areas that
appear in the universal case, are ruled out because of a too efficient
co-annihilation of the LSP.\s

The same situation occurs in the scenario {\bf 200}, in which the
weak-scale gaugino mass parameters are given by $M_1\!:\!M_2\!:\!M_3
\sim 2.4\!:\!1\!:\!1.9$ leading to $\chi_1^0$ and $\chi_1^\pm$ states
which are wino- or higgsino-like and almost mass-degenerate in most of
the parameter space. In this case, co-annihilation with $\chi_1^\pm$
and $\chi_2^0$ is too efficient except for the band with
$\mu=900$--1100 GeV and $M_2 \gsim 1.2$ TeV as in the scenario {\bf
75}. In this case the band is extended at larger $\mu$ values by a
strip in which the LSP is a wino--higgsino mixture, but, again, the
bulk of the relic cosmological density $\Omega h^2$ is generated
through co-annihilation of the LSP with the lightest chargino
$\chi_1^\pm$. \s

Finally, in the case of the {\bf OII} string model, which leads to a
weak-scale gluino mass parameter that is smaller than the wino and
bino mass parameters, $M_1\!:\!M_2\!:\!M_3 \sim 1.4\!:\!1.3\!:\!1$ for
$M_S=10^4$ GeV, the gluino is the LSP in a large part of the parameter
space and this blue (dark grey) area should therefore be excluded.
For $M_2$ smaller than 1 TeV the LSP is higgsino-like at low $\mu$
values and co-annihilation of the LSP with $\chi_1^\pm$ and $\chi_2^0$
is too efficient, while at high $\mu$ values the LSP is the
gluino. Only in a relatively narrow band, $\mu \sim 800$--1100 GeV,
similar to the ones observed in the scenarios {\bf 75} and {\bf 200},
does one obtain the $\Omega h^2$ range required by WMAP, with the LSP
being an almost pure higgsino state with a relatively large mass; for
instance, for $M_2 \sim 2.5\, \mu \sim 2.5$ TeV, one has $m_{\chi_1^0}
\sim m_{\chi_2^0} \sim m_{\chi_1^\pm} \sim 1$ TeV and $\Omega h^2$ is
almost exclusively generated by co-annihilation (95\%). A very
interesting feature occurs at the border between the gluino-LSP area
and the band with $\mu \sim 0.8$--1 TeV: the main channel that leads
to the required range for the relic density is gluino annihilation
into gluon and quark pairs, $\tilde g \tilde g \to q\bar q, gg$ (the
co-annihilation $\tilde g \tilde \chi \to q\bar q$ is suppressed as it
is mediated by the super-heavy squarks).  For instance, in the point
$M_2 \simeq 1.5$ TeV and $\mu\simeq 900$ GeV one obtains the value
$\Omega h^2 \sim 0.1$ with the $\tilde g \tilde g \to gg$ (45\%) and
$\tilde g \tilde g \to q\bar q$ (35\%) reactions. This is one of the
rare examples within constrained SUSY models where gluino
co-annihilation is at work.\s

\newpage

\subsection{The gluino lifetime}

In this section we summarize the constraints that can be obtained on
the parameter space, and in particular on $\msusy$, from the
requirement that the gluino lifetime does not exceed the age of the
Universe, $\tau_{\tilde g} \lsim 14$ Gyear, a possibility that is
excluded from the absence of anomalous isotopes. We extend the
discussions of Refs.~\cite{Toharia,ADGPW,GGS}, held in the context of
universal gaugino masses, to the various scenarios with non-universal
boundary conditions at $M_{\rm GUT}$.\s

The total decay width of the gluino, $\Gamma_{\tilde g}= \hbar
/\tau_{\tilde g}$, has been calculated following ref.~\cite{GGS},
where the large logarithmic corrections that appear for very heavy
scalars and are controlled by the strong coupling $\alpha_s$ and by
the top Yukawa coupling are resummed with an effective Lagrangian
approach. Both the three-body decays into charginos or neutralinos and
a quark--antiquark pair through the exchange of heavy squarks, $\tilde
g \to q \tilde q^* \to q \bar q \chi_i$, and the loop induced decay
into a neutralino and a gluon, $\tilde g \to g \chi_i^0$ have been
included (see section \ref{sec:gludec} for more details). The gluino
lifetime is approximately given by \cite{GGS}
\beq 
\tau_{\tilde g} = \frac{\hbar}{\Gamma_{\tilde g}} =
\frac{4\,{\rm sec.}}{N} \left( \frac{M_S} {\rm 10^9~GeV} \right)^4 \,
\left( \frac{\rm 1~TeV}{ m_{\tilde g} } \right)^5 ,
\eeq
where $N$ is a normalization factor which is generally of order unity
if phase-space effects are ignored. This equation exhibits the main
trend: the gluino lifetime is larger for higher values of the scalar
mass $M_S$ and smaller values of the gluino mass $m_{\tilde g}$. In
Refs.~\cite{Toharia,ADGPW,GGS} it has been shown in the universal
scenario that for $M_S={\cal O}(10^{13})$ GeV the gluino is almost
stable as its lifetime is larger than the age of the Universe.\s

\begin{figure}[!t]
\vspace*{3mm}
\begin{center} 
\rotatebox{-90}{\epsfig{figure=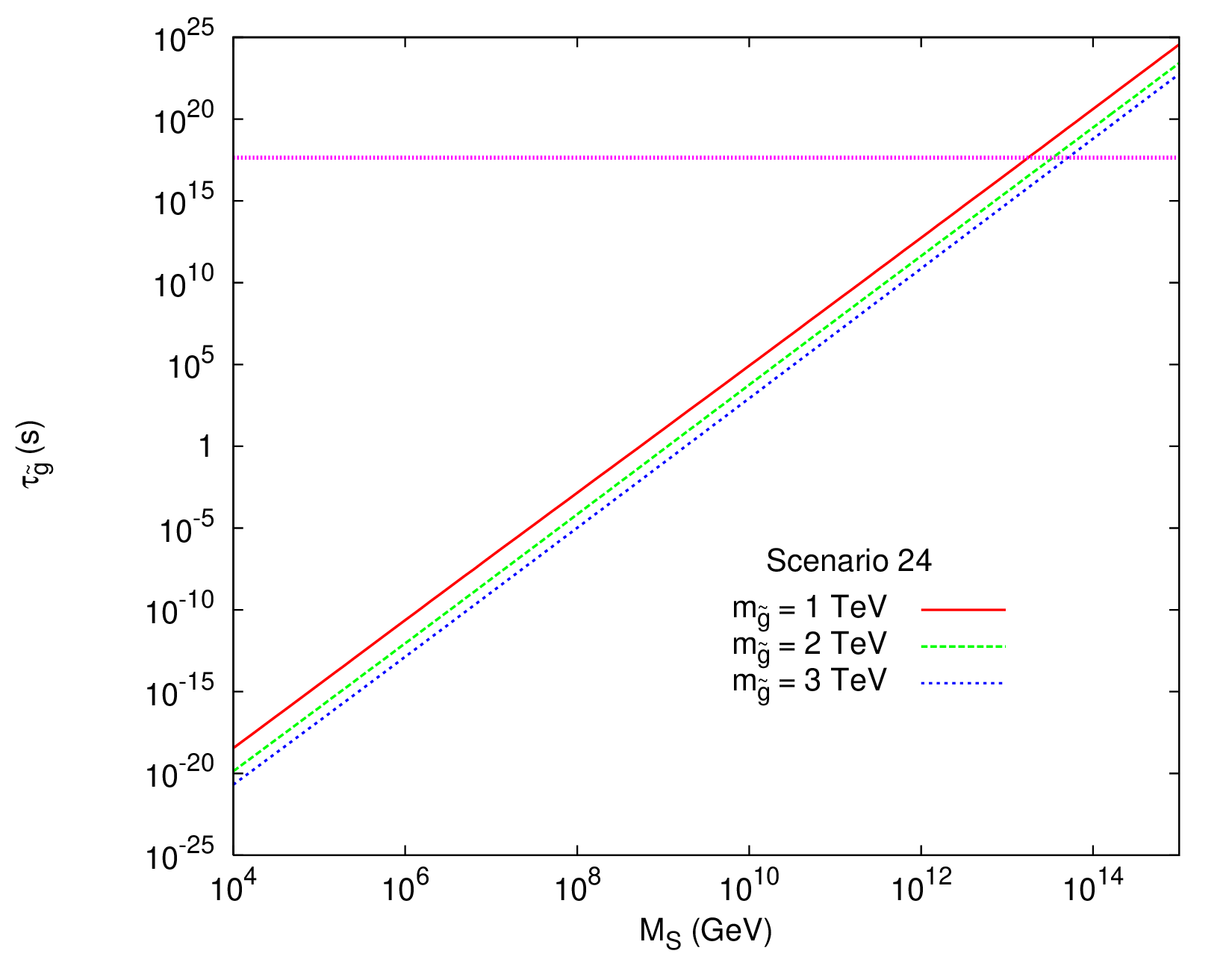,width=8.2cm,height=12.5cm}}\\[0.7cm] 
\rotatebox{-90}{\epsfig{figure=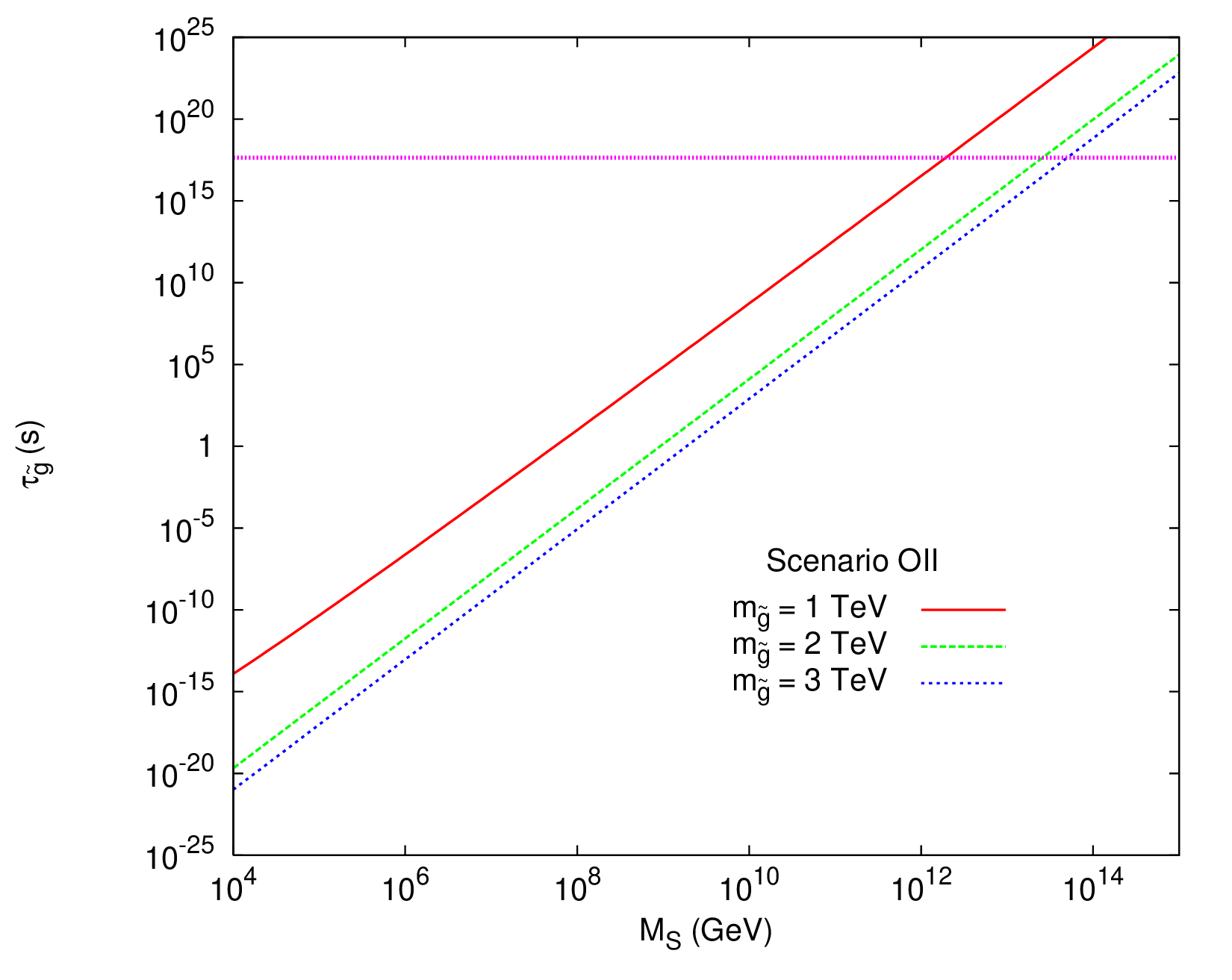,width=8.2cm,height=12.5cm}} 
\caption{The gluino lifetime (in seconds) as a function of the common
scalar mass $M_S$ for the two non-universal scenarios {\bf 24} and
{\bf OII} for various values of $m_{\tilde g}$ and the parameter $\mu$
fixed from the DM requirement; the horizontal line is for the upper
bound of 14 Gyear.}  \label{fig:TauG}
\end{center} 
\end{figure}

The gluino lifetime $\tau_{\tilde g}$ is displayed in
fig.~\ref{fig:TauG} as a function of $M_S$ for various values of the
gluino mass in the two non-universal scenarios {\bf 24} and {\bf OII};
for each value of $m_{\tilde g}$ and hence of $M_2$ (the latter
obtained via RG evolution from the boundary conditions specific to the
scenario) the relevant value of $\mu$ is obtained by requiring that
the LSP relic abundance $\Omega_\chi h^2$ falls in the range allowed
by WMAP (see fig.~\ref{fig:DM3}).  As can be seen, in the scenario
{\bf 24} a gluino lifetime of the order of the age of the Universe,
$\tau_{\tilde g}=14$ Gyear (the horizontal line), is obtained for
$M_S= 10^{13}$ to $10^{14}$ GeV for $m_{\tilde g} =1$--3 TeV. The
results are thus similar to the universal scenario discussed in
Refs.~\cite{wells,GGS}. In fact, the same results are also obtained in
the non universal scenarios {\bf 75} and {\bf 200} and we refrain from
showing them again.\s

A slightly different situation occurs in the scenario {\bf OII}, in
which the correct DM relic density can be generated when the gluino is
close in mass to the higgsino-like neutralino LSP. In this case the
only allowed decays of the gluino are the three-body decays into two
light quarks and the higgsino-like chargino and neutralinos, and the
loop-induced two-body decay into the higgsino-like neutralinos and a
gluon (see section 4.3).  These decays are suppressed by phase space:
$\Gamma \propto (m_{\tilde g} - m_\chi)^5$ for the three-body decays
and $\Gamma \propto (m_{\tilde g} - m_\chi)^3$ for the radiative
decay. Therefore, smaller values of $M_S$ are required for the gluino
lifetime to be compatible with the age of the Universe. This is
exemplified in the right-hand side of fig.~\ref{fig:TauG}, where one
can see that for $m_{\tilde g} = {\cal O}(1$ TeV) (with $\mu \sim 1$
TeV for $\Omega_\chi h^2$ to fall in the WMAP range) a gluino lifetime
of $\tau_{\tilde g}=14$ Gyear is already reached for $M_S \lsim
10^{12}$ GeV.

\section{Decays of the Higgs and SUSY particles}
\label{section4}

\subsection{Higgs boson decays}

As discussed in section 2.1, in the MSSM with heavy scalars the Higgs
boson is SM-like, except that its mass is constrained to be in the
range $M_H \sim 130$--160 GeV for $\msusy=10^ 4$--$10^{14}$ GeV and 
$\tb=30$ (as will be assumed throughout this section). It
will thus decay mostly like the SM Higgs particle in this mass range
\cite{anatomy,HSMdecays}. For not too large $\msusy$ values for which
$M_H \lsim 130$ GeV, the Higgs boson decays into a large variety of
channels, the main modes being by far the decay into $b\bar{b}$ pairs
with a branching ratio of $\sim 90\%$ followed by the decays into
$c\bar{c}$ and $\tau^+\tau^-$ pairs with branching ratios of $\sim
5\%$.  Also of significance, the top-loop mediated Higgs decay into
gluons which for $M_H$ around 130 GeV occurs at the level of few
percent. The top- and $W$-loop mediated $\gamma\gamma$ and $Z \gamma$
decay modes are very rare, with branching ratios of ${\cal O
}(10^{-3})$. However, these decays lead to clear signals and are
theoretically interesting, being sensitive to new electrically charged
particles such as charginos.  For values of $\msusy$ large enough that
$M_H \gsim 140$ GeV the Higgs bosons decay into $WW$ and to a lesser
extent $ZZ$ pairs, with one of the gauge bosons being virtual below
the threshold. For $\msusy \gsim 10^{10}$ GeV, which leads to $M_H
\gsim 150$ GeV, the Higgs boson decays almost exclusively into two
real $W$ bosons; the decay $H \to ZZ^*$ is strongly suppressed as one
of the $Z$ boson must be virtual.  In all cases, the Higgs boson is
very narrow, as its total decay width does not reach the 1 GeV
level.\s

\begin{figure}[t] 
\begin{center}
\mbox{ \hspace*{-9mm}
\epsfig{figure=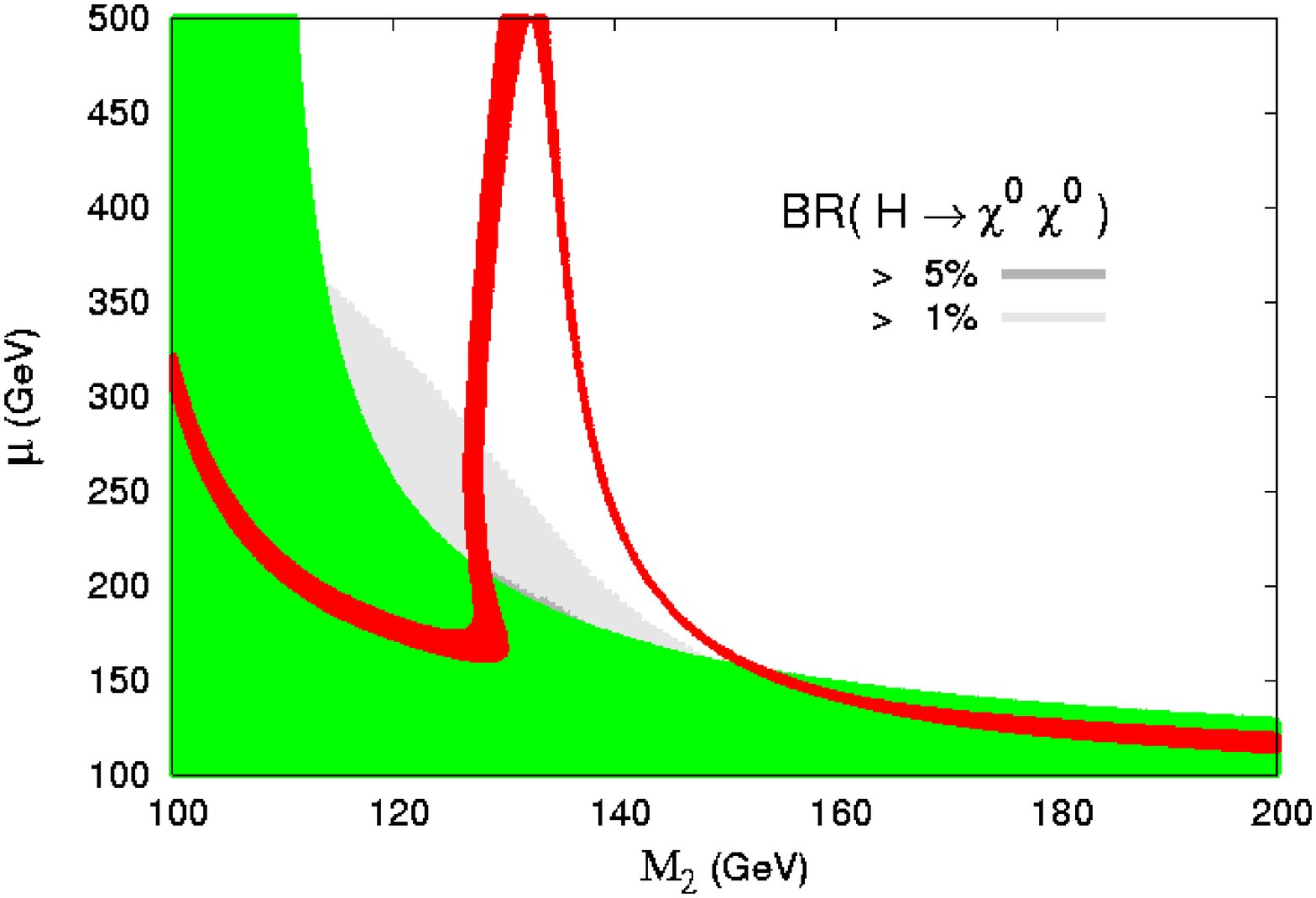,width=9.2cm}
\hspace{-9mm}
\epsfig{figure=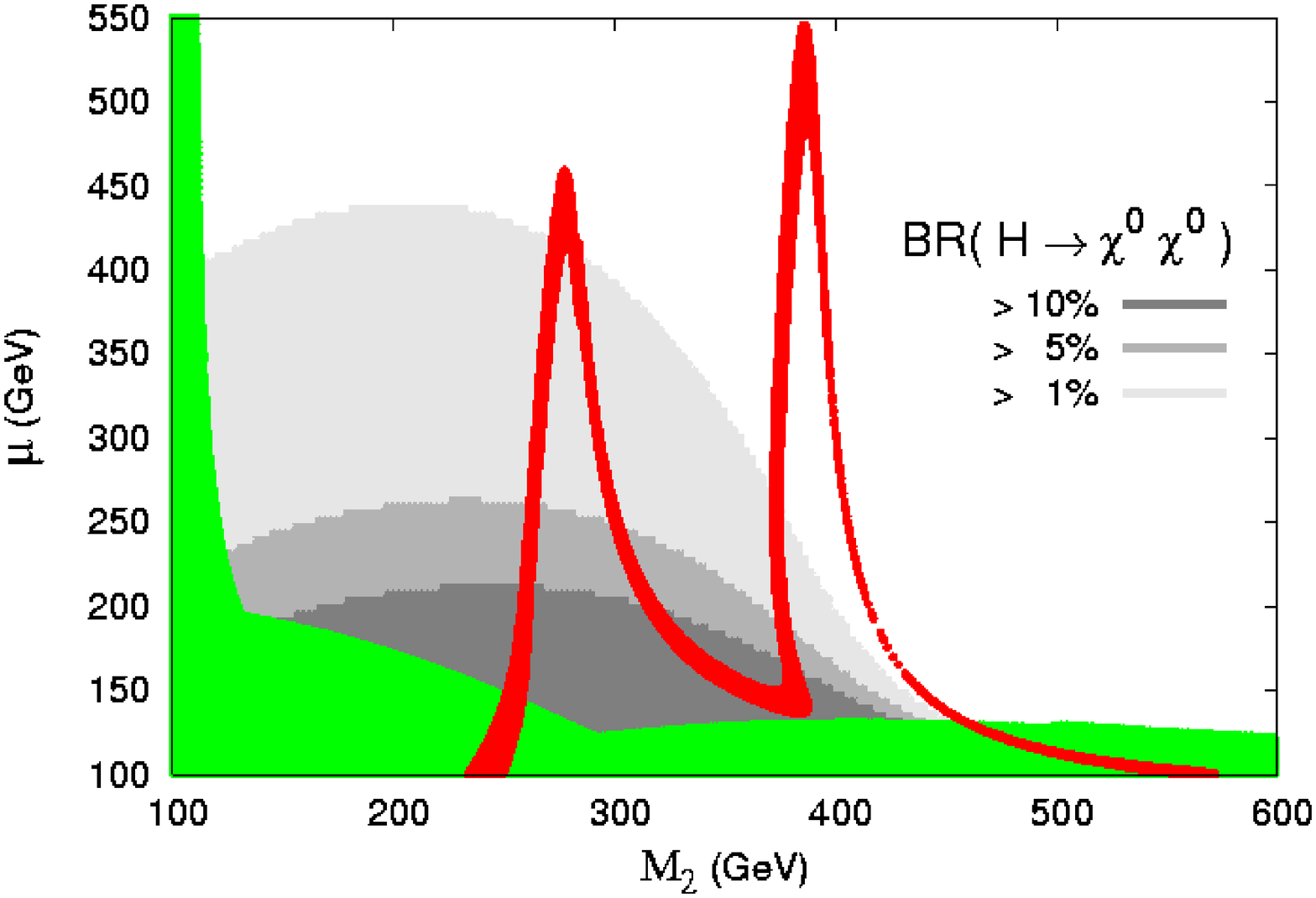,width=9.2cm}}
\end{center}
\vspace*{-7mm}
\caption{The areas in the $[M_2,\mu]$ plane in which the invisible
Higgs branching ratios BR$(H \to \chi_1^0 \chi_1^0)$ in larger than
1\%, 5\% and 10\% in the scenarios {\bf 1} and {\bf 24} with
$\msusy=10^4$ GeV and $\tb=30$; the region in which the relic density
is compatible with WMAP is also displayed (red region). The program
{\tt HDECAY} \cite{HDECAY}, adapted to deal with heavy scalars, has
been used.}
\label{dec:Higgs}
\end{figure}

There are, however, two situations in which the Higgs boson might have
decays that are slightly different from those of the SM Higgs
particle. First, for very light LSP neutralinos, the invisible decay
$H \to \chi_1^0 \chi_1^0$ \cite{Hinvisible} might be kinematically
accessible. In the scenario with universal gaugino masses leading to
$M_2 \sim 2\,M_1$ at the weak scale, this decay occurs for values of
$M_2$ small enough to have the phase space needed for the decay to
occur, $M_H \gsim 2 m_{\chi_1^0} \sim 2\,M_1$. In the left-hand side
of fig.~\ref{dec:Higgs} the areas in the $[M_2,\mu]$ parameter space
in which the branching ratio BR$(H \to \chi_1^0 \chi_1^0)$ is larger
than 1\%, 5\% and 10\% are shown for $\msusy=10^4$ GeV and $\tb=30$;
the area in which the neutralino relic density is in the range
required by WMAP is also displayed, as well as the area excluded by
collider bounds. For $M_2$ values in the range 120--150 GeV and small
$\mu$ values, the branching ratio is of the order of 5\% and sometimes
10\% and is therefore measurable at the ILC. The branching ratio drops
with increasing $\mu$ values since for $\mu \gg M_2$ the Higgs--LSP
coupling becomes too small.  In the non-universal scenario {\bf 24}
with the weak-scale relation $M_2\sim 6 M_1$ between the wino and bino
masses, fig.~\ref{dec:Higgs} (right), the corresponding areas in the
$[M_2,\mu]$ parameter space are larger as a result of a larger phase
space allowed for the invisible Higgs decays. In particular, branching
ratios larger than 10\% are possible in a significant portion of the
parameter in which the DM constraint is also fulfilled. Thus, if by
chance it is the $H$ or $Z$ boson pole which provides the correct
value of $\Omega h^2$, the invisible decay branching ratio BR$(H \to
\chi_1^0 \chi_1^0)$ could be measured at the ILC and would allow to
access directly to the Higgs--LSP couplings.\s

The decays of the Higgs boson into the heavier neutralinos and the
charginos are in general kinematically closed if one takes into
account the LEP2 bounds on the masses of these particles. There is
however one possible exception: in the scenario {\bf 24}, because the
absolute lower limit on the LSP mass is only $m_{\tilde \chi_1^0}
\gsim 17$ GeV, the possibility of the decay $H\to \chi_1^0 \chi_2^0$
is still open. This occurs for values of $M_1$ and $M_2$ that lie very
close to those ruled out by the experimental constraints and, for such
values, the requirement that the LSP provides the correct relic
density is not fulfilled.\s

Another possibility for a deviation from the SM predictions for the
Higgs decays is when the charginos are light enough to contribute to
the loop decay $H \to \gamma \gamma$ (contributions to the $H \to Z
\gamma$ decays are in general much smaller) \cite{Hgamma}. Indeed,
besides top quark and $W$ boson loops, one has to include also the
contribution of the chargino loops. However, in contrast to the SM
particles, the charginos do not couple to the Higgs boson
proportionally to their masses and the amplitudes are damped by
inverse powers of the $\chi^\pm$ masses.  The chargino contributions
are thus sizable only for relatively small masses; see also
ref.~\cite{SP-Hgamma} in which this topic has been discussed in the
\spsd\ scenario. \s

\begin{figure}[t] 
\begin{center}
\mbox{
\hspace*{-1.5cm}
\epsfig{figure=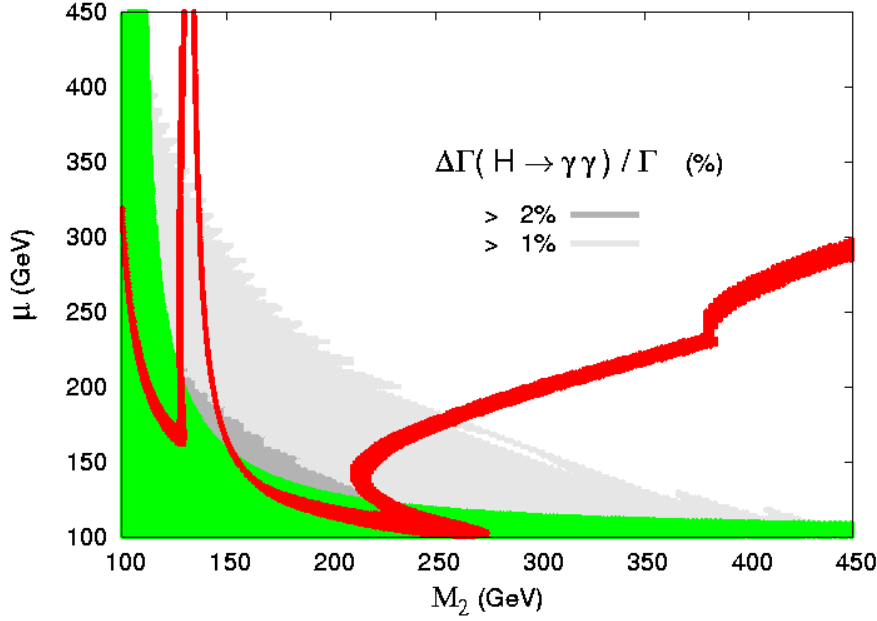,width=14cm} }
\end{center}
\vspace*{-8mm}
\caption{The areas of the $[M_2,\mu]$ plane in which the deviation of
$\Gamma(H \to \gamma \gamma)$ from the SM prediction is larger than
1\% and 2\% for $M_S=10^4$ GeV; the area where the WMAP DM constraint
in fulfilled is also shown. The program {\tt HDECAY} \cite{HDECAY},
adapted to deal with heavy scalars, has been used.}
\label{dec:Higgs2}
\vspace*{-1mm}
\end{figure}

This is exemplified in fig.~\ref{dec:Higgs2}, where the regions in the
$[M_2,\mu]$ plane in which the deviation of $\Gamma(H\to \gamma
\gamma)$ from SM prediction is larger than 1\% and 2\% are displayed
in the universal gaugino mass scenario with $M_S=10^4$ GeV; the area
where the WMAP DM constraint in fulfilled has been superimposed.  Only
for small $M_2$ values, and thus rather light chargino states, does
the deviation from the SM prediction exceed the level of 2\%, which
makes it potentially observable at the $\gamma \gamma$ option of the
ILC (where one expects the $H\gamma\gamma$ coupling to be measured at
the two-percent level). The sign of the contribution is controlled by
the sign of $\mu$, thus for $\mu<0$ one would have a negative shift in
$\Gamma(H\to \gamma \gamma)$. The chargino contribution drops for
higher $\mu$ values as, in this case, the chargino $\chi_1^\pm$ which
gives the most important contribution becomes more wino-like and has a
weaker coupling to the Higgs boson.  \s

Since the chargino masses depend only on $M_2$ and $\mu$, the same
figure holds for the non-universal scenarios; the only difference is
that the WMAP-allowed areas for the LSP relic density, which have been
given in fig.~\ref{fig:DM3}, would be different. In fact, one can see
from these figures that in the scenarios {\bf 75}, {\bf 200} and {\bf
OII} the DM constraint is fulfilled only for large $\mu, M_2$ values,
thus for charginos too heavy to contribute to the $\gamma \gamma$
decay of the Higgs boson. Furthermore, in these scenarios neutralino
LSPs that are compatible with WMAP data are in general also too heavy
to allow for the occurrence of the $H\to \chi_1^0 \chi_1^0$ invisible
decays discussed earlier.\s

Note that for very large $\msusy$ values, $\msusy \gsim 10^{10}$ GeV
when the Higgs mass $M_H \sim 160$ GeV becomes close to the $WW$
threshold, the branching ratios for both the invisible decay and the
$\gamma \gamma$ decay of the Higgs boson become smaller than for
$\msusy= 10^4$ GeV, as these decays have to compete with the $H\to
WW^*$ channel, which has a much larger decay rate than the $H \to
b\bar b$ channel.

\subsection{Chargino and neutralino decays}

In most cases, the charginos and the neutralinos (except for the LSP)
will decay into lighter $\chi$ states and $V=W/Z$ gauge bosons which
can possibly be virtual and subsequently decay into two massless
fermions, $\chi_i \to \chi_j V^{(*)} \to \chi_i f\bar f$
\cite{chi-decays}. As the scalar fermions are very heavy, their
virtual exchange $\chi_i \to f \tilde f^* \to f \bar f\chi_j$ is
strongly suppressed and they do not participate in the decay
processes. The branching ratio for the full final states will thus
essentially follow that of the gauge bosons, therefore the branching
ratios into $\ell=\mu, \nu$ leptons are rather small: 20\% for the
charged and 6\% for the neutral decays.  There are however two
interesting features which might occur and which will be briefly
discussed below: $i)$ decays of the heavier charginos/neutralinos into
lighter ones and the Higgs boson, $\chi_i \to \chi_j H$
\cite{chiHdecays} and $ii)$ the loop-induced decay of some neutralinos
into the LSP and a photon, $\chi_i^0 \to \chi_1^0 \gamma$
\cite{chi-gamma}.  The former process would allow to access directly
the $H\chi \chi$ couplings and the latter has an interesting
experimental signature, a monochromatic photon.\s

\begin{figure}[p]
\begin{center}
\rotatebox{-90}{\epsfig{figure=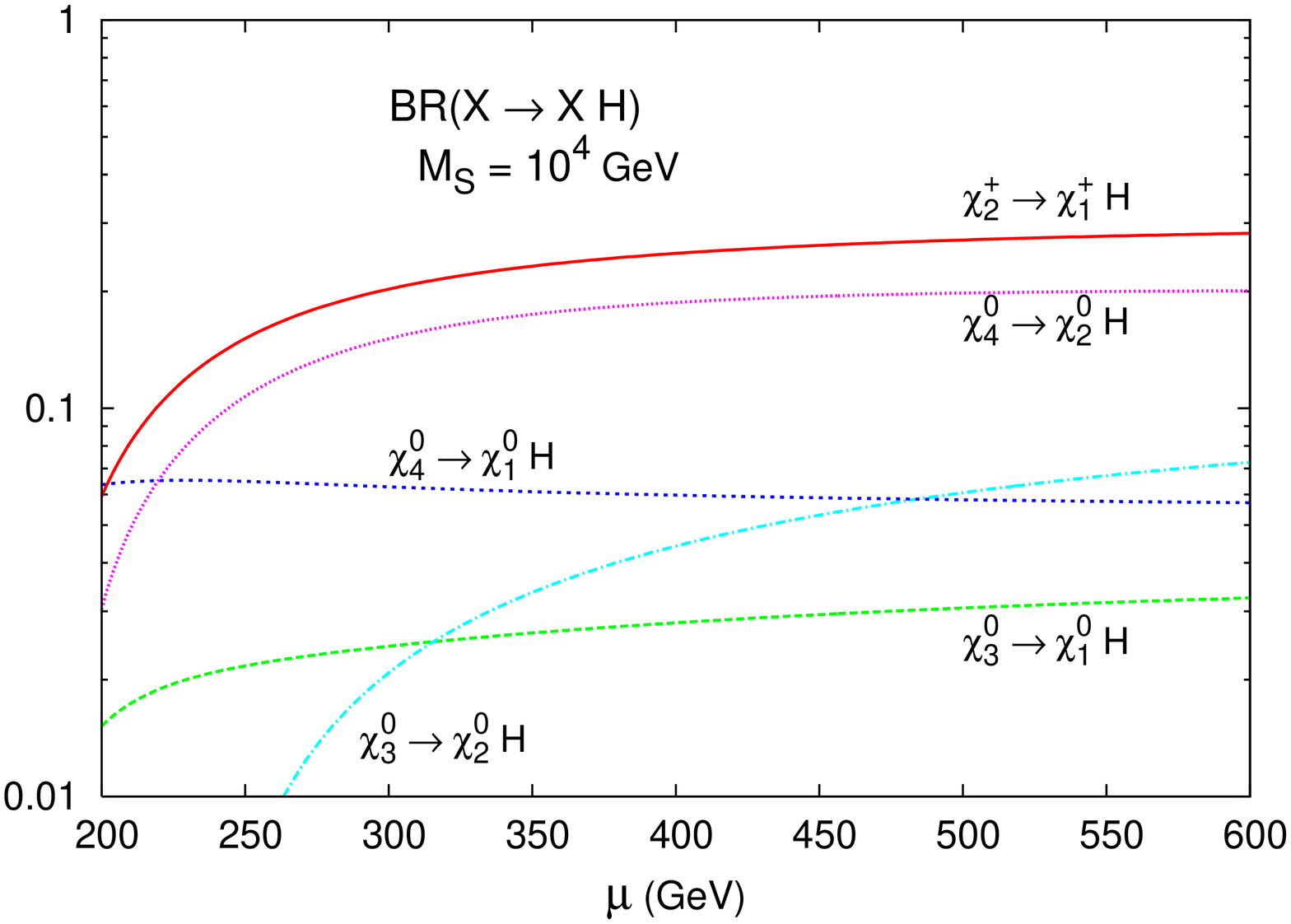,width=8.9cm,height=13.5cm}}\\[0.4cm]
\rotatebox{-90}{\epsfig{figure=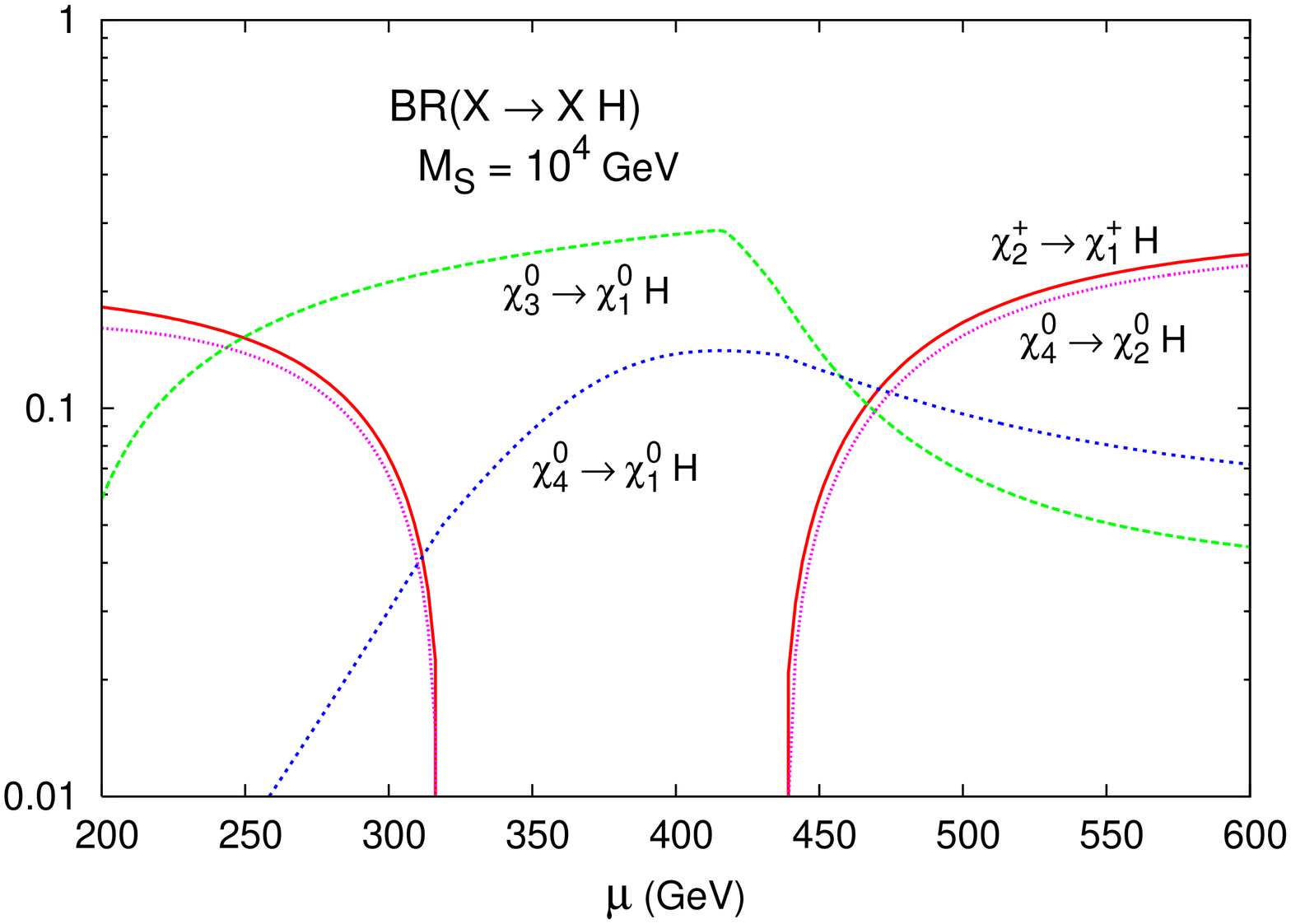,width=8.9cm,height=13.5cm}}
\end{center}
\vspace*{-0.3cm} 
\caption{The branching ratios for the decays of the heavier charginos
and neutralinos into lighter ones and Higgs bosons, BR$(\chi_i \to
\chi_j H)$ as a function of $\mu$ in the scenarios {\bf 1} (top) and
{\bf 24} (bottom); $\msusy=10^4$ GeV is assumed and $M_2$ is chosen
such that the relic density is compatible with WMAP. The program {\tt
SDECAY} \cite{SDECAY}, adapted to deal with heavy scalars, has been
used.}
\label{dec:inosH}
\end{figure}

Adapting the program {\tt SDECAY} \cite{SDECAY} to the case of heavy
scalars, we have calculated the branching ratios for the decays of the
charginos and heavier neutralinos into lighter states and the Higgs
boson. The result is illustrated in fig.~\ref{dec:inosH}, where the
branching ratios BR$(\chi_i \to \chi_j H)$ are shown as a function of
$\mu$ in the two scenarios {\bf 1} (top) and {\bf 24} (bottom) for
$\msusy=10^4$ GeV; for a given $\mu$, the value of $M_2$ is chosen in
such a way that we sit on the left band of the Higgs peak in the
WMAP-allowed region of the $[M_2,\mu]$ plane (see figs.~\ref{fig:DM1}
and \ref{fig:DM3}).  Two main ingredients control the size of the
$\chi_i \to \chi_j H$ branching fractions:\s

 $i)$ The mass difference between the initial and final $\chi$ states
and hence the importance of the phase space; in fact, the
$\chi_i\!-\!\chi_j$ mass difference needs to be larger than $M_H$ for
the Higgs boson to be on-shell, as the $H^* \to f \bar f$ virtuality
would be strongly suppressed by the small $Hf\bar f$ couplings.\s

 $ii)$ The initial and final neutralino and chargino states should
have different textures as to maximize the $H\chi_i \chi_j$
coupling.\s

 As can be seen from the figures, some decays such as $\chi_2^\pm \to
\chi_1^\pm H$ and $\chi_4^0 \to \chi_2^0 H$ can reach the 20\% level.
In the scenario {\bf 1} the other decay modes involving the Higgs
boson are below the 10\% level as they are suppressed either by phase
space or by the smaller $H\chi \chi$ couplings. In the scenario {\bf
24} an interesting feature occurs for intermediate $\mu$ values,
$\mu=300$--450 GeV, where all charginos and neutralinos are mixed
gaugino--higgsino states and have masses of the same order. In this
range, the dominant decay channels mentioned above are kinematically
closed, allowing for the phase-space favored decays $\chi_{3,4}^0 \to
\chi_1^0 H$ to dominate. \s

For the radiative and loop-induced decay $\chi_i^0 \to \chi_j^0
\gamma$ (which is of higher order in perturbation theory and thus
suppressed by additional powers of the electroweak coupling) to occur
with a substantial rate, the standard decay modes $\chi_i^0 \to
\chi_j^0 Z,\, \chi_j^0 H$ need to be strongly suppressed. This occurs
when the $\chi_i^0\!- \!\chi_j^0$ mass difference is smaller than
$M_Z$ and thus $M_H$, so that the decay is a three-body process with a
partial width that is suppressed by the virtuality of the $Z$ boson
and by the additional $Zf\bar f$ coupling. Furthermore, the $Z\chi_i^0
\chi_j^0$ coupling needs to be strongly suppressed, thus the initial
and final neutralinos need to be either pure gauginos or pure
higgsinos.  However, as discussed in section \ref{sec:DM}, for the LSP
to form the DM in the Universe in such conditions its mass should be
in the TeV range, see figs.~\ref{fig:DM1}--\ref{fig:DM2}. Therefore,
for relatively light neutralinos that lead to the cosmological relic
density favored by WMAP (and, hence, have at least a small higgsino
component leading to a non-negligible coupling to the $Z$ boson), the
branching ratio BR$(\chi_i^0 \to \chi_1^0 \gamma)$ is expected to be
very small.\s

\begin{figure}[t]
\begin{center}
\rotatebox{-90}{\epsfig{figure=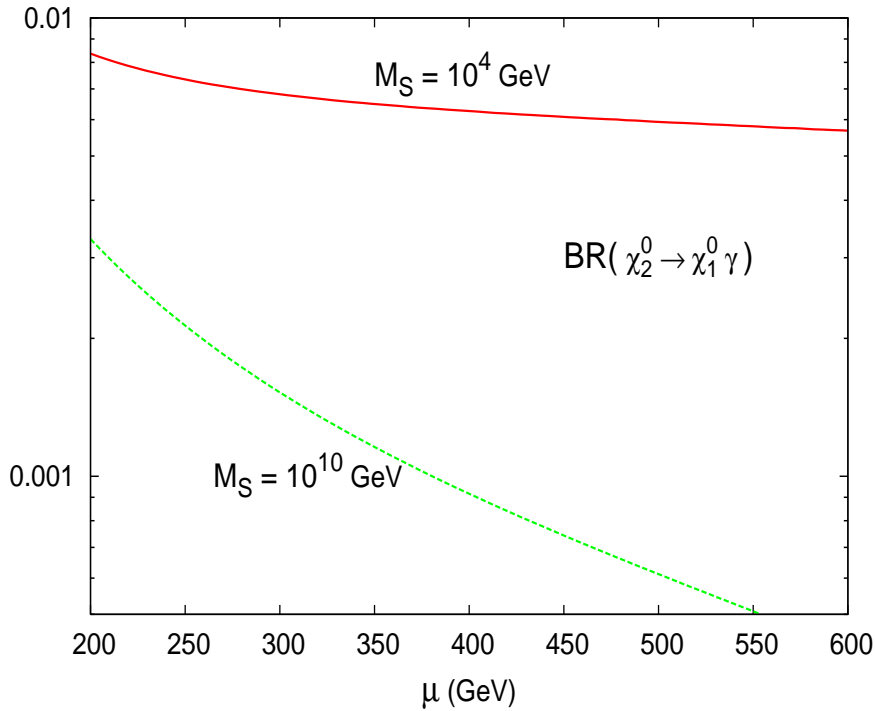,width=10cm,height=12.5cm}}
\end{center}
\vspace*{-0.5cm} 
\caption{The branching ratios for the radiative decay $\chi_2^0 \to
\chi_1^0 \gamma$ as a function of $\mu$ in the scenario {\bf 1} with
$\msusy=10^4$ GeV and $10^{10}$ GeV, and $M_2$ chosen such that the
relic density is compatible with WMAP. The program {\tt SDECAY}
\cite{SDECAY} has been used.}
\label{fig:decN}
\end{figure}

This is exemplified in fig.~\ref{fig:decN} where the branching ratio
for the decay of the next-to-lightest neutralino into the LSP and a
photon, BR$(\chi_2^0 \to \chi_1^0 \gamma)$, is displayed in the
universal scenario {\bf 1} as a function of $\mu$ for $M_S=10^4$ GeV
and $10^{10}$ GeV; as usual, the value of $M_2$ is adjusted in such a
way that the WMAP DM constraint is fulfilled. As can be seen, the
branching ratio hardly reaches the level of 1\% for $M_S=10^4$
GeV. For $M_S=10^{10}$ GeV, the branching fraction is even smaller as
the splitting $M_2\!-\!M_1$ is larger, see table 1, leading to a more
favored phase space for $\chi_2^0 \to \chi_1^0 Z^*$ which controls the
total decay width.\s

\subsection{Gluino decays}
\label{sec:gludec}

As already discussed in section 3.3, the gluinos decay through virtual
heavy squark exchange either into neutralinos or charginos and a
quark--antiquark pair \cite{chi-decays}, $\tilde g \to \chi_i^0 q \bar
q$ or $\tilde g \to \chi_i^\pm q\bar q'$, or into the two--body
neutralino--gluon final state \cite{Glu-g-dec}, $\tilde g \to \chi_i^0
g$, which is mediated by loops involving quarks and heavy squarks.
Thus, the final state topology will consist of a neutralino or a
chargino (which, if it is not the LSP neutralino, will subsequently
decay according to the discussion held in the previous subsection) and
one or two hard jets.\s

\begin{figure}[p] 
\vspace*{-5mm}
\begin{center}
\mbox{\hspace{-8mm}
\epsfig{figure=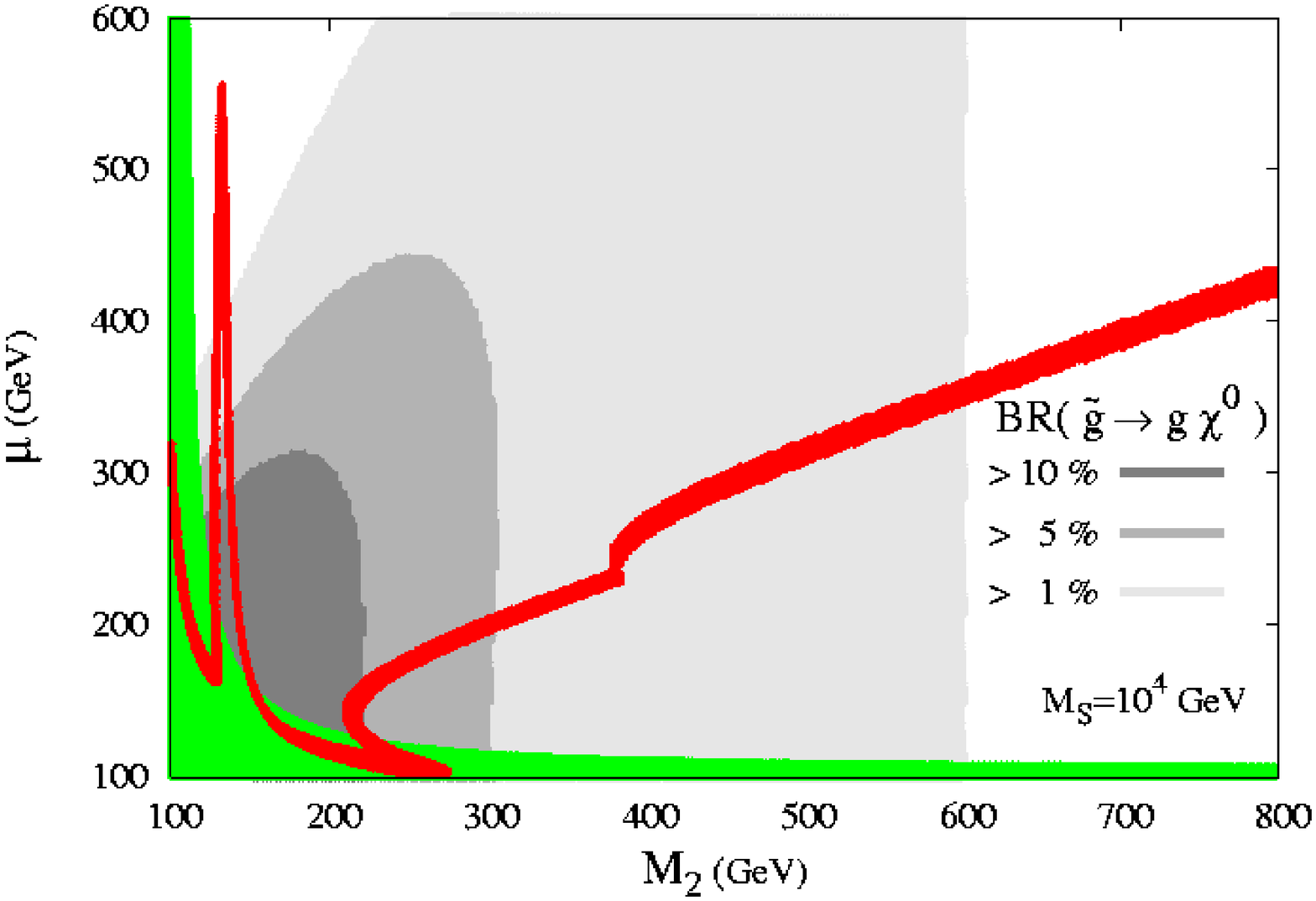,width=9.cm,height=9.6cm} \hspace{-6mm}
\epsfig{figure=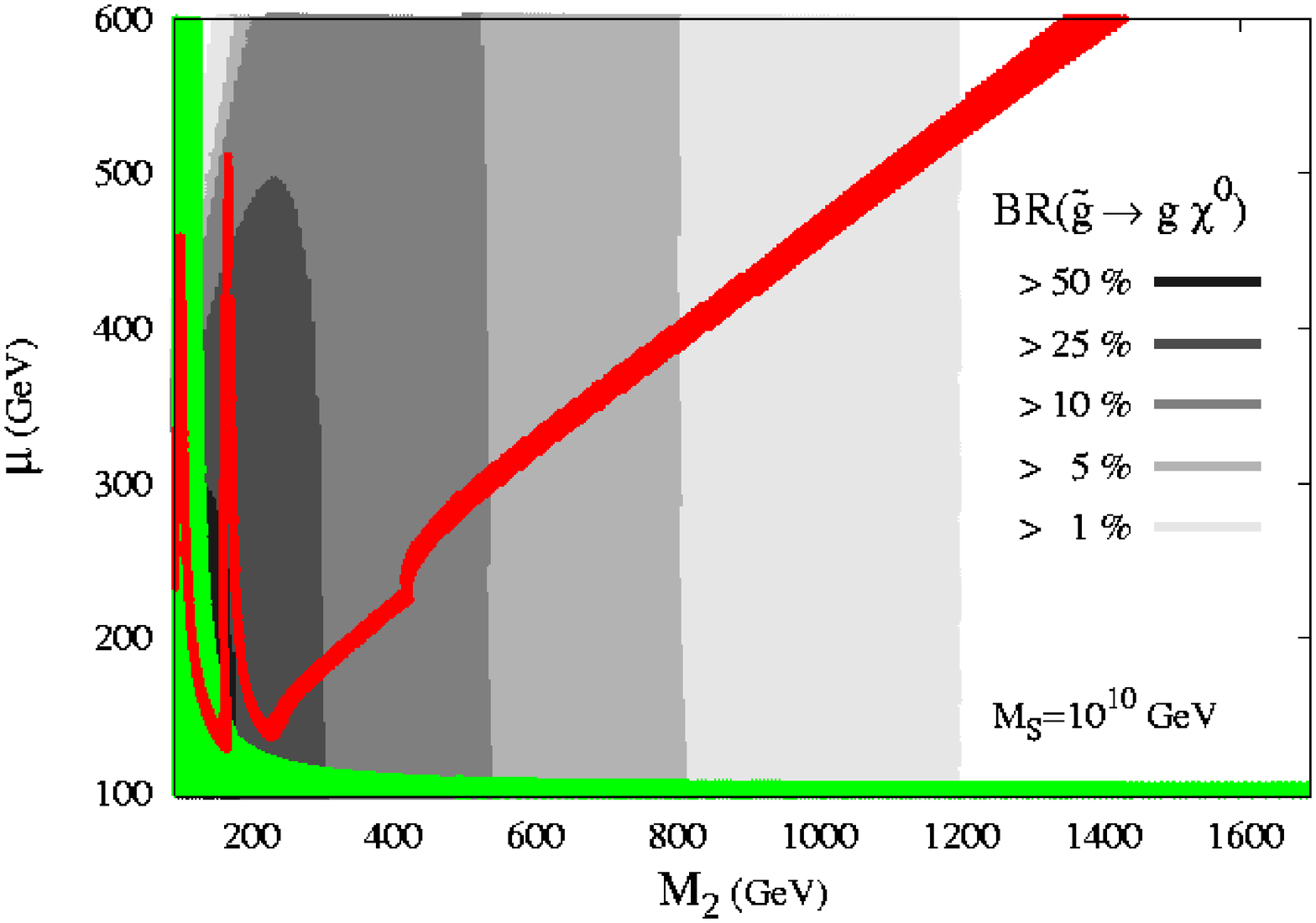,width=9cm,height=9.6cm} }
\end{center}
\vspace*{-8mm}
\caption{The branching ratios for the decays of gluinos into
neutralinos and gluons, BR$(\tilde g \to g \sum_i \chi_i^0)$ in the
$[M_2,\mu]$ plane in the scenario {\bf 1} with $\msusy=10^4$ GeV
(left) and $10^{10}$ GeV (right); the region in which $\Omega h^2$ 
is compatible with WMAP is also shown.}
\label{dec:gluino1}
\vspace*{-3mm}
\end{figure}

It is experimentally important to know the number of final-state jets
and, thus, the relative magnitude of the branching fractions for
the-loop induced decays and the tree-level three-body decays.  We have
thus scanned the usual $[M_2,\mu]$ parameter space and delineated the
areas in which the branching ratio BR($\tilde g \to g \sum_i
\chi_i^0)$, summed over all four neutralinos, is larger than
$1,5,10,25$ and 50\% (from lighter colors to darker ones). The results
for the universal gaugino mass scenario are shown in
fig.~\ref{dec:gluino1} for a common scalar mass values of $M_S=10^4$
GeV (left) and $M_S=10^{10}$ GeV (right); as usual the areas in which
the WMAP constraint is satisfied and those excluded by the LEP bounds
are also displayed.\s

As can be seen, the branching ratio BR($\tilde g \to g \sum_i
\chi_i^0)$ is larger for lighter gluinos and thus smaller values of
$M_2$, and for larger values of the scalar quark mass $M_S$. For
instance, the $g \sum_i \chi_i^0$ branching ratio exceeds the level of
50\% for $M_2\lsim 200$ GeV and $M_S=10^{10}$ GeV. As discussed in
Refs.~\cite{wells,GGS}, this is due to the fact that the main
contribution to the radiative decay originates from loops involving
top quarks and squarks (which have large couplings, $\propto m_t$, to
the higgsino components) and the ratio between this decay and the
three-body decay scales as
$$m_t^2/m_{\tilde g}^2\,[1-\log (M_S^2/m_t^2)]^2.$$ In particular, the
branching ratio is substantial in the region where the cosmological
relic density is generated by LSP annihilation through Higgs boson
exchange, in which all neutralinos are relatively light and thus
kinematically accessible.\s

\begin{figure}[p] 
\begin{center}
\mbox{\hspace{-8mm}
\epsfig{figure=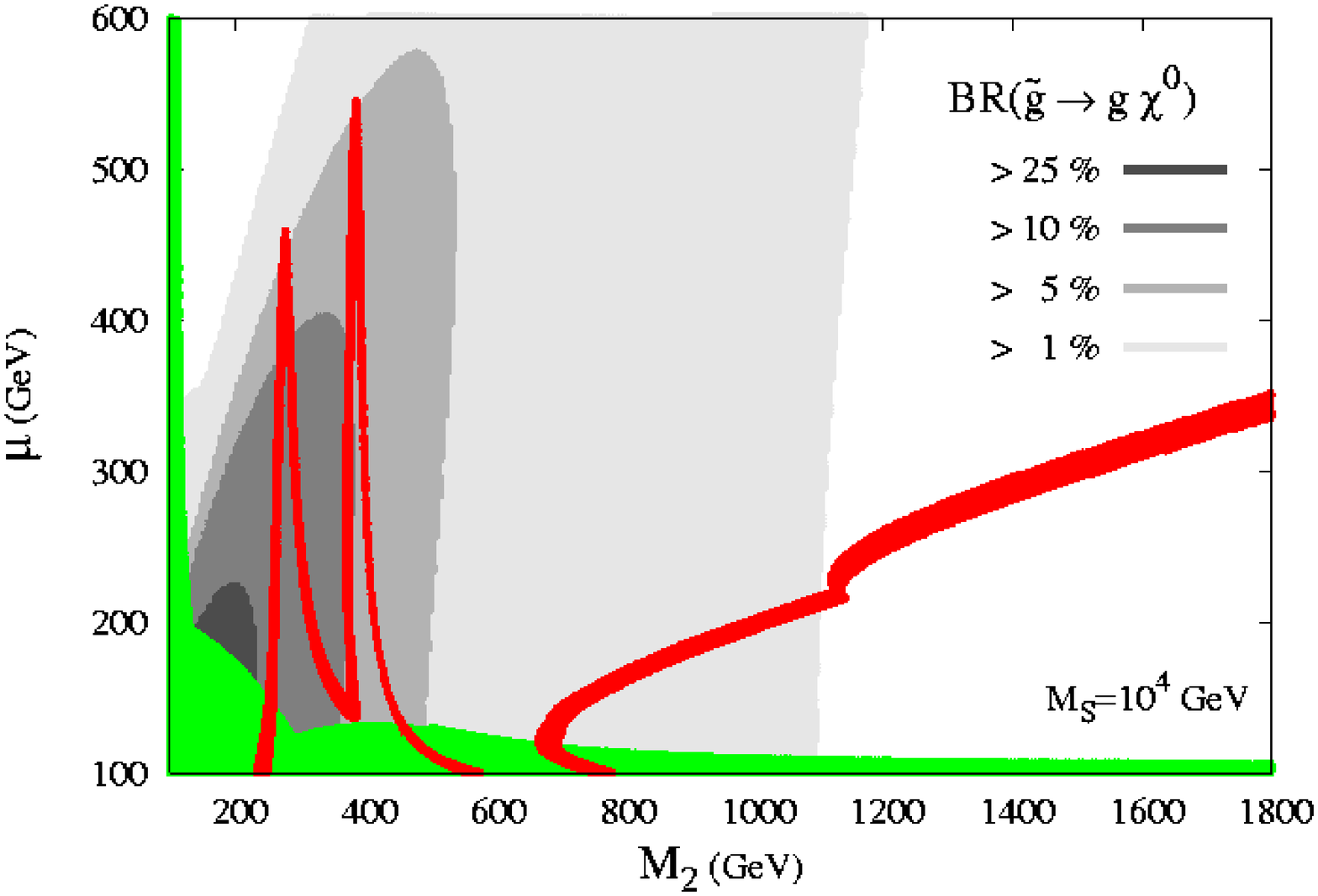,width=9.cm,height=9.6cm} \hspace{-6mm}
\epsfig{figure=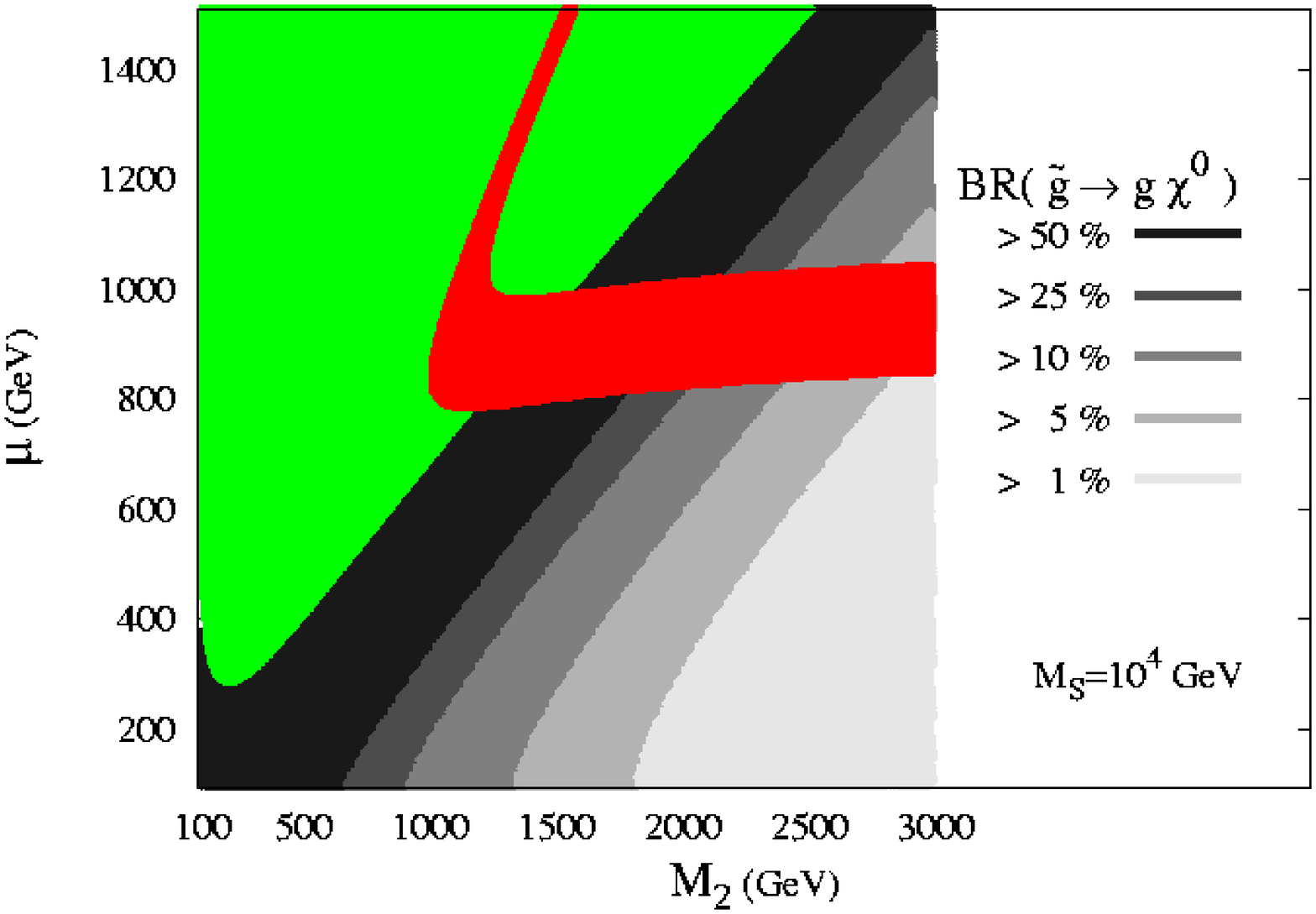,width=9cm,height=9.6cm} }
\end{center}
\vspace*{-7mm}
\caption{The same as in fig.~\ref{dec:gluino1} for the scenarios {\bf
24} (left) and {\bf OII} (right) with $\msusy\,=\,10^4$\,GeV.}
\label{dec:gluino2}
\end{figure}

Similar results are obtained in the case where SUSY breaking occurs
through an $F$--term that is not an SU(5) singlet.  As an
illustration, the areas of the $[M_2,\mu]$ plane in which BR$(\tilde g
\to g \sum_i \chi_i^0)$ is larger than $1,5,10,25$ and 50\% are shown
in the left-hand side of fig.~\ref{dec:gluino2} for the scenario {\bf
24} with $\msusy=10^4$ GeV. As can be seen, for the same value of
$M_2$ the branching ratio is significantly larger than in the scenario
{\bf 1}. For $M_S=10^{10}$ GeV, which is not shown, the decay $\tilde
g\to g \sum_i \chi_i^0$ is by far dominating compared to the
three-body decay for low $M_2$ values. The branching ratio for the
string-inspired {\bf OII} model is shown in the right-hand side of
fig.~\ref{dec:gluino2}. In this case the radiative $\tilde g \to g
\sum_i \chi_i^0$ decay is dominant in a large band in which the mass
of the gluino is close to that of the LSP neutralino, even for
$\msusy=10^4$ GeV.  This band intersects the one in which the LSP has
the correct density.  Thus, there are sizable areas of the parameter
space in which the radiative gluino decay is significant and even
dominant.

\section{Conclusions}

We have performed a comprehensive analysis of the MSSM in the scenario
where all the scalars, except for the SM-like Higgs boson, are very
heavy. This model, commonly known as Split Supersymmetry, is
phenomenologically viable and much more predictive than the usual MSSM
with light scalars. Indeed, besides the three soft-SUSY breaking mass
parameters for the bino, wino and gluino, one has as basic inputs only
the common scalar soft-SUSY breaking mass parameter $M_S$, the
higgsino mass parameter $\mu$, which is not fixed by the requirement
of proper radiative electroweak symmetry breaking, and the parameter
$\tb$. The model retains the interesting features of the usual MSSM,
that is, it leads to a consistent unification of the gauge coupling
constants at the GUT scale and provides a solution to the Dark Matter
problem in the Universe. Nevertheless, as the scalars are extremely
heavy, a large amount of fine tuning in the Higgs sector is present.\s

In a first step, we presented the model and summarized our precise
determination of the masses of the SM-like Higgs boson, the charginos,
neutralinos and gluino, and of the different couplings of these
particles. In our computation we resum large logarithmic corrections
by means of the appropriate RG evolution, and we also include finite
one-loop corrections to the particle masses.  We performed an analysis
of the variation of these masses with respect to the renormalization
scale, which can be viewed as as rough estimate of the higher-order
corrections. We then discussed the boundary conditions for the soft
SUSY-breaking gaugino mass parameters and considered, besides the
universal scenario in which they are set to a common value at the GUT
scale, two representative sets of scenarios in which the GUT values
are non-universal: one where SUSY breaking occurs via an F-term that
is not an SU(5) singlet and another based on an orbifold string
model. Finally, we briefly described how this model is implemented in
the RGE Fortran code {\tt SuSpect} which calculates the SUSY and Higgs
particle spectra in the MSSM.\s

In a second step, we summarized the available constraints on the
model, first from collider searches and high-precision measurements
and then from the requirement that the cosmological relic density of
the lightest neutralino, which is expected to form the Dark Matter in
our Universe, is compatible with the measurements made by the WMAP
satellite. These analyses have been performed in the cases of
universal and non-universal gaugino mass parameters. It turns out that
several new features appear in the non-universal scenarios: in the
case of the DM constraint, new channels such as neutralino
annihilation through the exchange of a $Z$ boson (which is ruled out
by collider constraints in the case of universal gaugino masses) and
gluino co-annihilation (which does not occur in the universal
scenario) are possible. Furthermore, the annihilation of neutralinos
through the exchange of a Higgs boson which then decays into a real
and virtual $W$ boson, leading to the three-body $H \to Wf\bar f$
final state, has to be taken into account. We also discussed the
constraints on the common scalar mass parameter $M_S$ from the
requirement that the gluino lifetime does not exceed the age of the
Universe, and showed that there are small differences in the
non-universal cases compared to the case of a universal gaugino mass
parameter.\s

We finally analyzed the phenomenology of the model, focusing on the
decays of the Higgs boson and those of the charginos, neutralinos and
gluinos. We emphasized the differences between the universal and
non-universal cases and, for the Higgs boson decays, the differences
between the SM and SUSY cases. For instance, we have shown that the
invisible Higgs-boson decay $H\to \chi_1^0 \chi_1^0$ can be
substantial, reaching branching ratios of the order of 10\% in models
where the bino mass parameter $M_1$ is much smaller that the wino mass
parameter $M_2$. The virtual contributions of charginos to the
two-photon decay of the Higgs boson have also been shown to reach the
level where they can be observable at the $\gamma \gamma$ option of
the future linear $e^+e^-$ collider. Decays of heavier charginos and
neutralinos into lighter ones and a Higgs boson have been shown to be
potentially large, opening the possibility of measuring the $H\chi
\chi$ couplings, while the loop-induced decays of heavier neutralinos
into lighter ones and a photon are very rare, barely reaching the 1\%
level. Finally, we have compared the three-body decays of gluinos
through the virtual exchange of heavy squarks, $\tilde g \to q \bar q
\chi$, with the loop-induced two-body decays $\tilde g \to g \chi$,
and have shown the latter to be potentially dominant in some
non-universal scenarios.

\subsection*{Acknowledgments:} We thank Manuel Drees for very useful 
discussions on the DM code of ref.~\cite{DN-routine}. This work is
supported by the French ANR project PHYS@COS\&COL.

\begin{appendletterA}
\section*{Appendix} 

\def\drbar{\overline{\rm DR}}

In this appendix we present for completeness all the formulae for the
one-loop radiative corrections that we include in the computation of
the mass spectrum of the low-energy effective theory of \sps. We
largely follow the notation and the results of ref.~\cite{PBMZ},
adapting when necessary the formulae of that paper to the case of
\sps.\s

\subsubsection*{Gauge and Yukawa couplings}
We start by summarizing how the minimally renormalized gauge and
third-family-Yukawa couplings, as well as the electroweak parameter
$\hat v$ (from now on a hat denotes minimally renormalized parameters
of the \spsd\ effective theory), are extracted at the scale $Q=M_Z$
from the set of SM input parameters [$\alpha_s(M_Z)\,, \alpha(M_Z)\,,
M_Z\,,$ $ G_F\,, M_t\,, M_\tau\,, m_b(m_b)$]. The strong gauge
coupling $\hat{\alpha}_s$ is extracted from the SM input
$\alpha_s(M_Z)$ as
\be
\label{gauges}
\hat{\alpha}_s(M_Z) ~=~ \frac{\alpha_s(M_Z)}{1-\Delta\alpha_s}~,
~~~~~~~~
\Delta\alpha_s = \frac{\alpha_s}{2\pi}\,\left[
\delta_s  -\frac23 \,\log\frac{M_t}{M_Z} - 2\,\log\frac{M_{\tilde g}}{M_Z}
\right]~,
\ee
where $\delta_s$ is a conversion factor that depends on the choice of
the minimal renormalization scheme, i.e.~ $\delta_s = 0$ in $\msbar$
and $\delta_s = 1/2$ in $\drbar$. Similarly, the electromagnetic
coupling $\hat{\alpha}$ is extracted from the SM input $\alpha(M_Z)$

\be
\label{gaugeew}
\hat{\alpha}(M_Z) ~=~ \frac{\alpha(M_Z)}{1-\Delta\alpha}~,
~~~~~~~~
\Delta\alpha = \frac{\alpha}{2\pi}\,\left[
\delta_e  -\frac{16}{9} \,\log\frac{M_t}{M_Z} - 
\frac43\,\sum_{i=1}^2 \log\frac{|m_{\chi^+_i}|}{M_Z}
\right]~,
\ee
where $\delta_e = 0$ in $\msbar$ and $\delta_e = 1/3$ in $\drbar$. The
weak mixing angle (we denote $\sin \theta_W$ and $\cos \theta_W$ as
$s$ and $c$, respectively; we thus define ${\hat s}^2 =
{\hat{g}}^{\prime\,2} / ( {\hat{g}}^{2}+ {\hat{g}}^{\prime\,2})$ and
$s^2 = 1- M_W^2/M_Z^2\,$) is computed from
\be
\label{ewmix}
\hat c^2 \,\hat s^2 ~=~ \frac{\pi\,\hat\alpha}{\sqrt{2}\,M_Z^2\,G_F\,
(1-\Delta r)}~,
~~~~~~~~
\Delta r = \hat\rho \,\frac{\Pi_{WW}(0)}{M_W^2} - 
\frac{\Pi_{ZZ}(M_Z^2)}{M_Z^2} + \delta_{VB}~,
\ee
where $\Pi_{WW}(p^2)$ and $\Pi_{ZZ}(p^2)$ are the transverse (and
real) part of the $W$ and $Z$ self-energy, respectively, and will be
explicitly given below. The $\hat\rho$ parameter is defined as
\be
\label{rhopar}
\hat\rho = \frac{1+ {\Pi_{ZZ}(M_Z^2)}/{M_Z^2}}{1+{\Pi_{WW}(M_W^2)}/{M_W^2}}~,
\ee
while the quantity $\delta_{VB}$, which accounts for the vertex and
box corrections to the muon decay, is:
\be
\label{deltaVB}
\delta_{VB} = \hat\rho\,\frac{\hat\alpha}{4\,\pi\,\hat s^2}\,
\left\{6 + \frac{\log c^2}{s^2}\,\left[
\frac72-\frac52\,s^2-\hat s^2\,\left(5-\frac32\frac{c^2}{\hat c^2}\right)
\right]\right\}~.
\ee
In \sps\ the non-SM contributions to $\delta_{VB}$ are suppressed by
the large sfermion masses and can be omitted. Once $\hat\alpha$ and
$\hat c^2 \hat s^2$ have been computed by means of
eqs.~(\ref{gaugeew}) and (\ref{ewmix}), it is trivial to extract the
low-energy boundary conditions on the running electroweak coupling
constants $\hat g(M_Z)$ and $\hat g^\prime(M_Z)$.\s

The top Yukawa coupling is extracted from the physical top mass $M_t$
and the running electroweak parameter $\hat v$ according to:
\be
\label{yukt}
\hat h_t(M_Z) \, \hat v(M_Z) = M_t + \Sigma_t(M_t^2) ~,
\ee
where $\Sigma_t(M_t^2)$ is real part of the top quark self-energy
(explicitly given below) computed with external momentum $p^2=M_t^2$,
and $\hat v$ is defined as:
\be
\label{vrun}
\hat v^2(Q) = 2 \,\frac{M_Z^2 + \Pi_{ZZ}(M_Z^2)}{\hat g^{\prime\,2}(Q)
+ g^{2}(Q)}~.
\ee
In the low-energy effective theory of \sps\ the couplings of the Higgs
boson to the down-type fermions are SM-like, thus they do not have a
large impact on the phenomenology. However, we include all the
third-family Yukawa couplings in our analysis for completeness. The
Yukawa couplings of the $b$ quark and of the $\tau$ lepton can be
extracted from the running masses evaluated at $Q=M_Z$ according to:
\be
\label{yukbt}
\hat h_{b,\tau}(M_Z) \, \hat v(M_Z) = \overline{m}_{b,\tau}(M_Z) 
+ \Sigma^{\rm th}_{b,\tau}({m}_{b,\tau}^2) ~,
\ee
where $\Sigma^{\rm th}_{b,\tau}({m}_{b,\tau}^2)$ contains the
weak-scale threshold contributions to the $b$ and $\tau$
self-energies. In order to resum to all orders the potentially large
logarithms of the ratio ${m}_{b,\tau}/M_Z$, the running masses
$\overline{m}_{b,\tau}$ are evolved up to $Q=M_Z$ by means of the RGE
of the model with five quarks and $SU(3) \times U(1)_{\rm em}$ gauge
symmetry:
\be
\label{mrun}
\overline{m}_{b,\tau}(M_Z) = {m}_{b,\tau}({m}_{b,\tau})
\,\left(1-\frac{g_3^2}{8\,\pi^2}\,\beta_s\,\log\frac{M_Z}{{m}_{b,\tau}}
\right)^{\frac{\gamma^{b,\tau}_s}{2\,\beta_s}}
\,\left(1-\frac{e^2}{8\,\pi^2}\,\beta_e\,\log\frac{M_Z}{{m}_{b,\tau}}
\right)^{\frac{\gamma^{b,\tau}_e}{2\,\beta_e}},
\ee
where $(\beta_s,\beta_e) = (23/3,-80/9)$, $(\gamma^b_s,\gamma^b_e) =
(8, 2/3)$, and $(\gamma^\tau_s,\gamma^\tau_e) = (0, 6)$. The running
bottom mass $m_b(m_b)$ is taken as input, while the running tau mass
$m_\tau(m_\tau)$ is extracted from the physical mass $M_\tau$
according to:
\be
\label{mtaurun}
m_\tau(m_\tau) = M_\tau\,\left(1-\frac{e^2}{16\,\pi^2}\,c_R\right)\,
\ee
where $c_R=4$ in $\msbar$ and $c_R=5$ in $\drbar$.\s

\subsubsection*{One-loop self-energies}
We now provide explicit formulae for the one-loop self-energies
appearing in eqs.~(\ref{ewmix}),(\ref{rhopar}) and
(\ref{yukt})--(\ref{yukbt}). The transverse parts of the gauge boson
self-energies $\Pi_{VV}$ (with $V=Z,W$) can be decomposed in a SM
contribution $\Pi_{VV}^{\rm SM}$ and a chargino/neutralino
contribution $\Pi_{VV}^{\chi}$. In the Feynman gauge, that we adopt
throughout this appendix, the SM contributions read:

\beq
\frac{16\,\pi^2\,\hat c^2}{\hat g^2}\, \Pi^{\rm SM}_{ZZ}(p^2)
&=&
\left[m_Z^2\,B_0(m_Z,m_H) - \widetilde{B}_{22}(m_Z,m_H)\right]
-\left[8\,\hat c^4 + (\hat c^2- \hat s^2)^2 \right]
\widetilde{B}_{22}(m_W,m_W)
\non\\
&&-2\,\hat c^4\,\left(2\,p^2+m_W^2-m_Z^2\,\frac{\hat s^4}{\hat c^2}\right)
B_0(m_W,m_W) ~+~ \Delta_Z
\non\\
\label{selfZSM}
&&+\sum_f \,N_c^f \biggr[
\left(g_{f_L}^2 + g_{f_R}^2\right) \,H(m_f,m_f)
- 4\,g_{f_L}\,g_{f_R}\,m_f^2\,B_0(m_f,m_f)\biggr]~,\\
\non\\
\frac{16\,\pi^2}{\hat g^2}\, \Pi^{\rm SM}_{WW}(p^2)
&=&
\left[m_W^2\,B_0(m_W,m_H) - \widetilde{B}_{22}(m_W,m_H)\right]
-(1 + 8\,\hat c^2)\,\widetilde{B}_{22}(m_Z,m_W)
\non\\
&&-\left[(4\,p^2+m_Z^2+m_W^2)\,\hat c^2-m_Z^2\,\hat s^4\right]
B_0(m_Z,m_W) ~ + ~ \Delta_W
\non\\
\label{selfWSM}
&&-\hat s^2\left[8\, \widetilde{B}_{22}(m_W,0)+
4\,p^2\,B_0(m_W,0)\right]
~+~\sum_f\,\frac{N_c^f}{2}\,H(m_{f_u},m_{f_d})~.
\eeq
The Passarino-Veltman functions $B_0, \widetilde B_{22}$ and $H$ are
defined as in the appendix B of ref.~\cite{PBMZ}.  The Higgs and gauge
boson masses appearing in the equations above are interpreted as
running masses. The summation in the last line of each equation is
over the fermion species, $N_c^f$ is the color number (3 for quarks
and 1 for leptons) and $g_{f} = I_3^{f} - e_{f}\,\hat s^2$ are the
weak neutral-current couplings.  $\Delta_{Z}$ and $\Delta_{W}$ are
$\drbar$--$\msbar$ conversion factors: they are both equal to zero in
the $\drbar$ scheme, while in the $\msbar$ scheme $\Delta_{Z} = -2/3\,
\hat c^4\,p^2$ and $\Delta_{W} = -2/3\,p^2$. \s

The chargino and neutralino contributions to the gauge boson
self-energies can be expressed as:
\beq
16\,\pi^2\, \Pi^{\chi}_{ZZ}(p^2) &=&
\frac12 \,\sum_{i,j}\,\left[
\left({a^0_{ijZ}}^2+{b^0_{ijZ}}^2\right)
H(m_{\chi^0_i},m_{\chi^0_j})
+4\,a^0_{ijZ}\,b^{0}_{ijZ}\,m_{\chi^0_i}\,m_{\chi^0_j}\,
B_0(m_{\chi^0_i},m_{\chi^0_j})\right]\non\\
&+&
\sum_{i,j}\,\left[
\left({a^+_{ijZ}}^2+{b^+_{ijZ}}^2\right)
H(m_{\chi^+_i},m_{\chi^+_j})
+4\,a^+_{ijZ}\,b^{+}_{ijZ}\,m_{\chi^+_i}\,m_{\chi^+_j}\,
B_0(m_{\chi^+_i},m_{\chi^+_j})\right],\non\\
\label{selfZchi}\\
16\,\pi^2\, \Pi^{\chi}_{WW}(p^2) &=& \sum_{i,j}\,\left[
\left({a_{ijW}}^2+{b_{ijW}}^2\right)
H(m_{\chi^0_i},m_{\chi^+_j})
+4\,a_{ijW}\,b_{ijW}\,m_{\chi^0_i}\,m_{\chi^+_j}\,
B_0(m_{\chi^0_i},m_{\chi^+_j})\right].\non\\
\label{selfWchi}
\eeq
In general, we write the Feynman rule for the chargino or neutralino
couplings to a gauge boson as $-i\,\gamma_\mu\,(a\,P_L+b\,P_R)$, the
rule for the couplings to a scalar as $-i\,(a\,P_L+b\,P_R)$ and the
rule for the coupling to a pseudoscalar as $(a\,P_L+b\,P_R)$. Under
this convention, the chargino and neutralino couplings to the gauge
bosons are
\be
a^0_{ijZ} ~=~ -b^0_{ijZ} ~=~ \frac{g}{2\,\hat c}\left(N_{i3}N_{j3}
-N_{i4}N_{j4}\right)~,
\ee
\be
a^+_{ijZ} ~=~ g\,\hat c\,V_{i1} V_{j1}+ \frac{g\,(\hat c^2 -\hat s^2)}
{2 \hat c}\,V_{i2} V_{j2}~,~~~~~~
b^+_{ijZ} ~=~ g\,\hat c\,U_{i1} U_{j1}+ \frac{g\,(\hat c^2 -\hat s^2)}
{2 \hat c}\,U_{i2} U_{j2}~,
\ee
\be
a_{ijW} ~=~ -g\,\hat c\,N_{i2}V_{j1} + \frac{g}{\sqrt{2}}N_{i4}V_{j2}~,~~~~~~
b_{ijW} ~=~ -g\,\hat c\,N_{i2}U_{j1} - \frac{g}{\sqrt{2}}N_{i3}U_{j2}~.
\ee
The matrices $N$ and $U,V$ rotate the neutralino and chargino states,
respectively, so that the mass matrices $N\,{\cal M}_N\,N^T$ and
$U\,{\cal M}_C V^T$ are diagonal. The matrices ${\cal M}_N$ and ${\cal
M}_C$ are given in eq.~(\ref{neutchar}). We assume that there are no
CP-violating phases in the higgsino and gaugino mass parameters, and
we choose $N,U$ and $V$ to be real, allowing for negative signs in the
chargino and neutralino masses.\s

The top quark self-energy in eq.~(\ref{yukt}) can be expressed as
\beq
\label{sigmat}
\frac{16\,\pi^2}{m_t}\,\Sigma_t(m_t^2) &=&
\frac{4\,g_3^2}{3}\,\left(3\,\log\frac{m_t^2}{Q^2}-c_R\right)
+\frac{4\,e^2}{9}\,\left(3\,\log\frac{m_t^2}{Q^2}-c_R\right)\non\\
&&+\frac{h_t^2}{2}\,\biggr[B_1(m_t,m_H)+B_0(m_t,m_H)
+B_1(m_t,m_Z)-B_0(m_t,m_Z)\biggr]\non\\
&&+\frac{h_t^2+h_b^2}2\,B_1(m_b,m_W)-\frac{h_b^2}{2}\,B_0(m_b,m_W)
+\frac{g^2}2\biggr[B_1(m_b,m_W) + \delta_R\biggr]\non\\
&&+\frac{g^2}{\hat c^2}\,\left\{
(g_{t_L}^2+g_{t_R}^2)\,\biggr[B_1(m_t,m_Z)+\delta_R\biggr]
+4\,g_{t_L}\,g_{t_R}\,\biggr[B_0(m_t,m_Z)+\delta_R\biggr]\right\}~,\non\\
\eeq
where $(c_R,\delta_R)$ are equal to $(4,-1/2)$ in $\msbar$ and to
$(5,0)$ in $\drbar$. The Passarino-Veltman functions $B_0$ and $B_1$
are defined as in the appendix B of ref.~\cite{PBMZ}.  The weak-scale
contribution to the bottom quark self-energy $\Sigma_b^{\rm th}(m_b)$,
appearing in eq.~(\ref{yukbt}), can be extracted from
eq.~(\ref{sigmat}) by omitting the first line and replacing everywhere
$t\leftrightarrow b$. The analogous quantity for the tau lepton,
$\Sigma_\tau^{\rm th}(m_\tau)$, can also be extracted from
eq.~(\ref{sigmat}): one has to omit the first line and the terms
proportional to $h_b^2$ in the third line, then replace everywhere
$t\rightarrow \tau$ and $m_b\rightarrow 0$.\s

\subsubsection*{Corrections to the Higgs mass}
We provide here the formulae for the Higgs mass corrections
$\delta^{\rm SM}(Q)$ and $\delta^{\chi}(Q)$ appearing in
eq.~(\ref{hmass}). The SM contribution reads \cite{SZ}
\be
\label{sirlin}
\delta^{\rm SM}(Q) =  - \frac{G_F}{\sqrt{2}}\,\frac{M_Z^2}{16\,\pi^2}\,
\biggr[\xi\,f_1(\xi,Q)+f_0(\xi,Q)+\xi^{-1}\,f_{-1}(\xi,Q)\biggr]~,
\ee
where $\xi = m_H^2/M_Z^2$ and the functions $f_k(\xi,Q)$ 
are defined as:
\beq
f_1(\xi,Q) & = &  6\,\log\frac{Q^2}{m_H^2} + \frac32\,\log\xi -
\frac12 \,Z(\xi^{-1})-Z(c^2\,\xi^{-1})-\log c^2 +
\frac92\left(\frac{25}{9}-\frac{\pi}{\sqrt 3}\right),\\
\non\\
f_0(\xi,Q) & = & -6\,\log\frac{Q^2}{M_Z^2}\,\left[
1 + 2\,c^2-2\,\frac{m_t^2}{M_Z^2}\right] +\frac{3\,c^2\,\xi}{\xi-c^2}
\,\log\frac{\xi}{c^2}+2\,Z(\xi^{-1})+ 4\,c^2\,Z(c^2\,\xi^{-1})\non\\
&&+\left(\frac{3}{s^2}+12\right)\,c^2\,\log c^2-\frac{15}{2}\,(1+2\,c^2)
-3\frac{m_t^2}{M_Z^2}\left[2\,Z\left(\frac{m_t^2}{m_H^2}\right)+4\,
\log\frac{m_t^2}{M_Z^2}-5\right]~,\non\\
\\
f_{-1}(\xi,Q) &=&
6\,\log\frac{Q^2}{M_Z^2}\,\left[1 + 2\,c^4-4\,\frac{m_t^4}{M_Z^4}\right]
-6\,Z(\xi^{-1})-12\,c^4\,Z(c^2\,\xi^{-1})-12\,c^4\,\log c^2\non\\
&&+8\,(1+2\,c^4)+24\,\frac{m_t^4}{M_Z^4}\,\left[\log\frac{m_t^2}{M_Z^2}
- 2 + Z\left(\frac{m_t^2}{m_H^2}\right)\right]~.
\eeq
In the equations above we define $c^2 = M_W^2/M_Z^2$ in terms of
physical masses. On the other hand, as described in section 2.2, we
explore the consequences of choosing $m_t$ as either the physical or
the running top mass. The auxiliary function $Z$ appearing in the equations
above is defined as:
\be
Z(x) = \left\{
\begin{array}{ll}
2\,A\,\tan^{-1}(A^{-1})& \left(x>\frac14\right)\\
&\\
A\,\log [(1+A)/(1-A)]& \left(x<\frac14\right)~,\end{array}\right.
\ee
where $A = |1-4\,x|^{1/2}$.\s

The results of ref.~\cite{SZ} are derived under the assumption that
the Higgs quartic coupling $\lambda$ appearing in eq.~(\ref{hmass}) is
expressed in the $\msbar$ renormalization scheme. If $\lambda$ is
expressed in the $\drbar$ scheme the Higgs mass correction in
eq.~(\ref{sirlin}) is modified as
\be
\label{deltaHdr}
\delta^{\rm SM}(Q) ~ \rightarrow ~ \delta^{\rm SM}(Q)
- \frac{g^2}{16\,\pi^2}\,\frac{M_W^2}{m_H^2}\,\left(1 + \frac{1}
{2\,c^{\,4}}\right)~.
\ee

The chargino and neutralino contributions to the Higgs boson
self-energy and tadpole, appearing in the correction term
$\delta^{\chi}(Q)$ defined in eq.~(\ref{deltachi}), read
\beq
16\,\pi^2\, \Pi^{\chi}_{HH}(p^2) &=&\!
\frac12 \sum_{ij} \left[  \left({a^{0}_{ijH}}^2+{b^{0}_{ijH}}^2\right)
\,G(m_{\chi^0_i},m_{\chi^0_j})
-4\,a^{0}_{ijH}\,b^{0}_{ijH}\,m_{\chi^0_i}\,m_{\chi^0_j}\,
B_0(m_{\chi^0_i},m_{\chi^0_j})
\right]\non\\
&+&\! \sum_{ij} \left[  \left({a^{+}_{ijH}}^2+{b^{+}_{ijH}}^2\right)
\,G(m_{\chi^+_i},m_{\chi^+_j})
-4 \, a^{+}_{ijH}\,b^{+}_{ijH}\,m_{\chi^+_i}\,m_{\chi^+_j}\,
B_0(m_{\chi^+_i},m_{\chi^+_j})\right],\non\\\\
16\,\pi^2\, T^{\chi}_{H} &=&
- \sum_i\,\left(a^{0}_{iiH}+b^{0}_{iiH}\right)\,m_{\chi^0_i}\,
A_0(m_{\chi^0_i})
-2\,\sum_i\,\left(a^{+}_{iiH}+b^{+}_{iiH}\right)\,m_{\chi^+_i}\,
A_0(m_{\chi^+_i})~.\non\\
\eeq
The Passarino-Veltman functions $G, B_0$ and $A_0$ in the equations
above are defined as in the appendix B of ref.~\cite{PBMZ}. The
chargino and neutralino couplings to the Higgs boson are
\be
a_{ijH}^0 ~=~  b_{ijH}^0 ~=~\frac12\,N_{ik}N_{j\ell}
\biggr[ - \tilde g_d^\prime\, \delta_{\{k1}\delta_{\ell 3\}}
+ \tilde g_u^\prime\, \delta_{\{k1}\delta_{\ell 4\}}
+ \tilde g_d\, \delta_{\{k2}\delta_{\ell 3\}}
- \tilde g_u\, \delta_{\{k2}\delta_{\ell 4\}}\biggr]~,
\ee
\be
a_{ijH}^+ ~=~  b_{jiH}^+ ~=~
\frac{\tilde g_d}{\sqrt 2}\,V_{i1}U_{j2} 
+ \frac{\tilde g_u}{\sqrt 2}\,V_{i2}U_{j1}~,
\ee
where $\tilde g_{u,d}$ and $\tilde g^\prime_{u,d}$ are the effective
Higgs--higgsino--gaugino couplings defined in eq.~(\ref{lagrangian}).
We define $\delta_{\{ki}\delta_{\ell j\}} =
\delta_{ki}\delta_{\ell j} +\delta_{kj}\delta_{\ell i}$, where
$\delta_{ij}$ is the Kronecker delta and summation over repeated
indices is understood.\s

\subsubsection*{Corrections to the chargino and neutralino masses}
We provide here the formulae for the radiative corrections to the
chargino and neutralino mass matrices, once again adapting to the
\spsd\ case the results of ref.~\cite{PBMZ}. The one-loop neutralino
mass matrix reads
\be 
\label{loopneut}
{\cal M}_N(p^2) = {\cal M}_N^{0} + \frac12 \,\left(\delta {\cal M}_N(p^2) 
+ \delta {\cal M}_N^T(p^2)\right)~,
\ee
where
\be
\delta {\cal M}_N(p^2) = -\Sigma_R^0(p^2) \,{\cal M}_N 
- {\cal M}_N \,\Sigma_L^0(p^2) - \Sigma_S^0(p^2)~.
\ee
The one-loop chargino mass matrix is instead
\be
\label{loopchar}
{\cal M}_C(p^2) = {\cal M}_C^{0} -\Sigma_R^+(p^2) \,{\cal M}_C 
- {\cal M}_C \,\Sigma_L^+(p^2) - \Sigma_S^+(p^2)~.
\ee
The tree-level mass matrices ${\cal M}_N^{0}$ and ${\cal M}_C^{0}$ are
given in eq.~(\ref{neutchar}) and are expressed in terms of minimally
renormalized parameters. The neutralino and chargino self-energies
$\Sigma_{L,S}^{0}(p^2)$ and $\Sigma_{L,S}^{+}(p^2)$ are $4\!\times\!4$
and $2\!\times\!2$ matrices, respectively, and they read
\beq
16\,\pi^2\,\Sigma^0_{Lij} &=& 
~~ \sum_{k=1}^2 \,\biggr\{
\tilde a^0_{ikG^+}\,\tilde a^0_{jkG^+}\,B_1(m_{\chi^+_k},m_W)
+2\,\tilde a^0_{ikW}\,\tilde a^0_{jkW}\,
\left[B_1(m_{\chi^+_k},m_W)+\delta_R\right]
\biggr\}\non\\
\label{sigma0L}
&+& \!\frac12\,\sum_{k=1}^4 \,\biggr\{
\tilde a^0_{ikG^0}\,\tilde a^0_{jkG^0}\,B_1(m_{\chi^0_k},m_Z)
+ \tilde a^0_{ikH}\,\tilde a^0_{jkH}\,B_1(m_{\chi^0_k},m_H)\non\\
&&~~~~~~~~~
+2\,\tilde a^0_{ikZ}\,\tilde a^0_{jkZ}\,
\left[B_1(m_{\chi^0_k},m_Z)+\delta_R\right]
\biggr\}~,\\
\non\\
16\,\pi^2\,\Sigma^0_{Sij} &=& 
2\, \sum_{k=1}^2 \,m_{\chi^+_k}\biggr\{
\tilde b^0_{ikG^+}\,\tilde a^0_{jkG^+}\,B_0(m_{\chi^+_k},m_W)
-4\,\tilde b^0_{ikW}\,\tilde a^0_{jkW}\,
\left[B_0(m_{\chi^+_k},m_W)+\delta_R\right]
\biggr\}\non\\
&+&~~\sum_{k=1}^4 \,m_{\chi^0_k}\,\biggr\{
\tilde b^0_{ikG^0}\,\tilde a^0_{jkG^0}\,B_0(m_{\chi^0_k},m_Z)
+ \tilde b^0_{ikH}\,\tilde a^0_{jkH}\,B_0(m_{\chi^0_k},m_H)\non\\
\label{sigma0S}
&&~~~~~~~~~~~~~~~
-4\,\tilde b^0_{ikZ}\,\tilde a^0_{jkZ}\,
\left[B_0(m_{\chi^0_k},m_Z)+\delta_R\right]
\biggr\}~,\\
\non\\
16\,\pi^2\,\Sigma^+_{Lij} &=& 
\!\frac12\,\sum_{k=1}^4 \,\biggr\{
\tilde a^+_{kiG^+}\,\tilde a^+_{kjG^+}\,B_1(m_{\chi^0_k},m_W)
+2\,\tilde a^+_{kiW}\,\tilde a^+_{kjW}\,
\left[B_1(m_{\chi^0_k},m_W)+\delta_R\right]
\biggr\}\non\\
&+&\!\frac12\, \sum_{k=1}^2 \,\biggr\{
\tilde a^+_{ikG^0}\,\tilde a^+_{jkG^0}\,B_1(m_{\chi^+_k},m_Z)
+ \tilde a^+_{ikH}\,\tilde a^+_{jkH}\,B_1(m_{\chi^+_k},m_H)\non\\
&&~~~~~~~
+2\,\tilde a^+_{ikZ}\,\tilde a^+_{jkZ}\,
\left[B_1(m_{\chi^+_k},m_Z)+\delta_R\right]
+2\,\tilde a^+_{ik\gamma}\,\tilde a^+_{jk\gamma}
\,\left[B_1(m_{\chi^+_k},0)+\delta_R\right]\biggr\}~,\non\\
\label{sigma+L}\\
16\,\pi^2\,\Sigma^+_{Sij} &=& 
\sum_{k=1}^4 \,m_{\chi^0_k}\,\biggr\{
\tilde b^+_{kiG^+}\,\tilde a^+_{kjG^+}\,B_0(m_{\chi^0_k},m_W)
-4\,\tilde b^+_{kiW}\,\tilde a^+_{kjW}\,
\left[B_0(m_{\chi^0_k},m_W)+\delta_R\right]
\biggr\}\non\\
&+& \sum_{k=1}^2 \,m_{\chi^+_k}\,\biggr\{
\tilde b^+_{ikG^0}\,\tilde a^+_{jkG^0}\,B_0(m_{\chi^+_k},m_Z)
+ \tilde b^+_{ikH}\,\tilde a^+_{jkH}\,B_0(m_{\chi^+_k},m_H)\non\\
&&~~~~~~~~~
-4\,\tilde b^+_{ikZ}\,\tilde a^+_{jkZ}\,
\left[B_0(m_{\chi^+_k},m_Z)+\delta_R\right]
-4\,\tilde b^+_{ik\gamma}\,\tilde a^+_{jk\gamma}\,
\left[B_0(m_{\chi^+_k},0)+\delta_R\right]
\biggr\}~,\non\\
\label{sigma+S}
\eeq
where $\delta_R=0$ in $\drbar$ and $\delta_R=-1/2$ in $\msbar$.  The
formulae for the self-energies $\Sigma^0_R$ and $\Sigma^+_R$ can be
obtained from those for $\Sigma^0_L$ and $\Sigma^+_L$, respectively,
by replacing $\tilde a \rightarrow \tilde b$ in eqs.~(\ref{sigma0L}) and
(\ref{sigma+L}). In the equations above we denote by $\tilde a_{ij\Phi}$
and $\tilde b_{ij\Phi}$ the couplings of a bosonic field $\Phi =
(Z,W,\gamma,H,G^0,G^+)$ with one rotated neutralino (or chargino) mass
eigenstate and one unrotated neutralino (or chargino) gauge
eigenstate.\s

%
In particular, the couplings of an unrotated neutralino $\psi^0_i$, a
rotated chargino $\chi^+_j$ and a charged pseudo-Goldstone boson read
\beq
\tilde a^0_{ijG^+} &=& U_{jk}\,\left(
\frac{\tilde g_d^\prime}{\sqrt 2}\,\delta_{i1}\delta_{k2}
+\frac{\tilde g_d}{\sqrt 2}\,\delta_{i2}\delta_{k2}
-\tilde g_d \,\delta_{i3}\delta_{k1}\right),\\
\tilde b^0_{ijG^+} &=& V_{jk}\,\left(
\frac{\tilde g_u^\prime}{\sqrt 2}\,\delta_{i1}\delta_{k2}
+\frac{\tilde g_u}{\sqrt 2}\,\delta_{i2}\delta_{k2}
+\tilde g_u \,\delta_{i4}\delta_{k1}\right).
\eeq
The couplings of an unrotated neutralino $\psi^0_i$, a rotated
chargino $\chi^+_j$ and a W boson read
\beq
\tilde a^0_{ijW} &=& g \, V_{jk}\,\left(
-\delta_{i2}\delta_{k1}+\frac{1}{\sqrt 2}\,\delta_{i4}\delta_{k2}\right),\\
\tilde b^0_{ijW} &=& g \, U_{jk}\,\left(
-\delta_{i2}\delta_{k1}-\frac{1}{\sqrt 2}\,\delta_{i3}\delta_{k2}
\right).
\eeq
The couplings of an unrotated neutralino $\psi^0_i$, a rotated
neutralino $\chi^0_j$ and a neutral pseudo-Goldstone boson read
\be
\tilde a_{ijG^0}^0 ~=~ -\tilde b_{ijG^0}^0 ~=~ \frac12 \,N_{jk}
\biggr( - \tilde g_d^\prime\, \delta_{\{i1}\delta_{k 3\}}
- \tilde g_u^\prime\, \delta_{\{i1}\delta_{k 4\}}
+ \tilde g_d\, \delta_{\{i2}\delta_{k 3\}}
+ \tilde g_u\, \delta_{\{i2}\delta_{k 4\}}\biggr).
\ee
The couplings of an unrotated neutralino $\psi^0_i$, a rotated
neutralino $\chi^0_j$ and a Higgs boson read
\be
\tilde a_{ijH}^0 ~=~ ~~~\tilde b_{ijH}^0 ~=~ \frac12 \,N_{jk}
\biggr( - \tilde g_d^\prime\, \delta_{\{i1}\delta_{k 3\}}
+ \tilde g_u^\prime\, \delta_{\{i1}\delta_{k 4\}}
+ \tilde g_d\, \delta_{\{i2}\delta_{k 3\}}
- \tilde g_u\, \delta_{\{i2}\delta_{k 4\}}\biggr).
\ee
The couplings of an unrotated neutralino $\psi^0_i$, a rotated
neutralino $\chi^0_j$ and a $Z$ boson read
\be
\tilde a_{ijZ}^0 ~=~ -\tilde b_{ijZ}^0 ~=~ \frac{g}{2\,\hat c}\, N_{jk}
\biggr(\delta_{i3}\delta_{k 3} - \delta_{i4}\delta_{k 4}\biggr).
\ee
%
%
The couplings of a rotated neutralino $\chi^0_i$, an unrotated
chargino $\psi^+_j$ and a charged pseudo-Goldstone boson read
\beq
\tilde a^+_{ijG^+} &=& N_{ik}\,\left(
\frac{\tilde g_u^\prime}{\sqrt 2}\,\delta_{k1}\delta_{j2}
+\frac{\tilde g_u}{\sqrt 2}\,\delta_{k2}\delta_{j2}
+\tilde g_u \,\delta_{k4}\delta_{j1}
\right),\\
\tilde b^+_{ijG^+} &=& N_{ik}\,\left(
\frac{\tilde g_d^\prime}{\sqrt 2}\,\delta_{k1}\delta_{j2}
+\frac{\tilde g_d}{\sqrt 2}\,\delta_{k2}\delta_{j2}
-\tilde g_d \,\delta_{k3}\delta_{j1}\right).
\eeq
The couplings of a rotated neutralino $\chi^0_i$, an unrotated
chargino $\psi^+_j$ and a W boson read
\beq
\tilde a^+_{ijW} &=& g \, N_{ik}\,\left(
-\delta_{k2}\delta_{j1}+\frac{1}{\sqrt 2}\,\delta_{k4}\delta_{j2}\right),\\
\tilde b^+_{ijW} &=& g \, N_{ik}\,\left(
-\delta_{k2}\delta_{j1}-\frac{1}{\sqrt 2}\,\delta_{k3}\delta_{j2}
\right).
\eeq
The couplings of an unrotated chargino $\psi^+_i$, a rotated
chargino $\chi^+_j$ and a neutral pseudo-Goldstone boson read
\beq
\tilde a_{ijG^0}^+ &=& \frac 1{\sqrt2}\,U_{jk}
\biggr( \tilde g_d\, \delta_{i1}\delta_{k2}
- \tilde g_u\, \delta_{i2}\delta_{k1}\biggr),\\
\tilde b_{ijG^0}^+ &=& \frac 1{\sqrt2}\,V_{jk}
\biggr( \tilde g_u\, \delta_{i1}\delta_{k2}
- \tilde g_d\, \delta_{i2}\delta_{k1}\biggr).
\eeq
The couplings of an unrotated chargino $\psi^+_i$, a rotated
chargino $\chi^+_j$ and a Higgs boson read
\beq
\tilde a_{ijH}^+ &=& \frac 1{\sqrt2}\,U_{jk}
\biggr( \tilde g_d\, \delta_{i1}\delta_{k2}
+ \tilde g_u\, \delta_{i2}\delta_{k1}\biggr),\\
\tilde b_{ijH}^+ &=& \frac 1{\sqrt2}\,V_{jk}
\biggr( \tilde g_u\, \delta_{i1}\delta_{k2}
+ \tilde g_d\, \delta_{i2}\delta_{k1}\biggr).
\eeq
The couplings of an unrotated chargino $\psi^+_i$, a rotated
chargino $\chi^+_j$ and a $Z$ boson read
\beq
\tilde a_{ijZ}^+ &=& g\,\hat c \, V_{jk}
\biggr(\delta_{i1}\delta_{k 1} + \frac{\hat c^2-\hat s^2}{2\,\hat c^2}
\delta_{i2}\delta_{k 2}\biggr),\\
\tilde b_{ijZ}^+ &=& g\,\hat c \, U_{jk}
\biggr(\delta_{i1}\delta_{k 1} + \frac{\hat c^2-\hat s^2}{2\,\hat c^2}
\delta_{i2}\delta_{k 2}\biggr).
\eeq
Finally, the couplings of an unrotated chargino $\psi^+_i$, a rotated
chargino $\chi^+_j$ and a photon read
\beq
\tilde a_{ij\gamma}^+ &=& e \, V_{jk}
\biggr(\delta_{i1}\delta_{k 1} + \delta_{i2}\delta_{k 2}\biggr),\\
\tilde b_{ij\gamma}^+ &=& e \, U_{jk}
\biggr(\delta_{i1}\delta_{k 1} + \delta_{i2}\delta_{k 2}\biggr).
\eeq

The self-energies in eqs.~(\ref{loopneut})--(\ref{loopchar}) induce an
external momentum dependence in the neutralino and chargino mass
matrices. We compute each of the physical neutralino and chargino
masses by diagonalizing the one-loop mass matrix with the external
momenta in the Passarino-Veltman functions set equal to the tree-level
mass of the corresponding particle.

\subsubsection*{More one-loop translations from $\drbar$ to $\msbar$}

Beyond tree level the boundary conditions on the quartic Higgs
coupling and on the Higgs--higgsino--gaugino couplings given in
eqs.~(\ref{boundlam})--(\ref{boundgt}) are valid only in the
$\overline{\rm DR}$ renormalization scheme. The $\msbar$ scheme breaks
supersymmetry, therefore in that scheme the SUSY relations between the
gauge couplings and the gaugino couplings and those between the gauge
couplings and the quartic scalar couplings are not preserved
\cite{mv}. In the $\msbar$ scheme the boundary conditions in
eqs.~(\ref{boundlam})--(\ref{boundgt}) are modified as~\footnote{{\bf
    Note Added:} in eq.~(\ref{lambdaMS}) the SU(2) gauge coupling
  $g(\msusy )$ entering the tree-level part of the boundary condition
  on $\lambda$ has to be interpreted as the $\drbar$
  coupling. By contrast, in eqs.~(\ref{guMS}) and (\ref{gdMS})
  $g(\msusy )$ has to be interpreted as the $\msbar$ coupling. If the
  coupling $g(\msusy )$ in the tree-level boundary condition on
  $\lambda$ is expressed in the $\msbar$ scheme, there is an additional
  shift in eq.~(\ref{lambdaMS}) amounting to
  $+\,g^4\cos^22\beta/(96\pi^2)\,$.}
\begin{eqnarray}
\lambda(\msusy) &=& \frac{1}{4}\left[ g^2(\msusy )+g^{\prime 2}(\msusy )
\right] \,\cos^22\beta
- \frac{g^4}{32\,\pi^2}\,\left( 1 + \frac{1}{2\,c^4}\right)
~ + \Delta_{\rm th} \lambda~, \label{lambdaMS}\\\nonumber\\
\gtilu (\msusy )&=& g (\msusy )\sin\beta\,
\left[1+\frac{g^2}{16\,\pi^2}\left(\frac{13}{12}-\frac{1}{8\,c^2}\right)\right]
~, \label{guMS}\\\nonumber\\
\gtild (\msusy ) &=& g (\msusy )\cos\beta\,
\left[1+\frac{g^2}{16\,\pi^2}\left(\frac{13}{12}-\frac{1}{8\,c^2}\right)\right]
~, \label{gdMS}\\\nonumber\\
\gtilup (\msusy ) &=& g^{\,\prime} (\msusy ) \sin\beta
\left[1-\frac{g^2}{16\,\pi^2}\left(\frac{1}{4}+\frac{1}{8\,c^2}\right)\right]
~,\\ \nonumber\\
\gtildp (\msusy )&=& g^{\,\prime} (\msusy )\cos\beta
\left[1-\frac{g^2}{16\,\pi^2}\left(\frac{1}{4}+\frac{1}{8\,c^2}\right)\right]
~,
\end{eqnarray}
where $\Delta_{\rm th} \lambda$ is the ${\cal O}(h_t^4)$ threshold correction
defined in eq.~(\ref{threshold}). \s

For completeness we conclude this section by providing the relations
between the $\drbar$ and $\msbar$ definitions of the gaugino and
higgsino mass terms:
\begin{eqnarray}
M_1^{\msbar} &=& M_1^{\drbar}~,\\
\nonumber\\
M_2^{\msbar} &=& M_2^{\drbar}\,\left[1+\frac{g^2}{8\,\pi^2}\right]~,\\
\nonumber\\
M_3^{\msbar} &=& M_3^{\drbar}\,\left[1+\frac{3\,g_s^2}{16\,\pi^2}\right]~,\\
\nonumber\\
\mu^{\msbar} &=& \mu^{\drbar}\,\left[1+\frac{g^2}{16\,\pi^2}
\,\left(\frac12 + \frac{1}{4\,c^2}\right)\right]~.
\end{eqnarray}
 
\end{appendletterA}

\end{document}